\documentclass[aps,pre,twocolumn,superscriptaddress,floats,floatfix,showpacs]{revtex4-1}
\usepackage{amssymb,amsmath}
\usepackage{physics}
\usepackage{mathrsfs}

\usepackage{float}
\usepackage{graphics}
\usepackage{subfigure}
\usepackage{epsfig}
\usepackage{rotating}
\usepackage{anyfontsize}
\usepackage{verbatim}
\usepackage{standalone}
\usepackage{hyperref}
\graphicspath{ {figures/} }

\newcommand{\floor}[1]{\lfloor #1 \rfloor}

\usepackage{xcolor}
\usepackage{color}

\usepackage{xr}

\begin{document}

\title{Anisotropy and multifractal analysis of turbulent velocity and temperature in the roughness sublayer of a forested canopy}
%

\author{Soumak Bhattacharjee}
    \email{soumakb@iisc.ac.in}
    \affiliation{Department of Physics,
Indian Institute of Science, Bangalore 560012, India.}
\author{Rahul Pandit}
    \affiliation{Centre for Condensed Matter Theory, Department of Physics,
Indian Institute of Science, Bangalore 560012, India.}
\author{Timo Vesala}
    \affiliation{Institute for Atmospheric and Earth System Research/Forest Sciences, Faculty of Agriculture and Forestry, University of Helsinki, Finland.}
    \affiliation{Institute for Atmospheric and Earth System Research/Physics, Faculty of Science, University of Helsinki, Finland.}
\author{Ivan Mammarella}
    \affiliation{Institute for Atmospheric and Earth System Research/Physics, Faculty of Science, University of Helsinki, Finland.}
\author{Gabriel Katul}
    \affiliation{Nicholas School of the Environment, Duke University, Durham, NC, USA.}
\author{Ganapati Sahoo}
    \email{ganapati.sahoo@gmail.com}
    \affiliation{Institute for Atmospheric and Earth System Research/Physics, Faculty of Science, University of Helsinki, Finland.}
    \affiliation{Department of Mathematics and Statistics, Faculty of Science, University of Helsinki, Finland.}

\begin{abstract} 
Anisotropy and multifractality in velocity and temperature time series
sampled at multiple heights in the roughness sublayer (RSL) over a boreal
mixed-coniferous forest are reported. In particular, a turbulent-stress
invariant analysis along with a scalewise version of it are conducted to
elucidate the nature of relaxation of large-scale anisotropy to quasi-isotropic
states at small scales. As the return to isotropy is linked to nonlinear
interactions and correlations between different fluctuating velocity components
across scales, we study the velocity and temperature time series by using
multifractal detrended fluctuation analysis and multiscale multifractal
analysis to assess the effects of thermal stratification and surface roughness
on turbulence in the RSL. The findings are compared so as to quantify the
anisotropy and multifractality ubiquitous to RSL turbulent flow. As we go up
in the RSL, (a) the length scale at which return to isotropy commences
increases because of the weakening of the surface effects and (b) the largest
scales become increasingly anisotropic. The anisotropy in multifractal
exponents for the velocity fluctuations is diminished when we use the
extended-self-similarity procedure to extract the multifractal-exponent ratios.   
\end{abstract}

\maketitle

\section{\label{sec:Introduction}Introduction}
The characterization of the statistical properties of boundary-layer turbulence
continues to be a challenging problem in a variety of areas including
geophysical fluid dynamics and micro-meteorology. Atmospheric boundary layers
provide natural laboratories for such characterization at high Reynolds number
that remain outside the reach of direct numerical simulations and laboratory
experiments. Because of the high Reynolds number expected, the statistical
properties of the flow depend, to some extent, on the surface
topography~\cite{PhysRevFluidsSurfTopo,chen2019effects,poggi2008turbulent},
vegetation cover~\cite{TurbPlantsI}, and thermal stratification~\cite{GK2011}.
Boundary layers over sand grains ~\cite{KurienSreenivasan,Brugger2018} can be
different from those above forested canopies~\cite{Brugger2018,TurbPlantsI} in
numerous ways that remain to be uncovered. It is important to understand such
differences because they have a bearing on diverse applications in a variety of
areas including wind-energy production~\cite{Ummels2007,energyproductionii},
air-pollution control~\cite{Demirci2000}, meteorological
research~\cite{Tesfuhuney2013}, civil engineering~\cite{Cermak2003}, and
ecological applications dealing with pollen and seed
spread~\cite{nathan2005foliage}. 

One area where atmospheric flows are contributing to a fundamental
understanding of turbulence is flow near roughness elements (hereafter labelled
the roughness sublayer or RSL), especially forests.  The reason flow in the RSL
over forests can enlighten basic turbulence research is the scale separation
characterizing the spectrum of eddies.  Atmospheric boundary layers over
forests are characterized by (i) large boundary depth (exceeding 1000 m), (ii)
canopy heights that readily exceed 10 m, (iii) Kolmogorov dissipation length
scales that are smaller than 1 mm.  Another aspect of RSL flows that can be
interrogated is thermal stratification.  Turbulence in the RSL involves the
interaction between mechanical generation of turbulent kinetic energy (TKE) and
buoyancy forces that can produce or dissipate TKE.  Hence, the role of thermal
stratification cannot be ignored when analyzing the structure of turbulence in
the RSL. 

It is known that the co-spectrum of the vertical velocity and the air
temperature is finite in the inertial subrange and exhibits a $-7/3$ scaling
range, suggesting that a finite interaction must exist between velocity and
temperature at inertial scales, contrary to expectations based on isotropic
turbulence. This finite co-spectrum is primarily because of finite mean
temperature gradients. Likewise, a finite co-spectrum of vertical and
longitudinal velocity components, because of finite mean velocity gradients,
implies a sustained interaction between the vertical velocity and longitudinal
velocities, within the inertial subrange, contrary to expectations from
isotropic turbulence.

It is also well established that ramp-cliff patterns lead to dissimilarity in
the multifractal properties of temperature (or any scalar) and the velocity
~\cite{Sreenivasan1979, Warhaft2000}; we refer the reader to
Ref.~\cite{Katul2006} for a comprehensive review on ramp-cliff patterns in
laboratory and field experiments, under different surface-roughness
conditions.

Return to isotropy, with decreasing length scales, deals primarily with energy
redistribution, among velocity components, and the destruction of turbulent
stress. Hence, return to isotropy has been studied by using second-order
statistics. Mindful of these findings and scalewise interactions between
vertical velocity, temperature, and longitudinal velocity, we seek to examine
to what extent scalewise return to isotropy, characterized by using the
two-point measures discussed below, is retarded by (a) excess intermittency in
the temperature or (b) multifractality in high-order velocity structure
functions. The focus of our study is on production-to-inertial length scales,
which is where the signature of TKE production and its relaxation to isotropy
are most apparent.

We carry out a detailed investigation of the statistical properties of
velocities and air temperatures at two different heights in the RSL above the
Hyyti\"al\"a forest in Southern Finland. The statistical properties considered
are (a) energy spectra, (b) second-order velocity structure functions, (c)
measures of flow anisotropy that use invariant analysis  ~\cite{LUMLEY1979123}
in the anisotropy-invariant-map (AIM) and the barycentric-map (BAM) frameworks
~\cite{Banerjee2007}, and (d) multifractal detrended fluctuation analysis
(MFDFA), multifractal spectra, and multiscale multifractal analysis (MMA) for
the temperature and velocity fields.  These properties are all examined for
different atmospheric stability conditions, which are presumed to be controlled
by the stability parameter $\zeta$ (Eq. \ref{eq:zeta_def}). These
different characterization tools have not been brought to bear simultaneously
on the analysis of turbulent flows in the RSL. Our overarching goal is the
quantification of the effects of surface roughness and stratification on the
nature of anisotropic fluctuations and multifractal correlations in such
turbulent flows. By using invariant analysis and multifractal analysis, the
effects of stratification on the degree and relaxation of large-scale
anisotropy at different heights in the roughness sublayer above the forest
cover are quantified.

The remaining part of this manuscript is organized as follows:  In
Sec.~\ref{sec:Experiments}, a brief description of the experiment, its
location, and the setup is provided. In Sec.~\ref{sec:Methods}, the techniques
and statistical measures that are used to characterize the flow anisotropy and
the multifractality of the velocity and temperature field are reviewed. In
Sec.~\ref{Results}, the connection between the overarching goal and the
application of the aforementioned methods is presented.
Section~\ref{Conclusions} is devoted to concluding remarks about the
significance of the results for future models of the RSL. 

\section{\label{sec:Experiments}Experiments}

\subsection{Research site and measurements}

\begin{figure}
\includegraphics[width =\columnwidth]{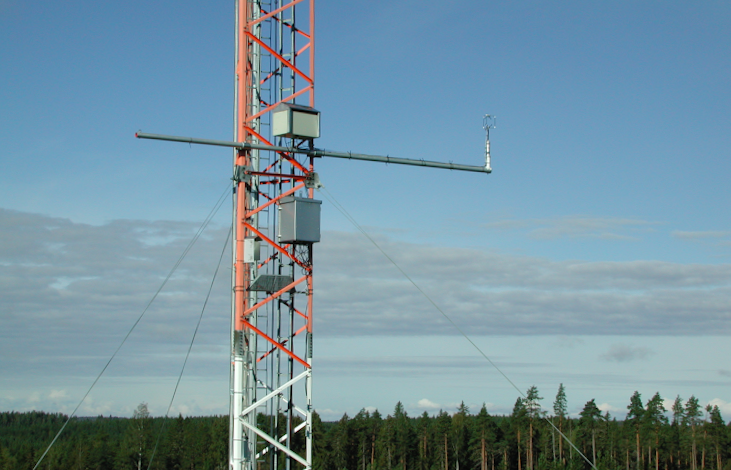}
\caption{The eddy-covariance tower in Hyyti\"al\"a forest. This tower is
equipped with three-dimensional (3D) sonic anemometers that measure all
velocity components and air temperature at high frequency (10 Hz).}
\label{fig:ec_schematic}
\end{figure}

The data sets were collected at the SMEAR (Station for Measuring
Ecosystem-Atmosphere Relations) - II station located within the Hyyti\"al\"a
forest in Southern Finland ($61^{\circ}51'$ N, $24^{\circ}17'$ E – $180$ m
above sea level). The station is equipped with a tall meteorological tower
where turbulence statistics are collected as part of a long-term monitoring
initiative. In our study, we focus on data sampled at two heights ($23.3$ m
and $33.0$ m above ground level) in the RSL, roughly $4$ m and $14$ m above
the canopy top.  The mean canopy height is approximately $h_{\rm canopy} \simeq
19$ m. The tower is located at the highest spot in the area ($181$ m above
sea level). In 1962, the area was regenerated by clear-cutting and sowing
Scots pine seeds. The ecosystem that the SMEAR II station is surrounded by is a
typical boreal forest, which is dominantly Scots pine with some spruce and
birch patches further away from the measurement tower (Fig.
\ref{fig:ec_schematic}). A notable spruce forest patch is located some $180$ m
south of the tower.

The wind velocity components and air temperature have been measured by
three-dimensional sonic anemometers from Gill Instruments Ltd. (Lymington,
Hampshire, England). For the $23.3$ m measurement set-up, the ultrasonic
anemometer model is Solent Research 1012 R2 (sampling frequency at $10.41$ Hz),
and for the $33.0$ m set-up, a Gill HS-50 ultrasonic anemometer (sampling
frequency at $10$ Hz) is in use.

\subsection{\label{sec:Postprocessing}Post-processing}

The measured time series of the 3 velocity components and the air temperature
are divided into non-overlapping 30-minute segments.  For each 30-minute
segment, the coordinate system used assumes $u_1(=u)$, $u_2(=v)$, and $u_3(=w)$
are the longitudinal (or streamwise), lateral (or spanwise), and vertical (or
wall-normal) velocity components aligned along $x_1$ (or $x$), $x_2$ (or $y$),
and $x_3$ (or $z$) with $z=0$ being the forest floor and $x$ aligned along the
mean wind direction.  Primed quantities are turbulent fluctuations around their
mean state defined by time-averaging over 30 minutes.  Throughout, overline
represents time-averaged quantities.  A standard double rotation was performed
to align the velocity vector along the mean-flow direction resulting in
$\overline{v}=\overline{w}=0$.  For notational convenience, $\overline{u}=U$
defines the mean longitudinal velocity.  A successive-difference-based
despiking scheme~\cite{ECBook} with a threshold of $10$ m/s for streamwise and
spanwise components, $5$ m/s for the wall-normal component of the velocity, and
$5$ K for the air temperature was then used to identify anomalous spikes. We
have not considered segments that display more than $5\%$ spikes in a given
30-minute. 

For the invariant and multifractal analyses, we have used data from these
30-minute segments. Each of these 30-minute periods is classified as stable,
unstable, or near-neutral using the atmospheric stability parameter $\zeta$:
\begin{eqnarray}\label{eq:zeta_def}
\zeta &=& \frac{z-z_d}{L_{obu}}, \quad L_{obu} = -\frac{u_{\ast}^3}{\kappa g\: \overline{w'\theta'_v}/\bar{\theta}_v},\nonumber  \\
u_{\ast}^2 &=& \sqrt{(\overline{u'w'})^2+(\overline{v'w'})^2}, \quad z_d =  \frac{2}{3} h_{\rm canopy}, 
\end{eqnarray}
where $L_{obu}$, $u_{\ast}$, $\kappa$, $g$, and $z_d$ denote the Obukhov
length, the friction velocity, the von K\'arm\'an constant, the gravitational
acceleration, and the estimated zero-displacement height, respectively (Table
\ref{table:all_params} for details). The $\theta'_v$ denotes the potential
temperature fluctuations; $\bar{\theta}_v$, $\overline{u'w'}$,
$\overline{v'w'}$, $\overline{w'\theta'_v}$ denote the mean virtual potential
temperature, the vertical component of the turbulent fluxes of horizontal
momenta and potential temperature, respectively, with temporal averaging,
indicated by an overline, performed over each 30-minute period; periods with
$\zeta > 0.01$, $\zeta < -0.01$ and $|\zeta| \leq 0.01$ are classified as
stable, unstable, and near-neutral, respectively. In data set presented here,
there are $771,\,733$, and $233$ $30$-minute periods of stable, unstable, and
neutral stratification at $z_{\rm bot}$; whereas for $z_{\rm top}$ we have
$956,\,683$, and $98$ $30$-minute periods of stable, unstable and neutral
stratification.  
\begin{table*}[ht]
\caption{Definitions and ensemble-averaged values across all runs and stability
conditions of measured and calculated variables: $\overline{a'b'}$ denotes
covariance between variables $a$ and $b$; $\kappa = 0.4$ and $g = 9.8$ ${\rm ms}^{-2}$
denote the von K\'arm\'an and the gravitational acceleration constants,
respectively;
$u_{\ast}=\left[{\overline{u'w'}^2+\overline{v'w'}^2}\right]^{1/4} \simeq
0.44$ ${\rm ms}^{-1}$ (averaged across all runs and the two heights),
$\overline{u'_iu'_i}=2k$ are the local friction velocity and twice the
turbulent kinetic energy ($=k$), respectively. The ensemble-averaged values are
featured only to illustrate order-of-magnitude estimates of key flow properties
for the entire experiment.  All calculations in the figures are based on
run-to-run flow statistics. \label{table:all_params}}
\small
\begin{adjustbox}{width=0.9\textwidth}
{
\renewcommand{\arraystretch}{1.0}
\begin{tabular}{llllll}\hline\hline
Variable        & Symbol         &Units         & Definition                                 & $z_{\rm bot}=23.3$m         & $z_{\rm top}=33$m  \\ \hline
Sampling frequency  & $f_s$      &  Hz        &                                            & 10.41                       & 10       \\  
Mean velocity     & $\bar{U}$  & m/s    &                                            & $2.39$  	                   & $3.15$ \\  
r.m.s velocity  &$\bar{u}_{\rm rms}$& m/s     & $ \qty({2 k}/3)^{1/2}$         &$0.81$  	                  &     $0.83$\\ 
energy dissipation rate & ${\epsilon}$ & ${\rm m}^2/{\rm s}^{3}$   & $\bar{u}^3_{\ast}\qty(\kappa (z-z_d))^{-1}$   & $4.24\times 10^{-2}$     & $2.16\times 10^{-2}$\\ 
Reynolds number & Re            &               & ${4 k^2}(9\epsilon \nu)^{-1}$      & $1.54\times 10^{6}$  	  &     $6.08\times 10^{6}$ \\ 
K41 length scale& $\eta$        & m           & $({\nu^3}/{\epsilon})^{1/4}$            & $7.91\times 10^{-4}$  	  &     $1.02\times 10^{-3}$\\ 
K41 time scale  & $\tau_\eta$   & s           & $({\nu}/{\epsilon})^{1/2}$              & $5.92\times 10^{-2}$  	  &     $1.01\times 10^{-1}$\\ 
Strain rate     & $S(z_{\rm bot})$     & ${\rm s}^{-1}$      & $\bar{U}_{23}(z_{\rm bot}-z_d)^{-1}$                   & $0.22$     &  \\ 
                & $S(z_{\rm top})$    & ${\rm s}^{-1}$      & $(\bar{U}_{33}-\bar{U}_{23})(z_{\rm top}-z_{\rm bot})^{-1}$       &                           & $7.85\times 10^{-2}$ \\ 
Shear parameter & $S^{\ast}$       &               & $ 2kS/\epsilon $               & $38.20$                    &     $48.17$ \\ 
                & $S_c^{\ast}$     &               & $ S\tau_{\eta}$                            & $9.84\times 10^{-3}$  	  &     $7.52\times 10^{-3}$  \\  
Integral length scales & $L_{u}$& m           & $ \bar{U}\int \rho_{uu}(\tau) \dd \tau$    & $33.1$  	                  &     $60.0$ \\  
         & $L_{v}$              & m           & $ \bar{U}\int \rho_{vv}(\tau) \dd \tau$    & $26.2$  	                  &     $54.1$ \\  
         & $L_{w}$              & m           & $ \bar{U}\int \rho_{ww}(\tau) \dd \tau$    & $6.74$  	                  &     $9.54$ \\  
         & $L_{T}$              & m           & $ \bar{U}\int \rho_{TT}(\tau) \dd \tau$    & $41.2$  	                  &     $72.3$ \\  
         & $ r_{\rm iso}/L_w $      &               & Eq. (\ref{eq:L_iso_def})                 & $ 0.71 $                   & $ 1.00 $  \\  
         & $ r_{\rm ani}/L_w $      &               & Eq. (\ref{eq:L_ani_def})                 & $ 18.56 $                  & $ 22.56 $  \\  \hline
\hline
\end{tabular}
}
\end{adjustbox}
\end{table*}

\section{\label{sec:Methods}Methods}

Kolmogorov's 1941 (K41) phenomenological theory~\cite{K41, Frisch96} of
statistically homogeneous, isotropic turbulence (assumed for small scales) does
not account for multifractality; and, of course, anisotropy in eddy sizes.  The
local-isotropy assumption of K41 implies that the signature of TKE generation
is lost as we small length scales in the energy cascade.  Furthermore, the K41
exponents for second-order structure functions are affected very marginally by
(a) deviations from local isotropy and (b) multifractality. It remains open to
what extent the effects of TKE generation are truly lost in the scalewise
turbulent cascade of each velocity component.  Mathematically, the scalewise
production term of TKE can be expressed in terms of the co-spectra and is given
by: $-E_{uw}(\mathsf{k}) \dd \bar{U}/\dd z +(g/T) E_{wT}(\mathsf{k})$, where $T$ denotes the air-temperature and $\mathsf{k}$ denotes the wave-number.\\ So, the strength of the
mean-velocity gradient $\dd \bar{U}/\dd z$ in the RSL and the shape of the
co-spectra of both $E_{uw}(\mathsf{k})$ and $E_{wT}(\mathsf{k})$, within the inertial subrange,
are likely to have an effect on the return to isotropy.  Differential
intermittency buildup, within the velocity components, entirely ignored by K41,
can slow the relaxation to local isotropy, in the inertial range.  For this
reason, we first discuss the spectra, co-spectra, and second-order structure
functions, prior to the scalewise return to isotropy. In the first two
sub-sections, an overview of the methods, which are used to characterize the
anisotropy of the flows, are described. The third sub-section features the
techniques with which multifractality~\cite{Frisch96, ParisiFrisch, Benzi84,
Halsey86, Argoul89, MenSreeni91, Boffetta08} is analyzed. 

\subsection{\label{sec:Invariant_Analysis}Invariant Analysis}

The anisotropy of the Reynolds stress tensor and the asymmetric distribution of
the turbulent kinetic energy among the different spatial components of the
velocity can be traced to stratification and surface roughness; these issues
have been explored, e.g., in
Refs.~\cite{Brugger2018},~\cite{Smalley2002},~\cite{Liu2017}. The relaxation of
anisotropy, at large scales, to quasi-isotropy, at the small scales (as
hypothesized in K41), e.g., via asymmetric correlations between the pressure
and the rate-of-strain tensor, remains an active research
area~\cite{Sarkar1990, Panda2017}. Such anisotropy can be quantified and
visualized by using the turbulent-stress invariant
analysis~\cite{LUMLEY1979123} and the scalewise invariant analysis of the
traceless deviator of the second-order velocity structure
function~\cite{Brugger2018}.

 In this invariant analysis, the invariants of the Reynolds stress anisotropy
tensor are determined from ~\cite{Banerjee2007}
\begin{equation}\label{eq:aij}
    a_{ij} = \frac{\overline{u'_iu'_j}}{2k} - \frac{1}{3} \delta_{ij} \text{ and } \quad k = \frac{\overline{u'_iu'_i}}{2}.
\end{equation}

For the scalewise invariant analysis, the traceless deviator of the velocity
structure-function tensor $D_{ij}(r) = \frac{1}{2} \overline{\delta u_i(r)
\delta u_j (r)} $ is used and is given by \cite{Brugger2018}:
\begin{align}\label{eq:aijr}
    A_{ij}(r) &= \frac{D_{ij}(r)}{D_{kk}(r)} - \frac{1}{3}\delta_{ij},
\end{align}
where $r$ is the separation distance (a surrogate for the eddy size) that may
be inferred from Taylor's frozen turbulence hypothesis. 

If the turbulence is statistically homogeneous and isotropic, $a_{ij}$ and
$A_{ij}$ are identically zero. Therefore, any deviation from these isotropic
forms for these tensors can be used to quantify the anisotropy of a turbulent
flow. Indeed, such invariant analysis~\cite{ Banerjee2007, Schmidt2007} has
been used extensively to describe the various limiting states in turbulent
flows and the route to the relaxation towards isotropy~\cite{ Sarkar1990,
Brugger2018,lumley1977}.

The state of turbulence and the relaxation to quasi-isotropic states at small
scales can be effectively visualized through the trajectories of the second and
the third invariants, $I_2$ and
$I_3$, respectively, (first invariant $I_1 = 0$ by construction). These are given by~\cite{lumley1977}:
\begin{eqnarray}\label{eq:i23l123}
    I_1 &=& \lambda_1+\lambda_2+\lambda_3, I_2 =  \lambda_1 \lambda_2 + \lambda_2 \lambda_3 + \lambda_3 \lambda_1, \nonumber \\
    I_3 &=& \lambda_1\lambda_2\lambda_3;  
    \lambda_i \geq \lambda_j,\textbf{ } i \leq j,\, i,j = 1,2,3,
\end{eqnarray}

where $\lambda_i, \, i = 1, 2, 3$ are the ordered eigenvalues of the tensors mentioned above.
This has been used extensively to visualize the flow
anisotropy in shear and wall-bounded flows~\cite{antonia1991, Krogstad2000}. 

\subsection{\label{sec:Anisotropy_Measures}Anisotropy Measures}

\begin{enumerate}

\item BAM \emph{framework}: In the BAM \emph{framework}, the measure of
flow anisotropy $C_{\rm iso}$ and scalewise anisotropy $C_{\rm iso}(r)$ are
defined from the eigenvalues $\lambda_i,\, i = 1,2,3$:
\begin{equation}\label{eq:cisodef}
C_{\rm iso} = 1 + 3 \lambda_3,\, C_{\rm iso}(r) = 1 + 3 \lambda_3(r).
\end{equation}
    
\item AIM \emph{framework}: In the AIM \emph{framework}, the measure of
flow anisotropy $F$ and scalewise anisotropy $F(r)$ are defined in terms of
the invariants  $I_2,I_3$ (Eq. (\ref{eq:i23l123})) as follows:
\begin{equation}\label{eq:fdef}
F = 1 + 27 I_3 + 9 I_2,\, F(r) = 1 + 27 I_3(r) + 9 I_2(r).
\end{equation}

\end{enumerate}

In the case of statistically homogeneous and isotropic
turbulence~\cite{Banerjee2007}, the eigenvalues and the invariants $I_2$ and
$I_3$, for $a_{ij}$ are identically $0$, so both the anisotropy measures
$C_{\rm iso}$ and $F$  are $1$.

\subsection{Correlation-significance testing}

Given two random variables $A$ and $B$, whose normalized fluctuations are
$A_n=A'/\sigma_A$ and $B_n=B'/\sigma_B$, the Pearson's correlation coefficient
$\rho(A,B)=\overline{A_nB_n}$. Given $\rho(A,B)$, the most conventional
significance testing relies on the $t$-statistic 
\begin{equation}\label{eq:t(A,B)_def}
t(A,B) = \frac{\sqrt{n-2}}{\sqrt{1-\rho(A,B)^2}} \lvert \rho(A,B) \lvert,
\end{equation}
where $n$ is the sample size.  This statistic is compared with the critical
value $t_{\alpha/2,n-2}$ of a $t$-distribution with a level of significance of
$\alpha/2$ and $n-2$ degrees of freedom. Here, $\alpha$ is the chosen
significance level for this test. The variables $A$ and $B$ are considered
statistically independent when $t(A,B) <t_{\alpha/2,n-2}$.

\subsection{Multifractal Analysis}

Multifractality and multiscaling corrections to K41 scaling in turbulence were
introduced by Frisch and Parisi~\cite{Frisch96, ParisiFrisch, Benzi84} and
studied extensively thereafter (see, e.g., Refs.~\cite{Frisch96, Halsey86,
Argoul89, MenSreeni91, Boffetta08}) including in the atmospheric surface layer
\cite{Katul2001,Katul2009,shi2005assessing}. Since then, multifractal
techniques have been used not only in turbulence (for recent trends see, e.g.,
Refs.~\cite{Boffetta08, Arneodo08, Ray08, Pandit2009, nairita2016}), but also
in diverse fields including atmospheric science~\cite{Zeng2016}, DNA
sequencing~\cite{DNAseq}, heart-rate dynamics~\cite{Gieraltowski2012,
Ivanov1999}, cloud structure~\cite{cloudmfdfa}, economics~\cite{ecophy1999},
and many parts of physics (see, e.g., Refs.~\cite{Amin, Sutradhar19}).  

The multifractal analysis adopted here studies fluctuations of different
non-stationary fields by using two extensions of the detrended fluctuation
analysis techniques, namely, Multifractal Detrended Fluctuation Analysis
(MFDFA) and Multiscale Multifractal Analysis (MMA). These techniques are
briefly described.  The code developed in Ref.~\cite{Ihlen2012} is modified and
used for this purpose. 

\subsubsection{Multifractal Detrended Fluctuation Analysis (MFDFA): } 

The MFDFA examines the scaling properties of fluctuations of a time-series
$a_i$, $i=1, \ldots,\, N$, where $N$ is, as before, the length of the time
series. The time-integrated series, which is referred to as a \emph{profile}
in such analysis, is obtained from:
\begin{equation}
    Y(i) = \sum_{k=1}^{i} \qty(a_k - \langle a \rangle), \, i=1,...,\, N ,
\end{equation}
where $ \langle \cdot \rangle$ denotes the overall mean. The \emph{profile}
is now partitioned into $N_s=\floor{N/s}$ segments each of length $s$. To
incorporate the extra floating segment at the tail of the series, if $s$ is
not a divisor of $N$, we perform a similar partitioning, by starting from the
end of the time series; this yields $2N_s$ bins. For each partitioned bin, we
calculate the order-$m$ polynomial trend and determine the variance of the
\emph{profile} as:
\begin{align} \label{eq:defF2mfdfa}
    F^2(o,s) &= \frac{1}{s} \sum_{i=1}^{s} \Big\{Y\qty(i+(o-1)s)-y^m_{o}(i)\Big\}^2, \nonumber \\
               & \hspace{8em} 1\leq o \leq N_s,  \\
    F^2(o,s) &= \frac{1}{s} \sum_{i=1}^{s} \Bigg\{Y\qty(N+i-(o-N_s)s)-y^m_{o}(i) \Big\}^2, \nonumber \\
               & \hspace{8em} N_s+1\leq o \leq 2N_s, \nonumber 
\end{align}
where $y^m_{o}$ is the order-$m$ polynomial trend for the segment indexed by
$o$. 

The order-$q$ fluctuation function $F_q(s)$ for the scale $s$ is the
generalized mean with exponent $q$:
\begin{align}
    F_q(s) &= \Bigg[ \frac{1}{2N_s} \sum_{o=1}^{2N_s} F^2(o,s)^{q/2} \Bigg]^{1/q}, \, q\neq 0 ; \\
    F_q(s) &= \exp\Bigg\{ \frac{1}{4N_s} \sum_{o=1}^{2N_s} \ln \qty[F^2(o,s)] \Bigg\}, \, q=0 .
\end{align}
The Hurst exponent $h(q)$ follows from the scaling form $F_q(s) \sim s^{h(q)}$
with the variable scale $s$. We calculate $h(q)$ from the (best-fit) slope of
the following line in the double-log plot of $F_q(s)$ versus $s$, over the
entire scale range of $s$:
\begin{equation}\label{eq:hqcalculation}
    \ln{F_q(s)} = h(q) \ln{s} + c ,
\end{equation}
where $c$, the intercept, is inconsequential in the calculations below. If
$h(q)$ is independent of $q$, the time series is labeled as monofractal, i.e.,
it is characterized by a single scaling exponent; $1>h(q)> 0.5$ indicates a
correlated time series, whereas $h(q)<0.5$ for an anti-correlated time
series~\cite{Kantelhardt2002,Gieraltowski2012,Zeng2016}.

Given $h(q)$, the singularity strength or H\"older exponent $\alpha$ and the
singularity spectrum $f(\alpha)$ can be computed by using
\begin{eqnarray}\label{eq:fdef_legendre}
f(\alpha) = \min_q \qty[q\alpha - \tau(q)];
\end{eqnarray}
i.e., $f(\alpha)$ is the Legendre transform of $\tau(q) = (q h(q)-1)$. The
$\alpha$ and $f(\alpha)$ are, respectively, conjugate variables of $q$ and
$\tau(q)$. For the time series under consideration, $\dv{\tau(q)}{q}$ exists,
so Eq. (\ref{eq:fdef_legendre}) is equivalent to:
\begin{eqnarray}\label{eq:falphadef}
\alpha = \dv{\tau(q)}{q}; \quad f(\alpha) = q\alpha - \tau(q).
\end{eqnarray}
The multifractal spectrum describing a plot of $f(\alpha)$ versus $\alpha$
reflects certain roughness characteristics of the time series. The width of the
multifractal spectrum, $ \alpha_{\rm max}- \alpha_{\rm min}$, as well as that
of the Hurst exponent, $h(q_{\rm max})-h(q_{\rm min})$, are measures of the
degree of multifractality (or deviations from monofractality).

To obtain $F_q(s)$, linear detrending is employed, i.e., $m=1$ and
correspondingly $y_{p}$s are order-$1$ polynomials. Although $q$ can take any
real value, in any practical calculation $q$ must be finite and discrete; we
use $q \text{ from } [-5,5]$ in steps of $0.1$. The partitioning scales $s$
have been chosen from the geometric sequence, $s \in \{s_j, j = 1,\dots n_{\rm
scales}\}$ defined in Eq. (\ref{eq:defscales}) below~\cite{Ihlen2012}:
\begin{align}\label{eq:defscales}
    s_j = \qty[2^{ s_{\rm min} + (j-1) \delta l }], \hspace{1em} \delta l  &= \frac{\log_2(s_{\rm max}/s_{\rm min})}{n_{\rm scales}-1},
\end{align}
where $[.]$ denotes rounding to the nearest integer. We have chosen $n_{\rm
scales}=19$; the choice of scales $s_{\rm min}$ and $s_{\rm max}$ is dictated
by (i) the limitations of the algorithm used and (ii) the physical processes of
interest. The former dictates a range $s_{\rm min}=32$ and $s_{\rm max}=1024$.
For $s> s_{19}= s_{\rm max} \simeq  N/18$, averages are unreliable (given our
data), whereas for $s<s_{1} = s_{\rm min}$, arithmetic underflow occurs, so the
values of the Hurst exponents are not
reliable~\cite{Kantelhardt2002,Gieraltowski2012,Zeng2016}. We are interested in
the scaling of structure functions in the inertial range, so we use $6 < s <
340$. Thus, for our analysis, we choose the scales commensurate with both these
criteria: $32 < s < 340$.

\subsubsection{Multiscale Multifractal Analysis (MMA): \label{sec:MMAdef}} 

In the MMA, a scale-dependent Hurst exponent $h(q,s)$ can be computed as \cite{Gieraltowski2012}: 
\begin{eqnarray}\label{eq:hqsdef}
    \ln{F_q(s)} = h(q,s)\ln{s} +c, s\in [s_{\rm low}(s),s_{\rm high}(s)], 
\end{eqnarray}
over a window $[s_{\rm low}(s),s_{\rm high}(s)]$; we choose $s_{\rm
lower}(s_j)=s_{j-2}$ and $s_{\rm upper}(s_j) = s_{j+2}$ for this linear fit, so
we have expanding windows $[s_{j-2},s_{j+2}],\,j=3,\dots, n_{\rm scales}-2$,
which are required for MMA calculations~\cite{Gieraltowski2012,Zeng2016}. Here
it should be noted that the scale axis in the Hurst-surface plots presented
later is calibrated so as to show the beginning of the fitting windows
$\{s'_j\}$.

For a multifractal time series, $h(q,s)$ exhibits intricate scale-dependent
properties~\cite{Zeng2016}. By comparing the Hurst surfaces $h(q,s)$ for
different velocity components, the anisotropy of a turbulent flow is expanded
to include scalewise measures of roughness and intermittency. 

\section{\label{Results}Results}
\begin{figure*}
	\includegraphics[width=0.26\textwidth]{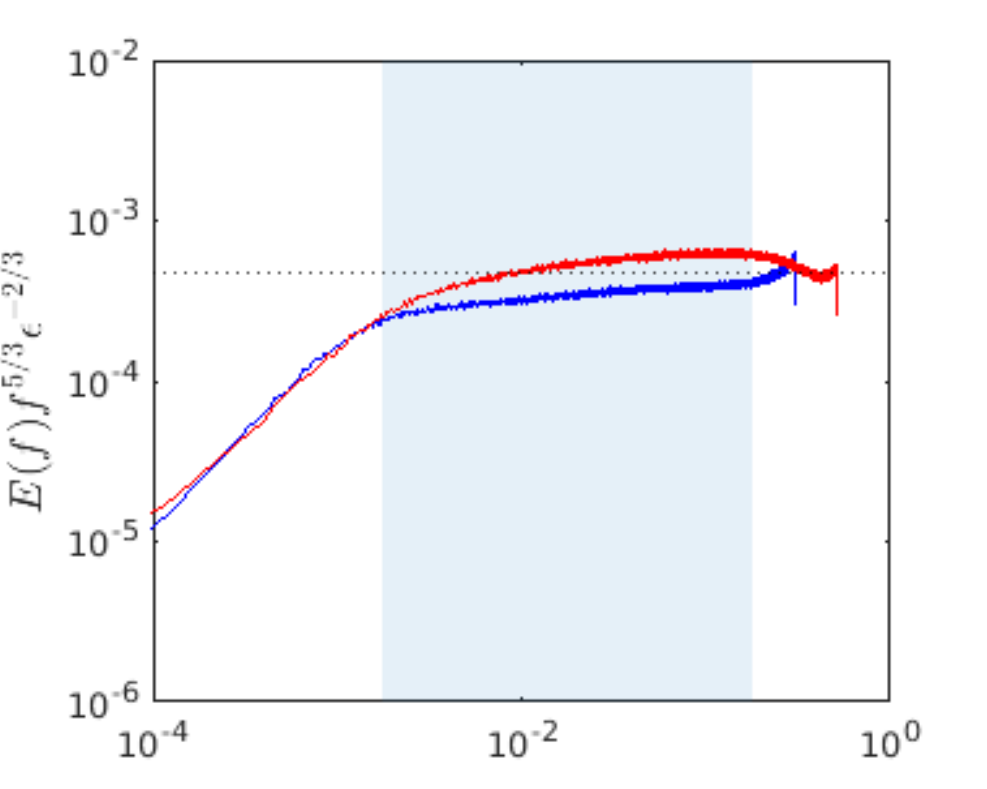}
	\put(-95,80){{\makebox[0.01\textwidth][r]{\bf \scriptsize{(a)}} }}
	\label{fig:E_tot}
	\hspace{-1.5em}
	\includegraphics[width=0.26\textwidth]{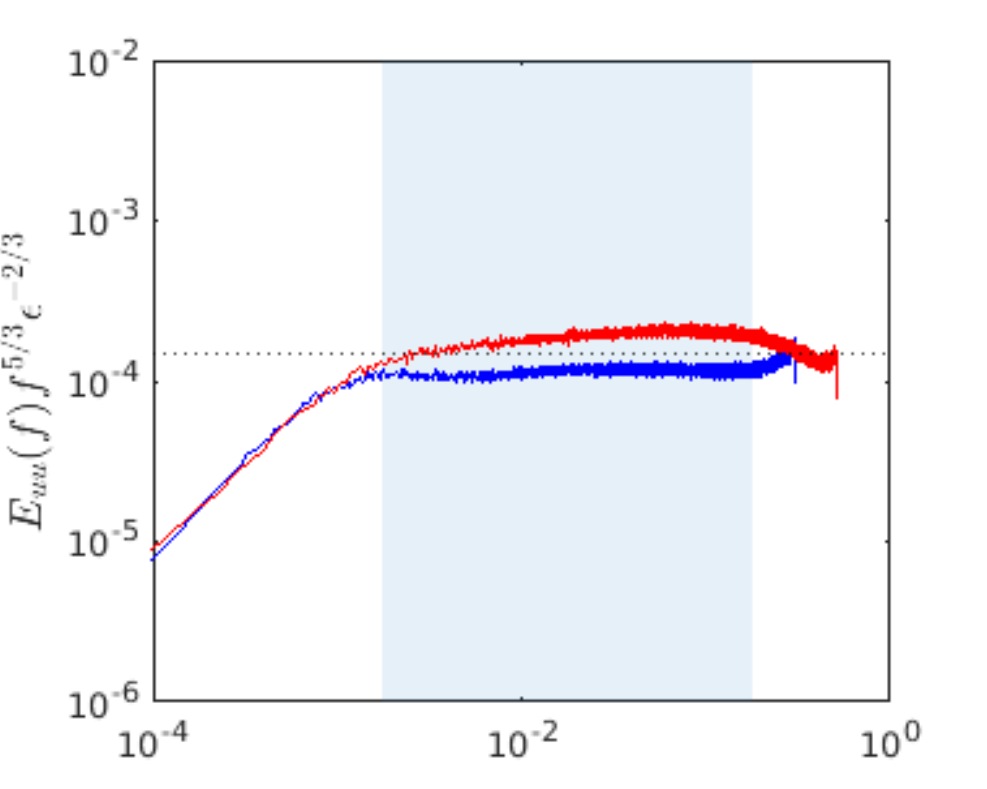}
	\put(-95,70){{\makebox[0.01\textwidth][r]{\bf \scriptsize{(b)}} }}
	\label{fig:E_11}
	\hspace{-1.5em}
	\includegraphics[width=0.26\textwidth]{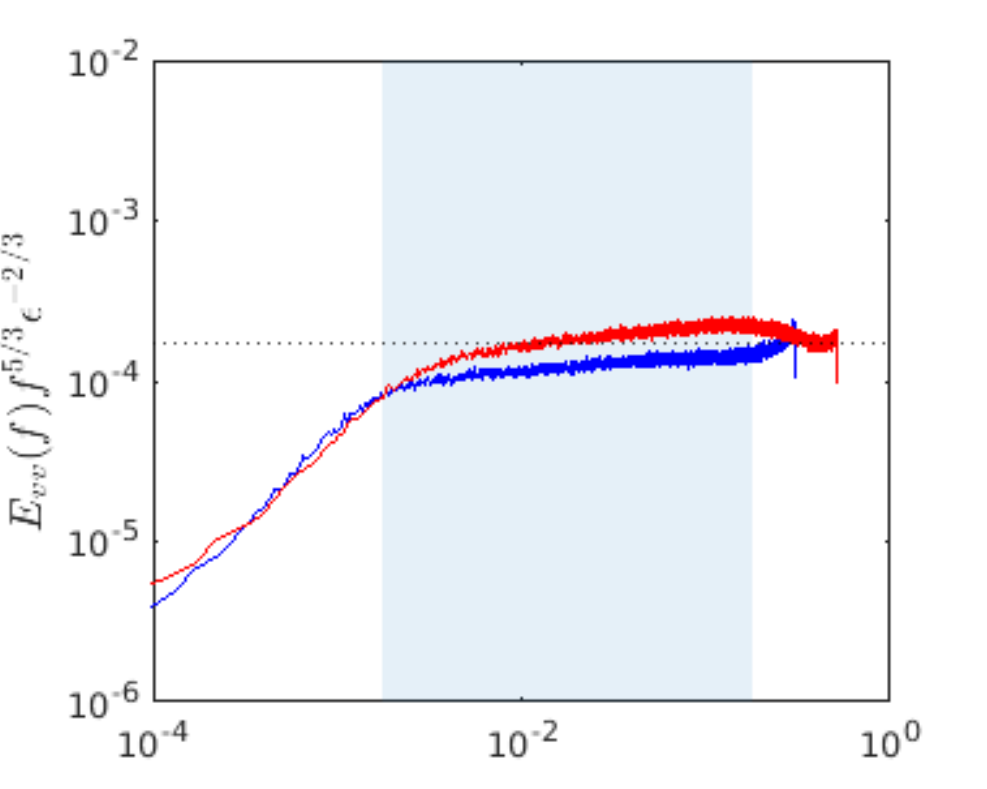}
	\put(-95,70){{\makebox[0.01\textwidth][r]{\bf \scriptsize{(c)}} }}
	\label{fig:E_22}
	\hspace{-1.5em}
	\includegraphics[width=0.26\textwidth]{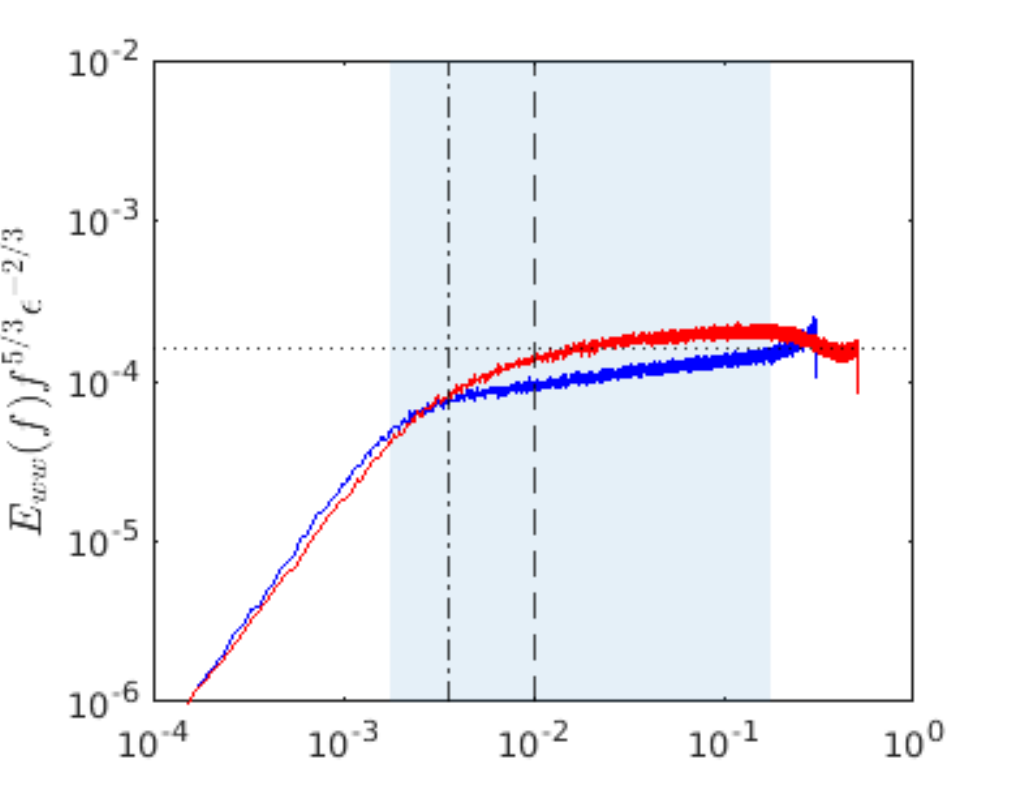}
	\put(-95,70){{\makebox[0.01\textwidth][r]{\bf \scriptsize{(d)}} }}
	\label{fig:E_33}\\
	\includegraphics[width=0.26\textwidth]{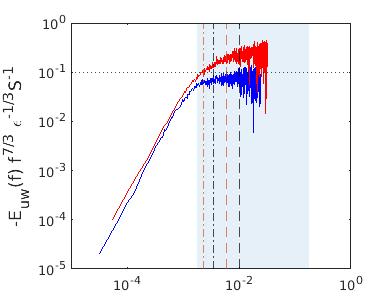}
	\put(-95,70){{\makebox[0.01\textwidth][r]{\bf \scriptsize{(e)}} }}
	\label{fig:shear_stress_cospectra}
	\hspace{-1.5em}
	\includegraphics[width=0.26\textwidth]{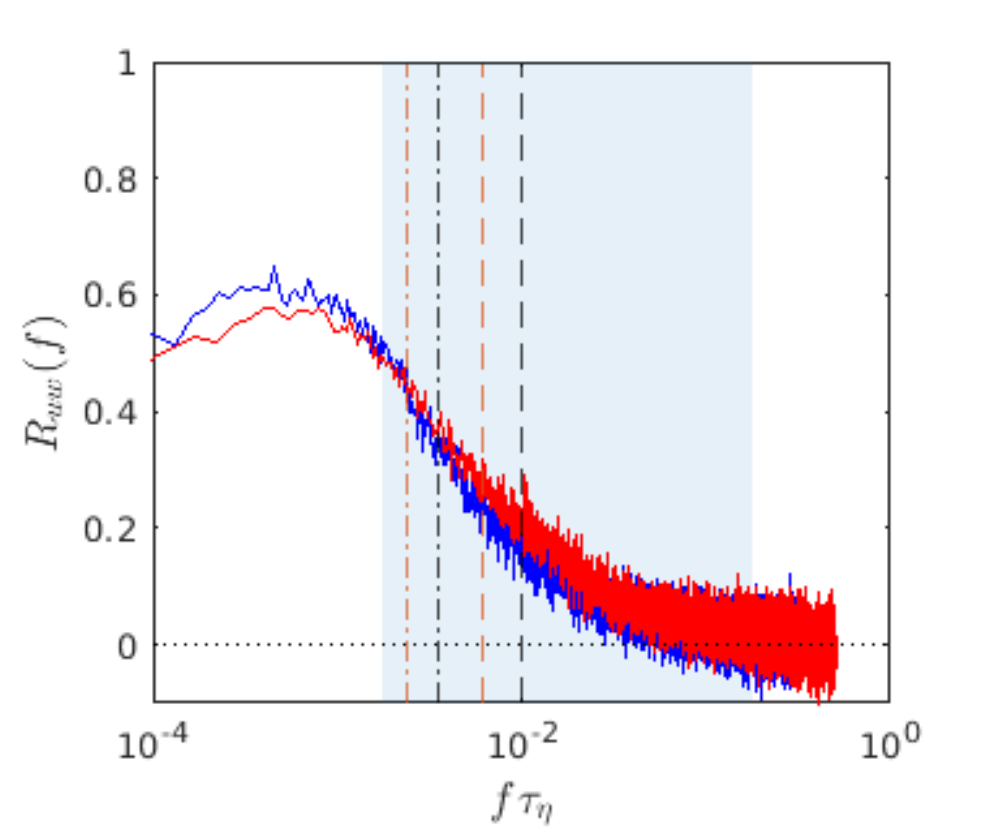}
	\put(-95,90){{\makebox[0.01\textwidth][r]{\bf \scriptsize{(f)}} }}
	\label{fig:correlation_cospectra}
	\hspace{-1.5em}
    \includegraphics[width =0.26\textwidth]{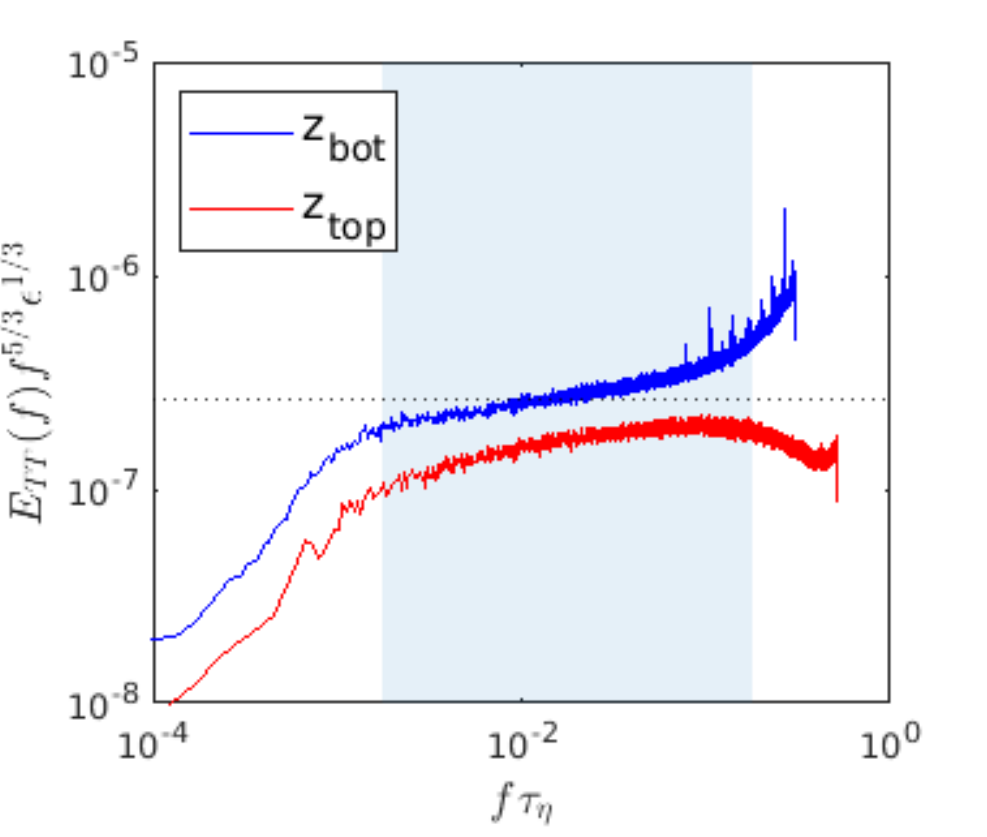}
    \put(-95,70){{\makebox[0.01\textwidth][r]{\bf \scriptsize{(g)}} }}
	\hspace{-1.5em}
    \includegraphics[width =0.26\textwidth]{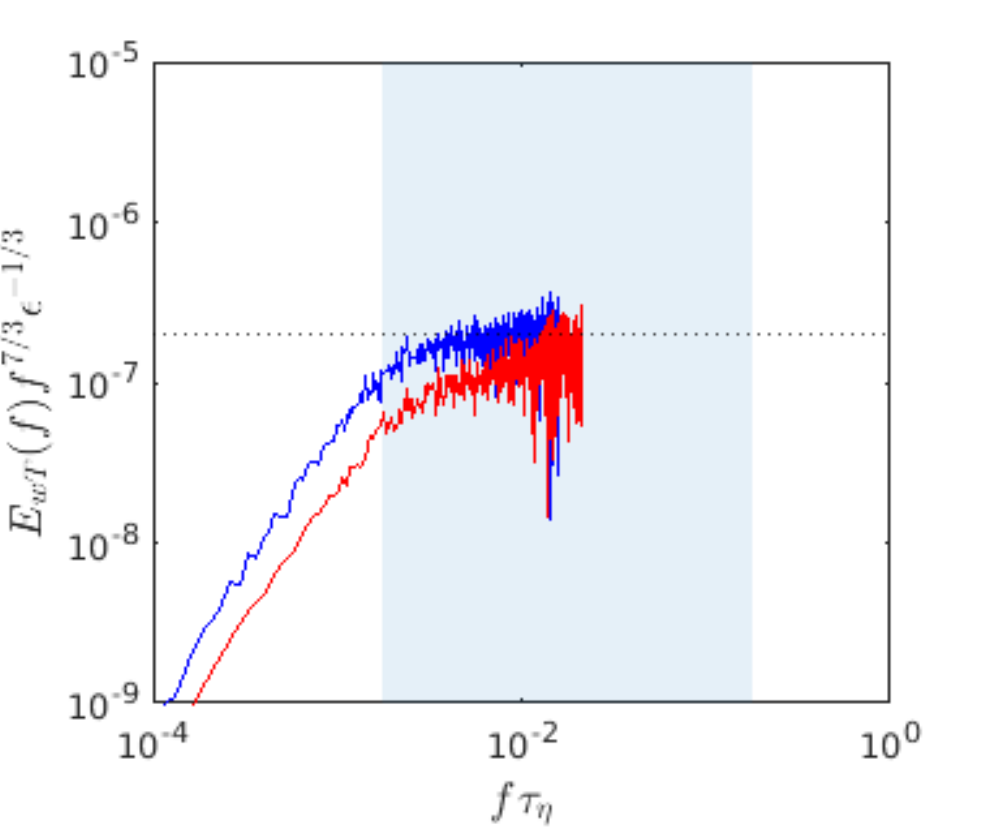}
    \put(-95,70){{\makebox[0.01\textwidth][r]{\bf \scriptsize{(h)}} }}\\

\caption{Log-log plots of spectral and co-spectral variables versus the
normalized frequency $f \tau_\eta$ (with $\tau_\eta$ the K41 dissipation time
scale), of ensemble-averaged and compensated (a) total-energy spectrum  $E(f)
f^{5/3}\epsilon^{-2/3}$ and componentwise energy spectra (b) $E_{uu}(f)
f^{5/3}\epsilon^{-2/3}$, (c) $E_{vv}(h) f^{5/3}\epsilon^{-2/3}$, (d) $E_{ww}(f)
f^{5/3}\epsilon^{-2/3}$, at $z_{\rm bot} = 23.3m$ (blue) , $z_{\rm top}=33m$
(red), (e) shear-stress co-spectra $E_{uw}(f) f^{7/3}\epsilon^{-1/3}S^{-1}$,
(f) correlation co-spectra $R_{uw}(f)$, with a $ -2/3$ scaling (not shown), (g)
compensated temperature power spectra $E_{TT} (f)f^{5/3}\epsilon^{1/3}$, and
(h) compensated (wall-normal) velocity-temperature co-spectra $E_{wT}(f)
f^{7/3}\epsilon^{-1/3}$. Regions shaded in blue show the spectral range in
which the energy spectrum shows an approximate $-5/3$ scaling for the power
spectra and $-7/3$ scaling for the co-spectra. Note that (e) is shown only in
the low-frequency range because, for $f>f_{\rm max}$ where isotropy is reached,
experimental values appear with both signs. Note that $f_{\rm max}^{z_{\rm
top}} < f_{\rm max}^{z_{\rm bot}}$. Orange lines (dashed for $z_{\rm top}$ and
dashed dots for $z_{\rm bot}$) mark the commencement of the $-7/3$ scaling
range for the shear-stress co-spectra; and black lines (dashed for $z_{\rm
top}$ and dashed dots for $z_{\rm bot}$) indicate the beginnings of the $-5/3$
scaling range for the wall-normal velocity spectra. We have averaged the
normalized variables over stable, unstable and near-neutral stratification for
the above graphs.  \label{fig:power_spectrum}}
\end{figure*}

We first present the findings from the invariant analysis in
sub-section~\ref{sec:Result_Anisotropy_Analysis}, followed by the results of
the MFDFA and MMA analyses. We also discuss the key features of certain bulk
and scalewise  changes as well as their connections and variations with
stability.

\subsection{\label{sec:Result_Anisotropy_Analysis}Anisotropy Analysis}

In Fig.~\ref{fig:power_spectrum} (a) log-log plots of the ensemble-averaged
and compensated total-energy spectrum  $E(f) f^{5/3}\epsilon^{-2/3}$; and
componentwise energy spectra (for the longitudinal, transverse, and
wall-normal components)  $E_{uu}(f) f^{5/3}\epsilon^{-2/3}$, $E_{vv}(f)
f^{5/3}\epsilon^{-2/3}$, and $E_{ww}(f) f^{5/3}\epsilon^{-2/3}$ are presented,
respectively, in Figs.(\ref{fig:power_spectrum}) (b), (c), and (d) for $z_{\rm
bot} = 23.3m$ (blue) and $z_{\rm top}=33m$ (red). A well-developed $-5/3$
scaling region is observed only for the longitudinal energy spectrum (in the
blue shaded region).

In terms of time-scale separation, the Corrsin-Uberoi criterion is satisfied
because the ratio of the Kolmogorov to the mean-shear time scale $S_c^{\ast} =
\eta/(S^{-1})^{-1} = S(\nu/\epsilon)^{1/2}$ is small (Table
\ref{table:all_params}); also, $\eta^{-1} \gg k \gg (S^3\epsilon^{-1})^{1/2}$~
so we can expect the flow, at small length scales, to be approaching a locally
isotropic state. However, because of large mean gradients within the RSL in
velocity~\cite{katul2013co} and temperature~\cite{katul2014two}, the
inertial-range shear-stress and heat flux co-spectra are not zero and exhibit a
$-7/3$ scaling exponent. This is indeed the case, as shown from the compensated
shear-stress co-spectra~\cite{saddoughi_veeravalli_1994} in
Fig.~\ref{fig:power_spectrum}(e). These co-spectra exhibit well-developed
$-7/3$ scaling regimes in the wave-number range where both the longitudinal and
wall-normal energy spectra exhibit power-law regions with exponents $\simeq
-5/3$. The corresponding correlation co-spectra $R_{uw} (f)$
(Fig.~\ref{fig:power_spectrum}(f)) also show power laws with exponents $\simeq
-2/3$ (not shown). We give plots of the shear co-spectra only in the
low-frequency range where the flow is anisotropic. These low frequencies are
determined as less than $f_{\rm max}^{z_{\rm bot}} = 0.48 Hz$ and $f_{\rm
max}^{z_{\rm top}} = 0.52 Hz$, for $z_{\rm bot}$ and $z_{\rm top}$,
respectively. The aforementioned frequencies correspond to the length scales
$l^{z_{\rm bot}} = 4.92m < l^{z_{\rm bot}} = 6.03m$, respectively, where
(quasi-) isotropy is attained and experimental values occur with both signs
(Fig.~\ref{fig:power_spectrum}(f)). It can be surmised that small-length-scale
quasi-isotropy is attained, when identified by the scaling laws, more so at
$z_{\rm top}$, in the RSL, than at $z_{\rm bot}$. Black dotted lines, parallel
to the  horizontal axis are shown for reference; in
Figs.\ref{fig:power_spectrum} (e) and (f), the orange lines (dashed, for
$z_{\rm top}$, and dashed-dotted, for $z_{\rm bot}$)  mark the commencement of
the $-7/3$ scaling regime for the shear-stress co-spectra. \\

With regard to temperature, Figs. \ref{fig:power_spectrum}(g) and (h) show
the compensated plot of the temperature power spectra $E_{TT}(f)
f^{5/3}\epsilon^{1/3}$ and (wall-normal) velocity-temperature co-spectra
$E_{wT}(f) f^{7/3}\epsilon^{-1/3}$, respectively. The vertical
velocity-temperature co-spectra is only shown in the low-frequency range for $f
< \tilde{f}_{\rm max}^{z_{\rm top}} = 0.26 Hz \simeq \tilde{f}_{\rm
max}^{z_{\rm bot}} = 0.21 Hz$, respectively, beyond which experimental values
occur with both signs. These frequencies $\tilde{f}$ correspond to length
scales at which the flow is anisotropic. For a detailed discussion on the other
possible power-laws and the implications of these scaling exponents, we refer
the reader to Ref.~\cite{Cava2012}.

\begin{figure*}
    \includegraphics[width=0.32\textwidth]{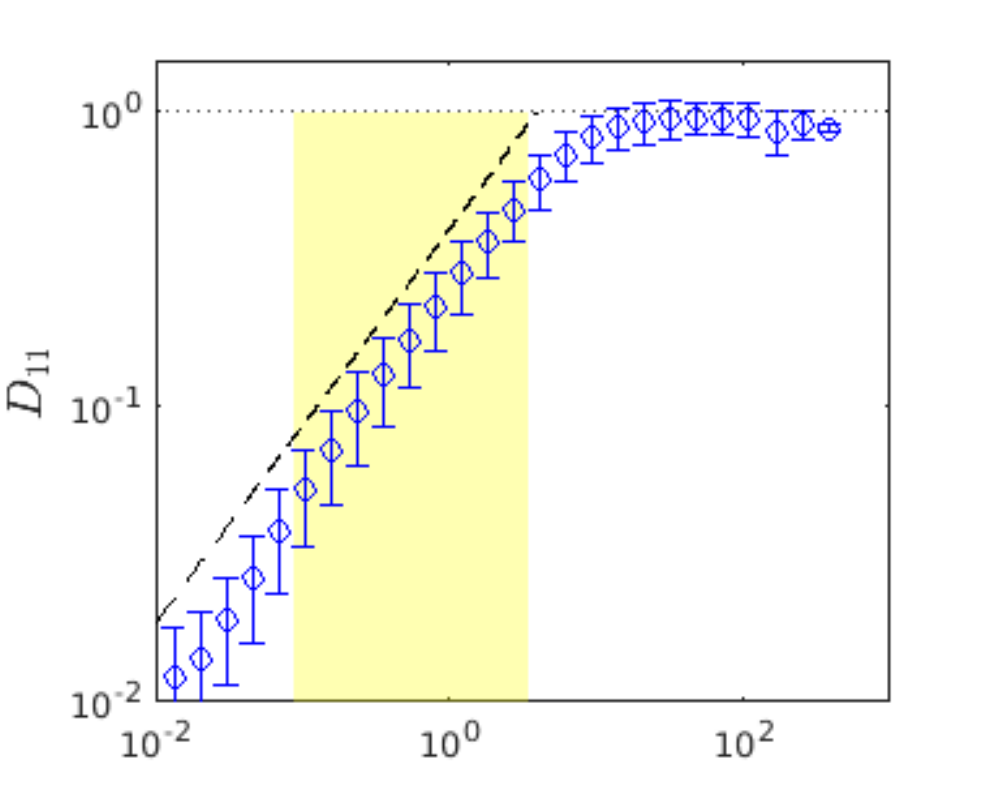}
      \put(-125,90){\bf \scriptsize{(a)}}
      \label{fig:struc_func_23_u}
    \includegraphics[width=0.32\textwidth]{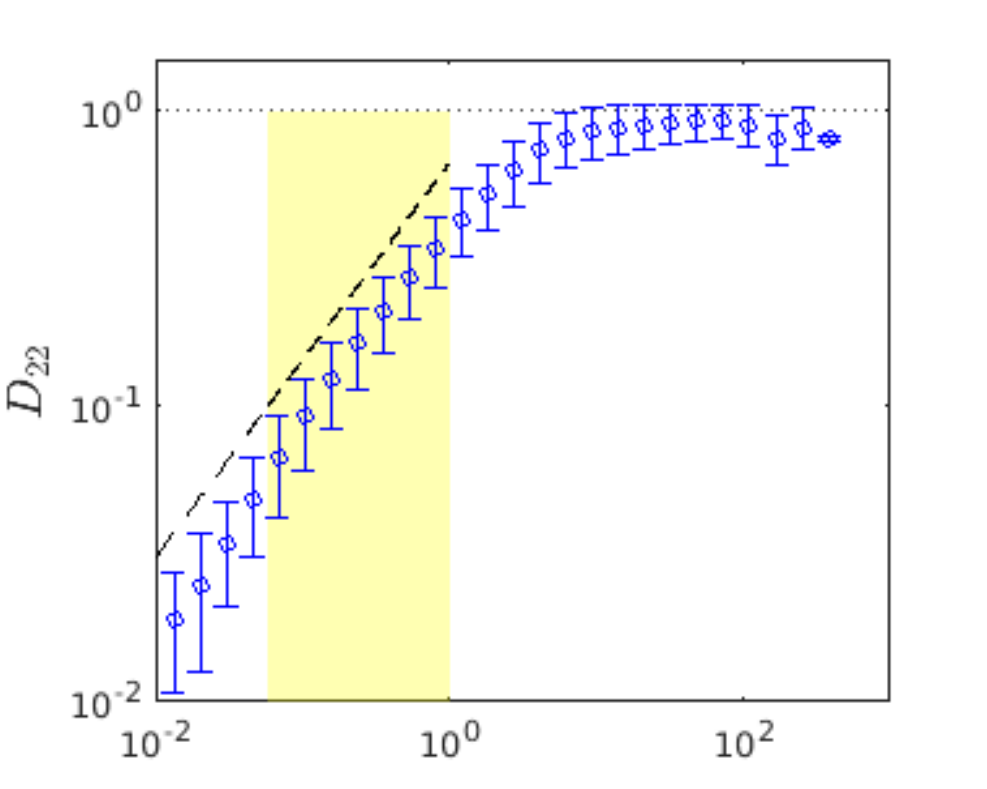}
      \put(-125,90){\bf \scriptsize{(b)}}
      \label{fig:struc_func_23_v}
    \includegraphics[width=0.32\textwidth]{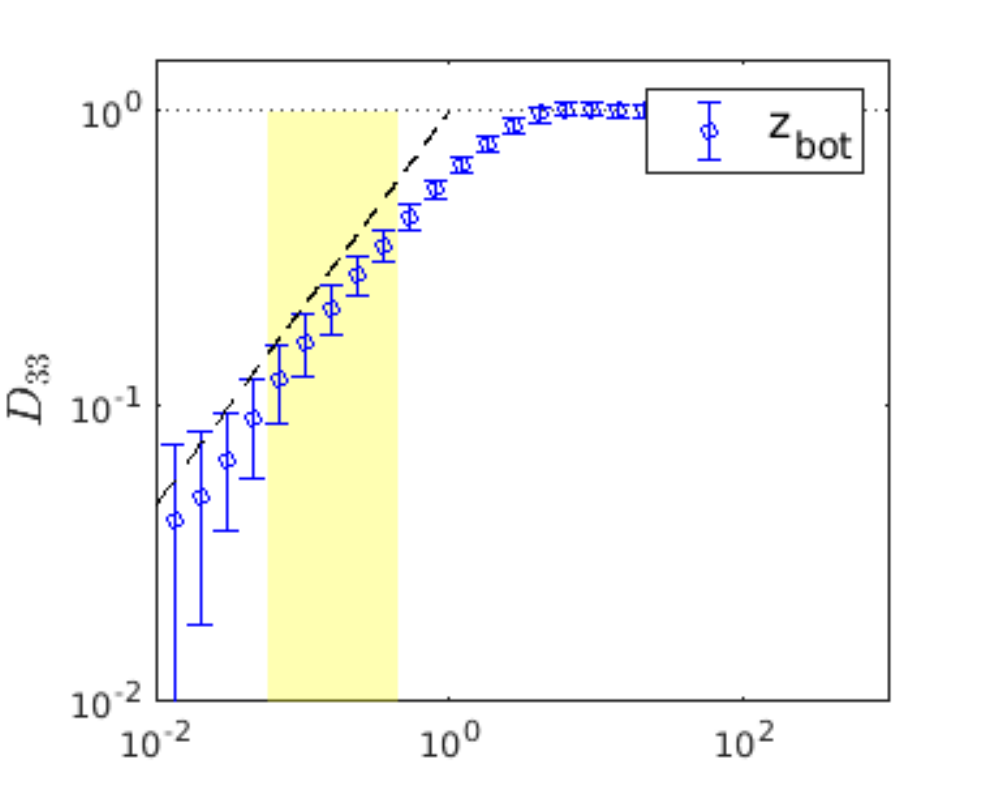}
      \put(-125,90){\bf \scriptsize{(c)}}
      \label{fig:struc_func_23_w}\\

    \includegraphics[width=0.32\textwidth]{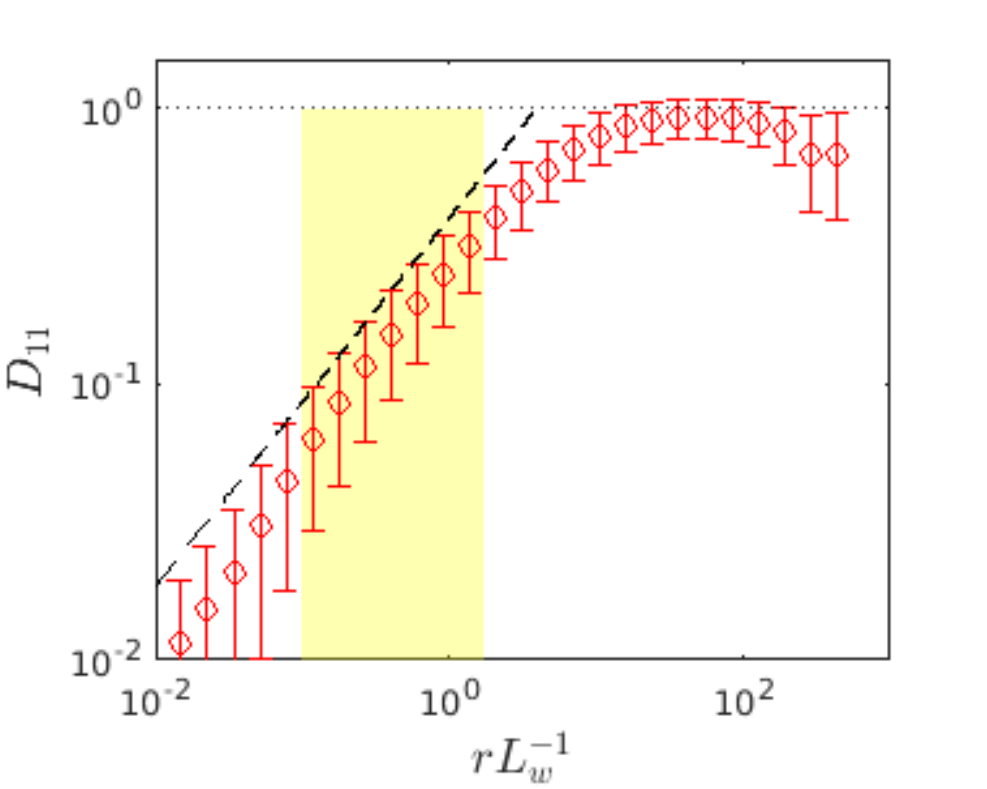}
      \put(-125,90){\bf \scriptsize{(d)}}
      \label{fig:struc_func_33_u}
    \includegraphics[width=0.32\textwidth]{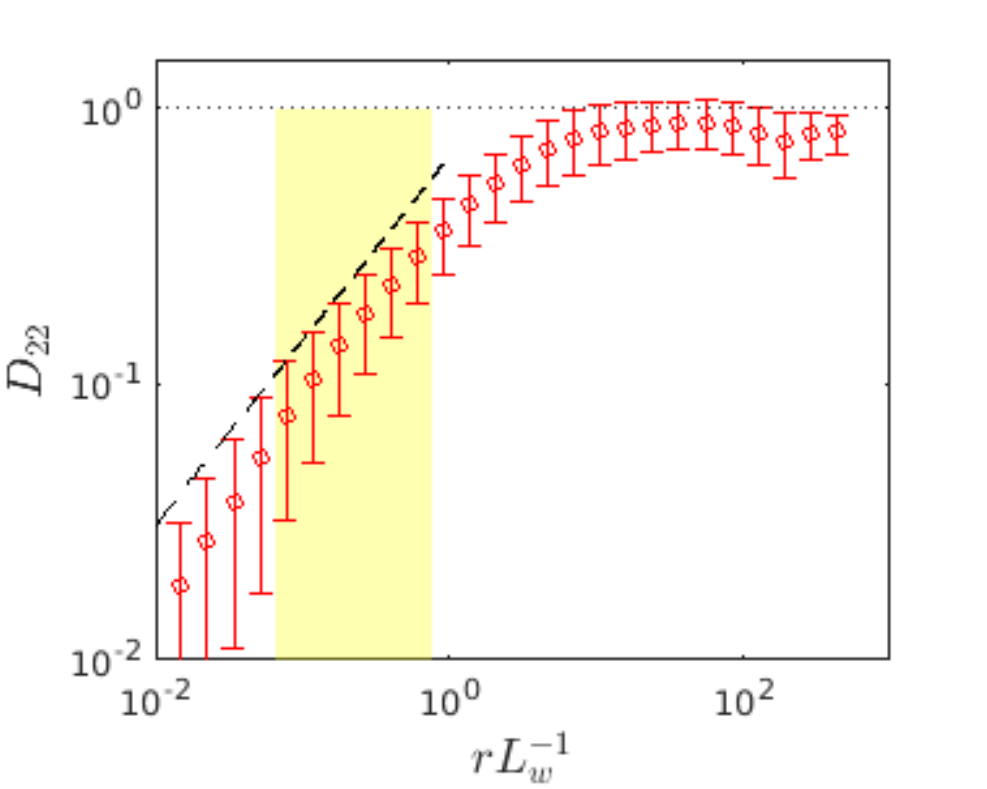}
      \put(-125,90){\bf \scriptsize{(e)}}
      \label{fig:struc_func_33_v}
    \includegraphics[width=0.32\textwidth]{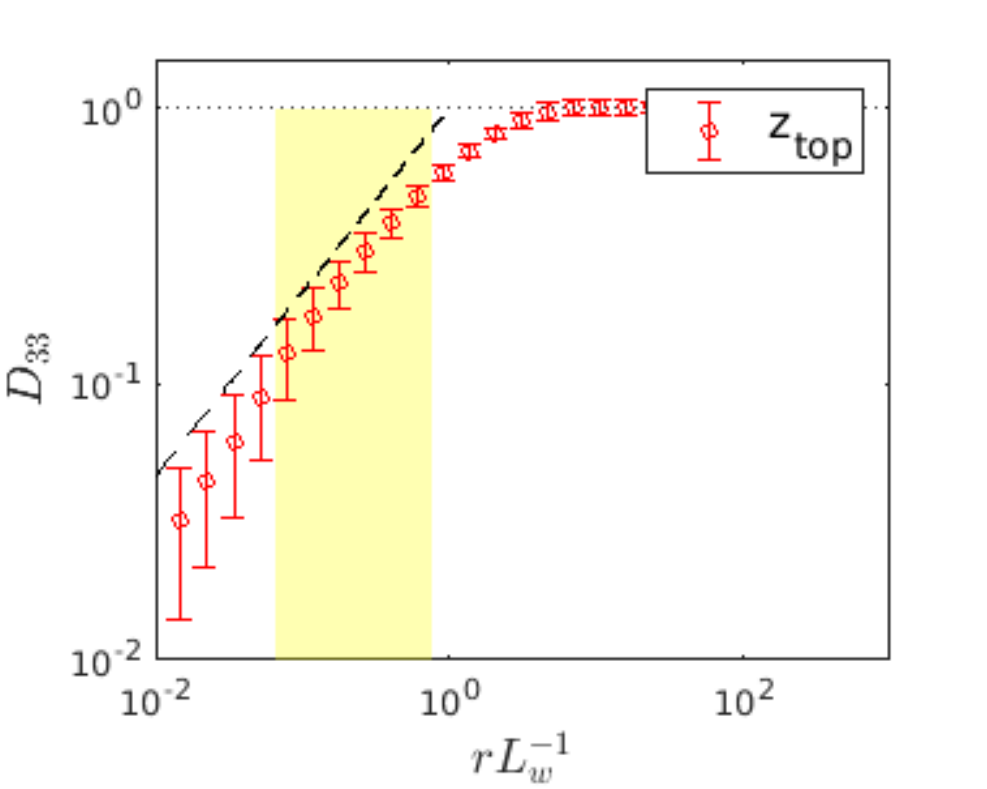}
      \put(-125,90){\bf \scriptsize{(f)}}
      \label{fig:struc_func_33_w }

\caption{Log-log plots of flow variables versus the scaled distance separation
$r L_w^{-1}$ (with $L_w$ the integral length scale
(Table~\ref{table:all_params})) of the ensemble-averaged normalized structure
functions $\frac{1}{2}D_{ii}(u_iu_i)^{-1}$ (with one-standard-deviation error
bars), for $z_{\rm bot}$ (blue) (a), (b), and (c) and $z_{\rm top}$ (red) (d),
(e), and (f); the range of approximate $r^{2/3}$ scaling is highlighted in
yellow; the dashed line shows the K41 result, i.e., scaling with a $2/3$
exponent.\label{fig:struc_func}}

\end{figure*}

In Fig.~\ref{fig:struc_func}, log-log plots of the  ensemble-averaged,
diagonal components of the structure functions $D_{u_i u_i}$ ,
$(u_1,u_2,u_3)=(u,v,w)$, normalized by $\overline{u_iu_i}$ against normalized
scales are shown. The K41 scaling form $r^{2/3}$ is indicated by black dashed
lines in the inertial range. We do not observe a distinct logarithmic
scaling~\cite{Davidson2006,Brugger2018} in the longitudinal (streamwise)
second-order structure function $D_{11}$ (Fig.~\ref{fig:struc_func}) or a
$f^{-1}$ scaling in longitudinal energy spectra $E_{uu}$ in
Fig.~\ref{fig:power_spectrum} (b) neither at $z_{\rm bot}$, nor at  $z_{\rm
top}$.

Yellow shading highlights regions where the structure functions show
approximate $r^{2/3}$ scaling; this K41-type range is smaller for the
wall-normal and the transverse components than for the streamwise component.
Within the roughness sublayer, the range of K41-type scaling is different at
different heights and for different velocity components.

\begin{figure*}
   \includegraphics[width=0.26\textwidth]{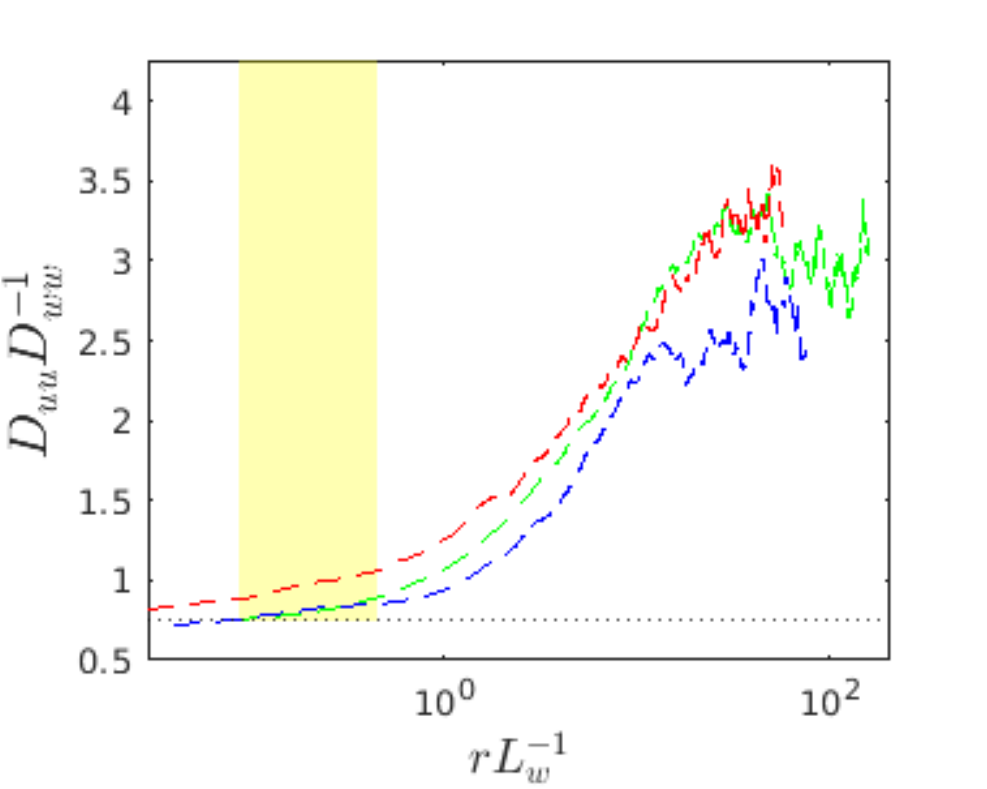}
   \put(-95,75){\bf \scriptsize{(a)}}
   \label{fig:duu_dww23}
\hspace{-1.75em}
   \includegraphics[width=0.26\textwidth]{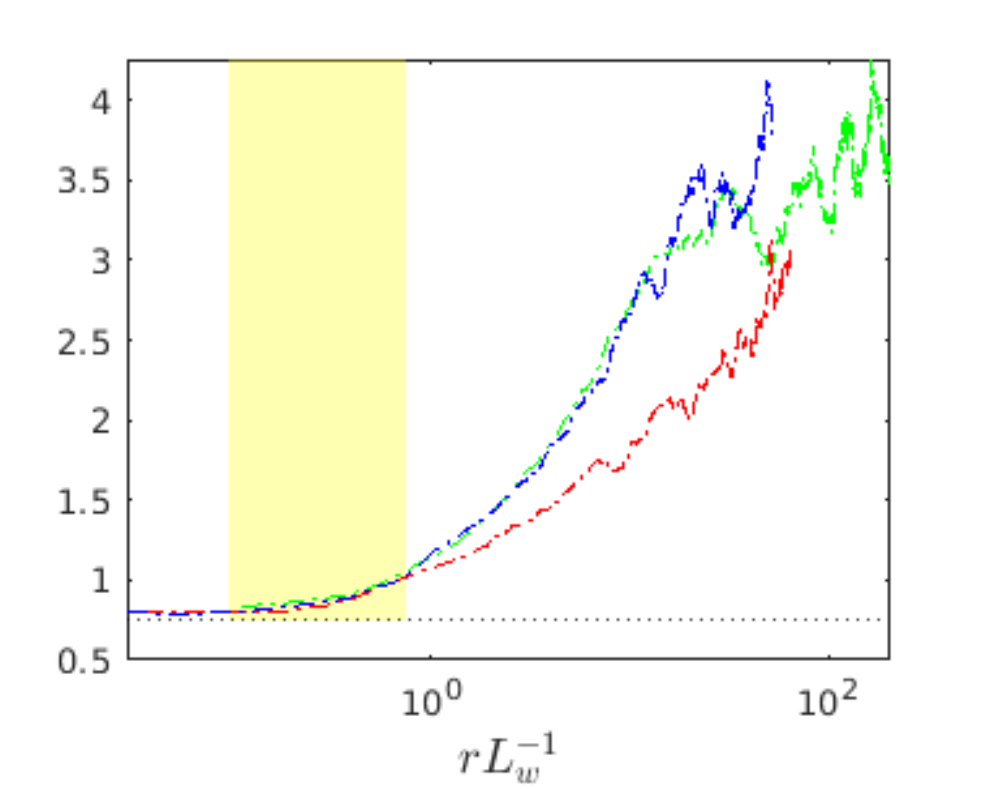}
   \put(-95,75){\bf \scriptsize{(b)}}
   \label{fig:duu_dww33}
\hspace{-0.5em}
   \includegraphics[width=0.26\textwidth]{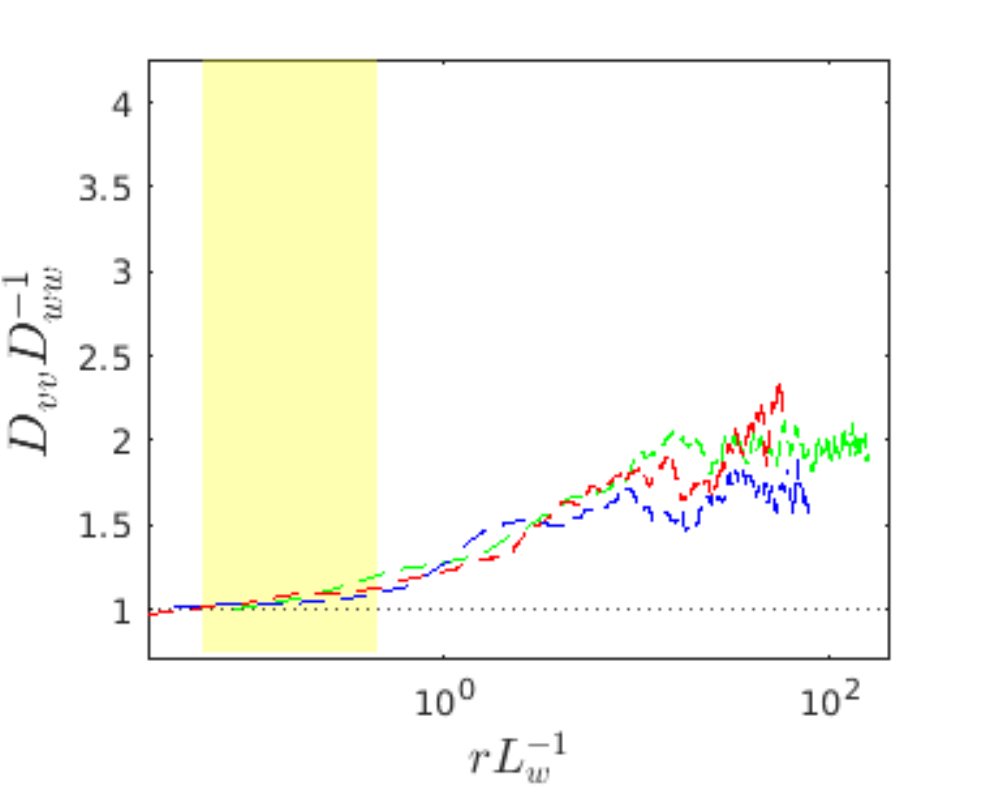}
   \put(-95,75){\bf \scriptsize{(c)}}
   \label{fig:dvv_dww23}
\hspace{-1.75em}
   \includegraphics[width=0.26\textwidth]{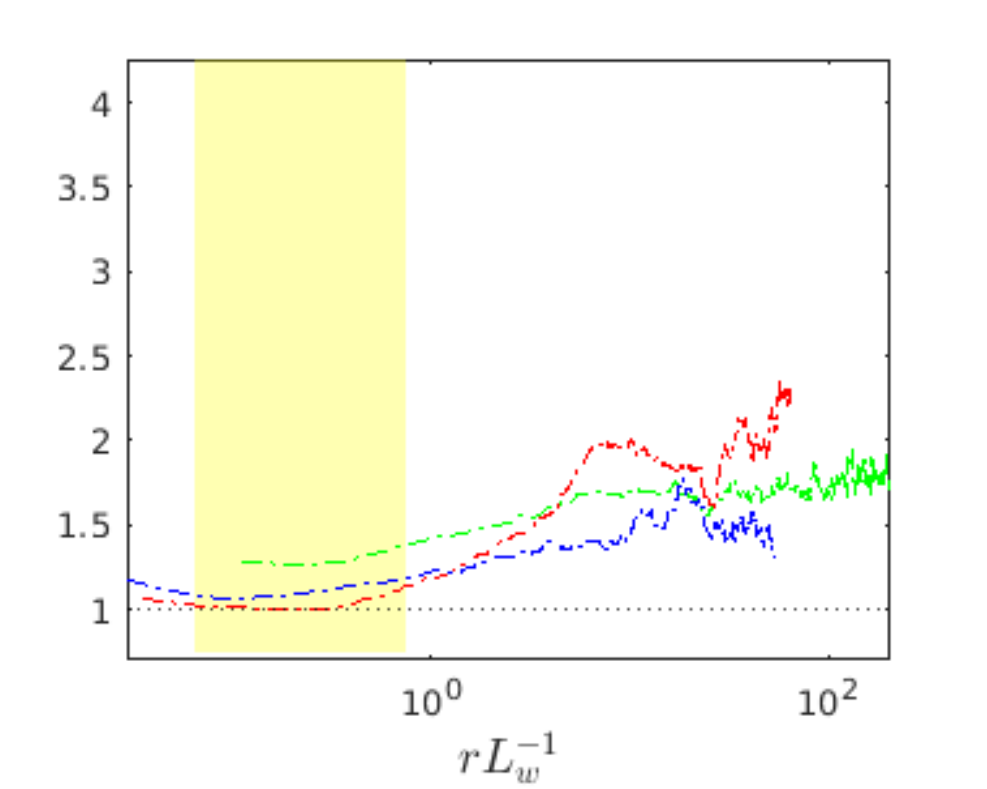}
   \put(-95,75){\bf \scriptsize{(d)}}
   \label{fig:dvv_dww33}
   
\caption{Linear-log plots of flow variables versus the scaled separation $r
L_w^{-1}$ for the ratios of the diagonal components of second-order structure
function; the K41 expectation is shown by the black dotted lines for $z_{\rm
bot}$ (a and c) and $z_{\rm top}$ (b and d) and for different stability
conditions: stable (green), unstable (red) and neutral (blue). The regions
highlighted in yellow indicate where these structure functions show  $r^{2/3}$
scaling. Taylor's frozen turbulence hypothesis was used to convert time to
space in both - the numerator and denominator of the abscissa. We expect some
distortions  due  to  the  use  of  frozen  turbulence; however,  the
normalized  scale  separation  is  likely  less  sensitive  to  those
distortions. \label{fig:dii_d33}}

\end{figure*}

Figure ~\ref{fig:dii_d33} shows the ratios of the diagonal components of the
second-order structure function. These ratios are more sensitive to anisotropy
than scaling exponents in structure functions.  The ratios exhibit some
deviations from the K41 expectation (shown by black dotted lines for low
turbulent intensity) for both $z_{\rm bot}$ (a), (c) and $z_{\rm top}$ (b), (d)
and for different stability conditions: stable (green), unstable (red), and
neutral (blue). The values in the region of a $2/3$ scaling (highlighted in
yellow) are close to the K41 result; deviation from this is significant at
larger length scales (as expected).  It is to be noted that for locally
isotropic conditions $D_{vv}(r)/D_{ww}(r) = 1$, whereas
$D_{uu}(r)/D_{ww}(r)=3/4$ although some deviations from those ratios are
expected because of the use of Taylor's frozen turbulence hypothesis as
discussed elsewhere~\cite{hsieh1997dissipation}.

\begin{figure*}
\centering
\includegraphics[width=0.25\textwidth]{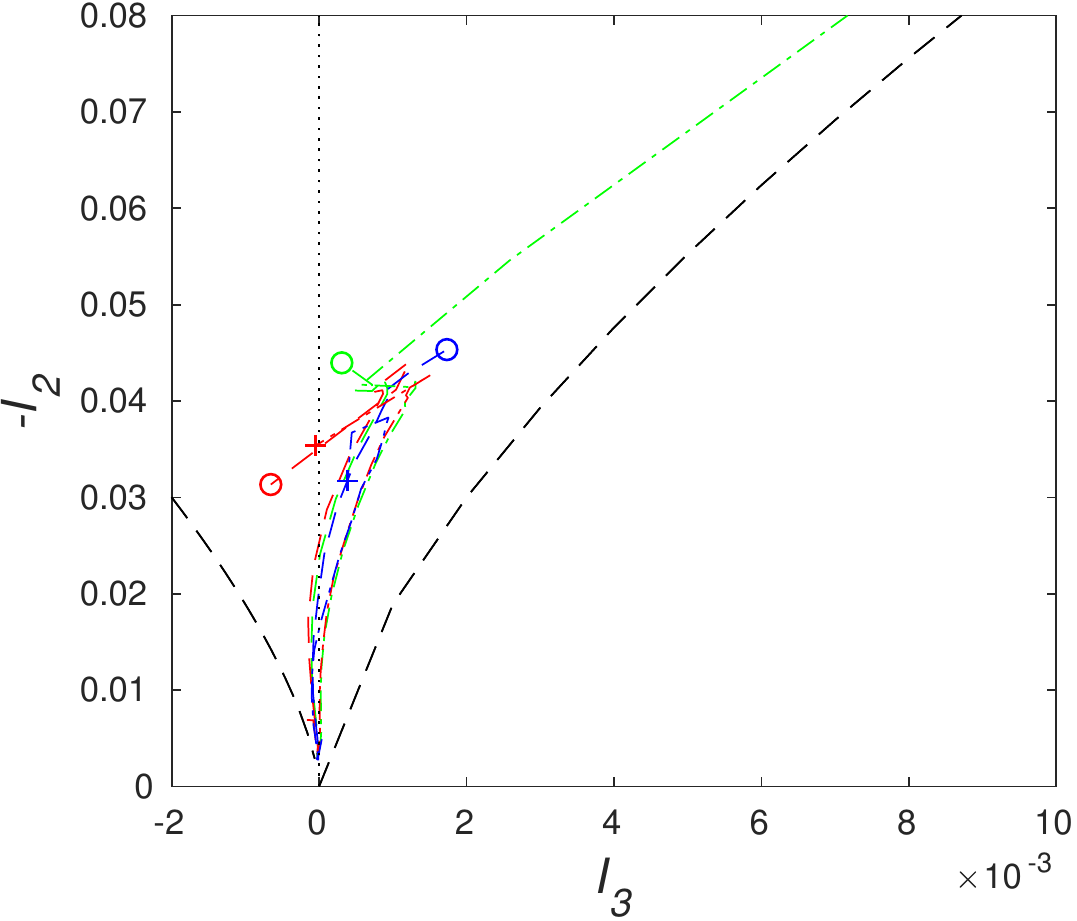}
\put(-95,90){\bf \scriptsize{(a)}}
\includegraphics[width=0.25\textwidth]{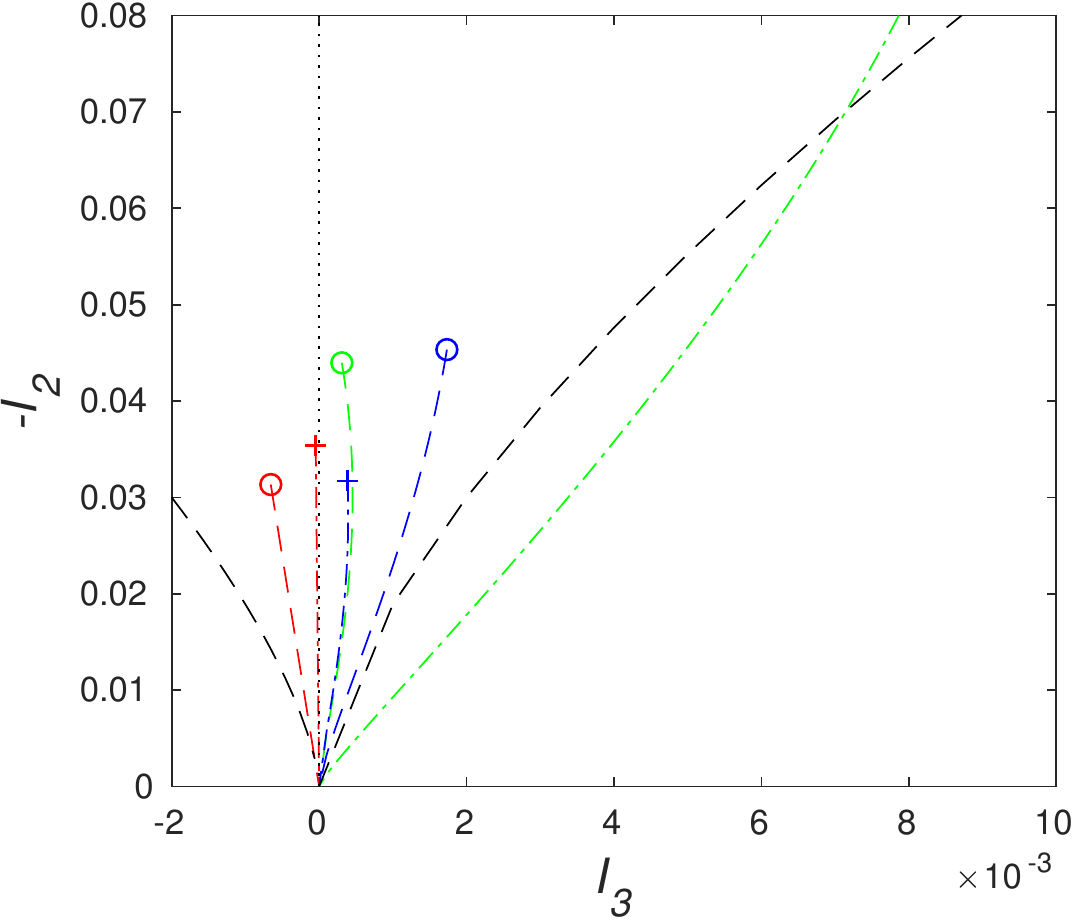}
\put(-95,90){\bf \scriptsize{(b)}}
\includegraphics[width=0.24\textwidth]{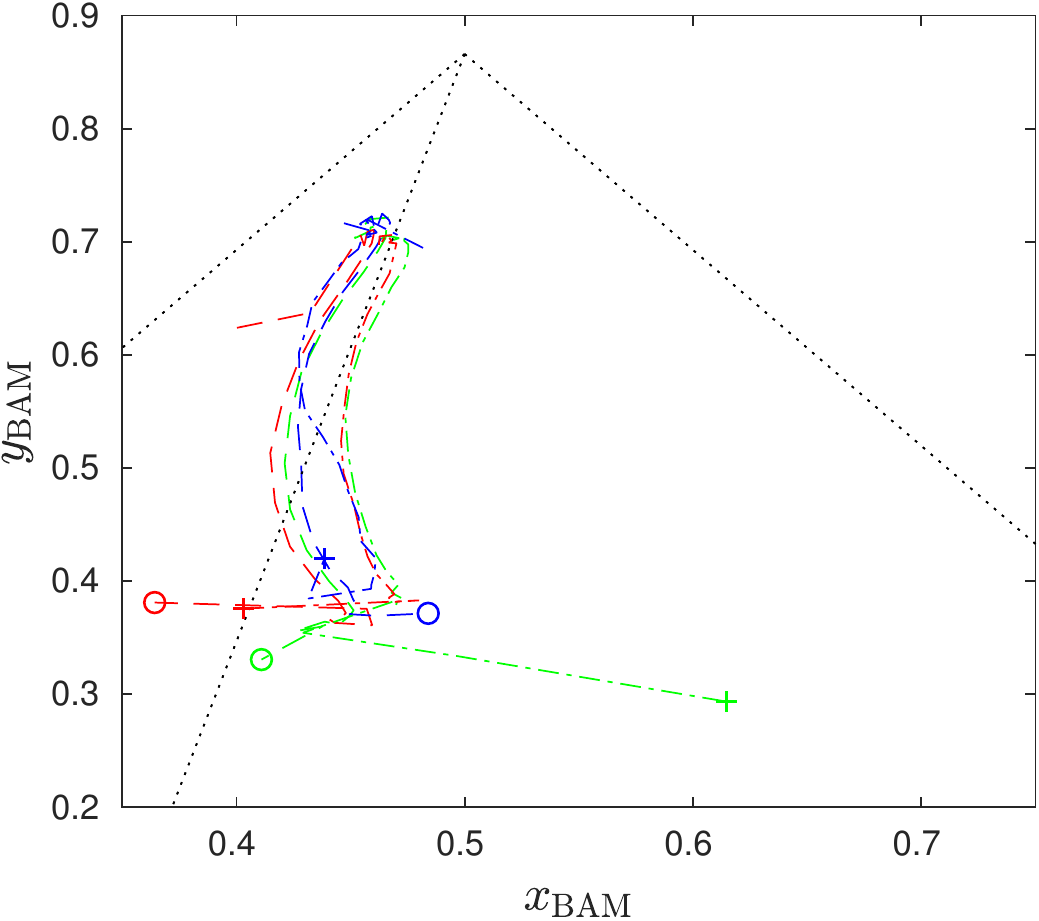}
\put(-95,90){\bf \scriptsize{(c)}}
\includegraphics[width=0.24\textwidth]{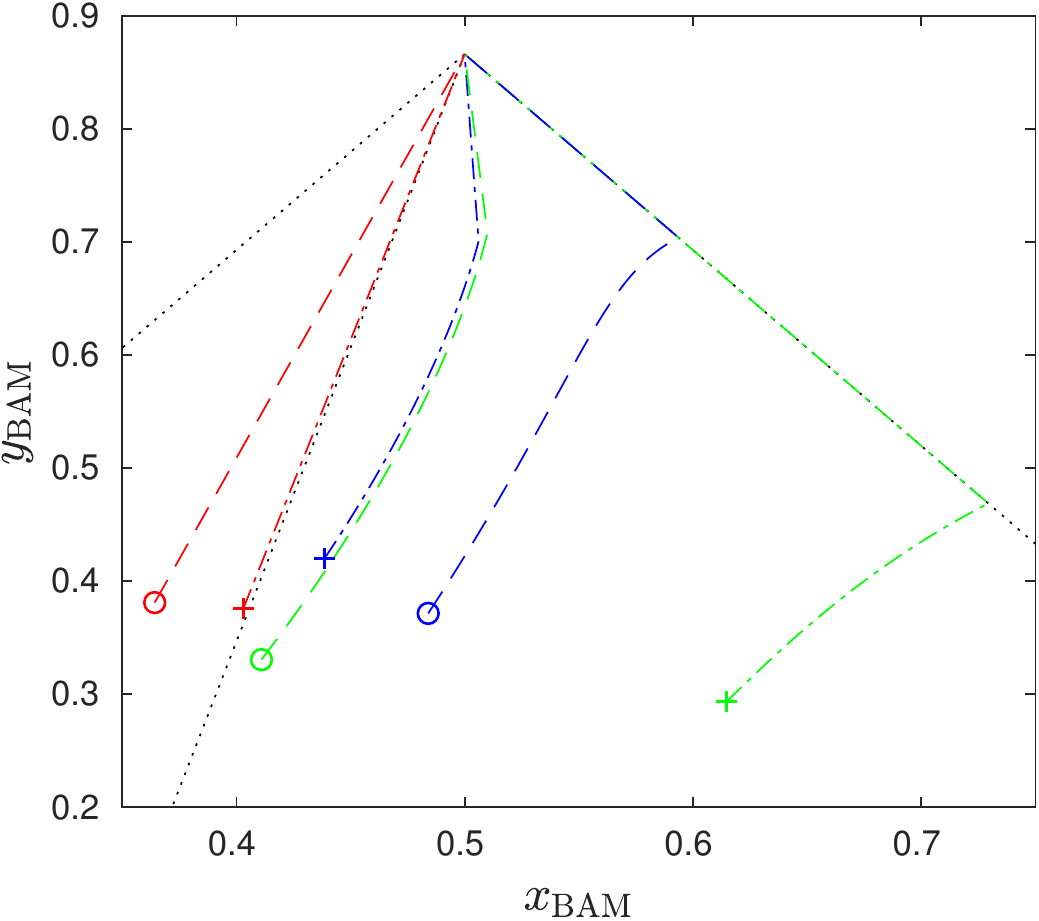}
\put(-95,90){\bf \scriptsize{(d)}}
\caption{Plots of the averaged relaxation-to-isotropy trajectories in the
$(-I_2, I_3)$ plane from observations (a) and (c) and those predicted (b) and
(d) by the quadratic Rotta model~\cite{Brugger2018}; (a) and (b) in AIM
framework and (c) and (d) in BAM framework. These plots are for different
stability conditions: stable (green); unstable (red); and neutral (blue); and
also for different heights: $z_{\rm bot}$ (dashed --) with circles; and $z_{\rm
top}$ (dashed dots $- \cdot$) with $+$. \label{fig:avg_pred_paths}}
\end{figure*}

\begin{figure*}  
\hspace{-2em}
    \includegraphics[width=0.25\textwidth]{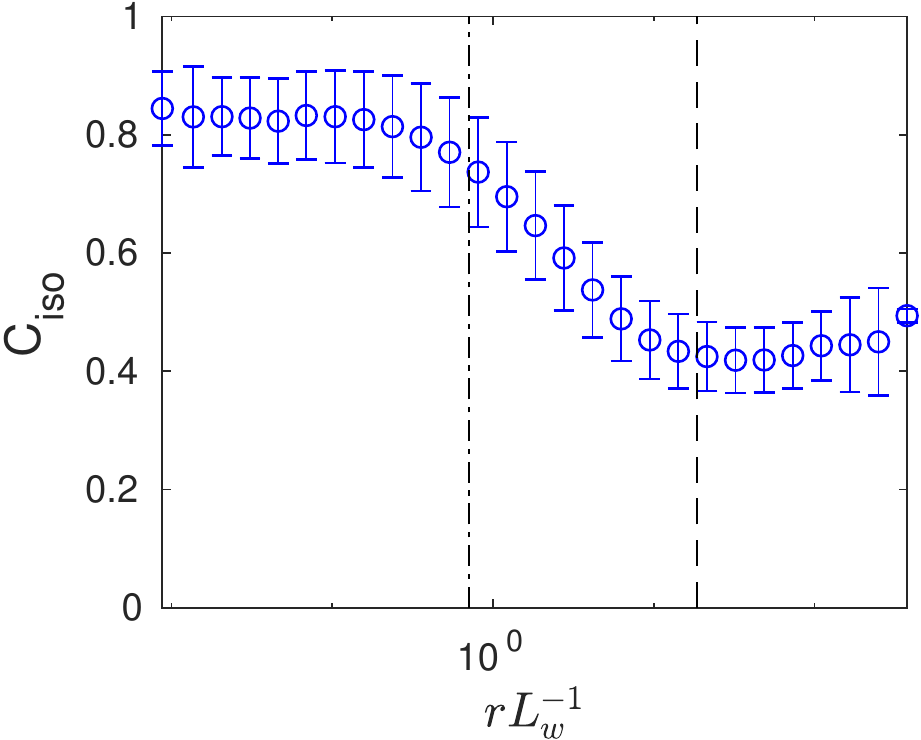}
    \put(-91,25){\bf \scriptsize{(a)}}
\hspace{0.25em}
    \includegraphics[width=0.23\textwidth]{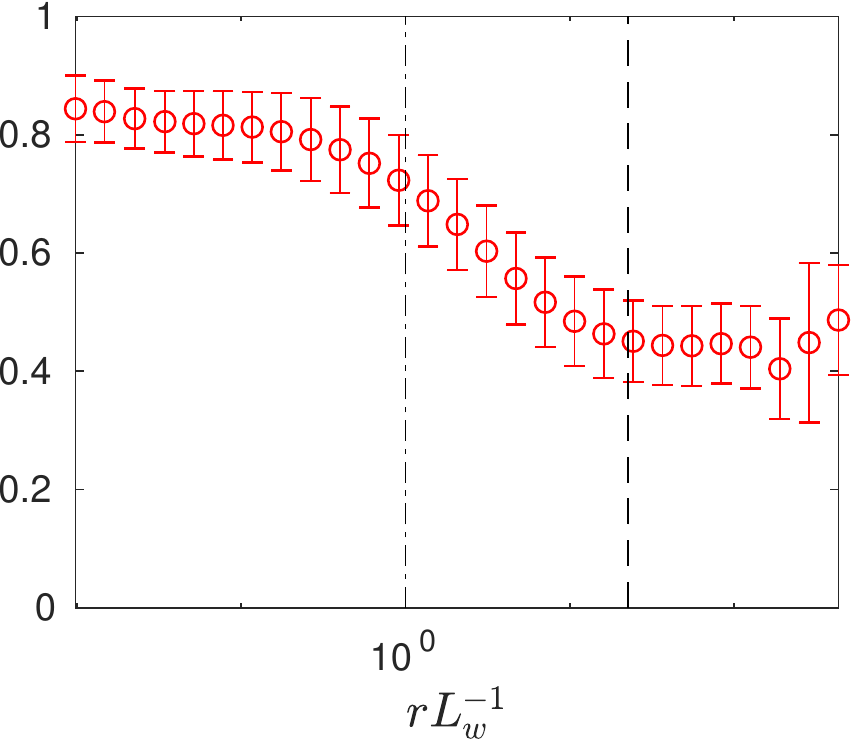}
    \put(-91,25){\bf \scriptsize{(b)}}
\hspace{0.5em}
    \includegraphics[width=0.24\textwidth]{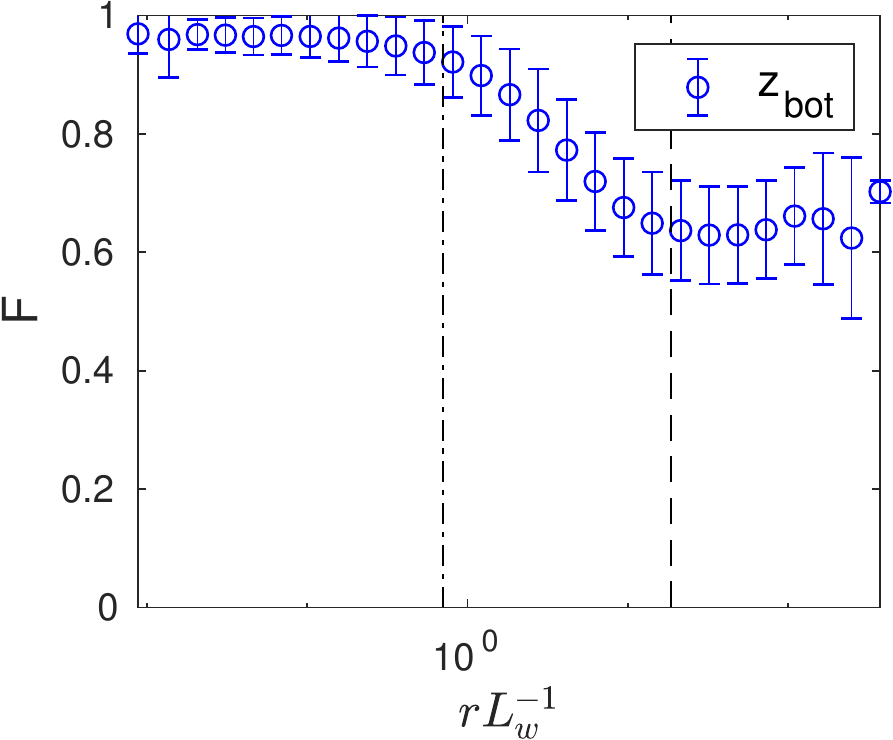}
    \put(-91,25){\bf \scriptsize{(c)}}
\hspace{0.25em}
    \includegraphics[width=0.23\textwidth]{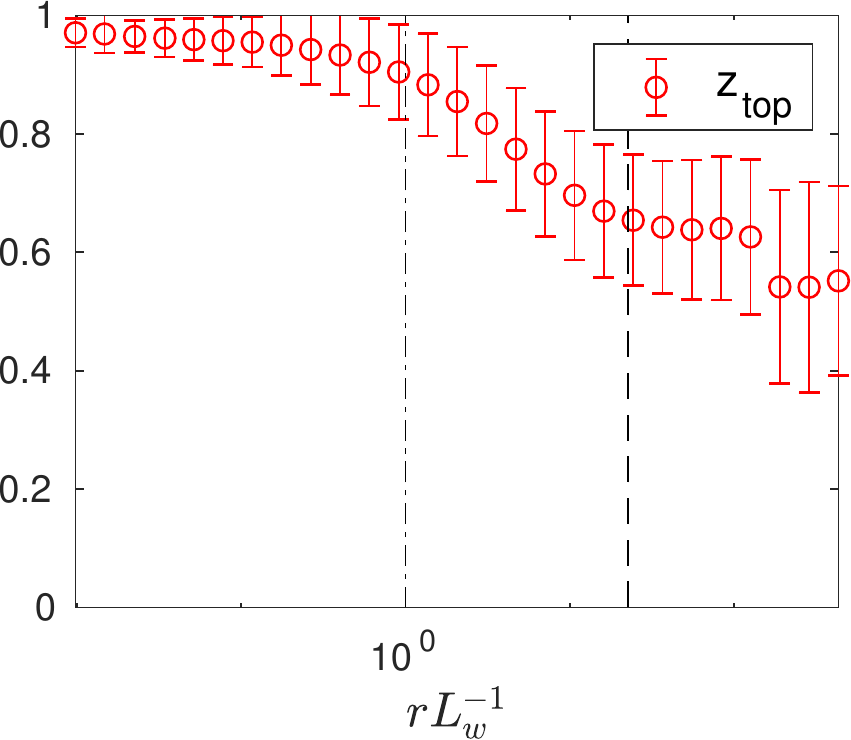}
    \put(-91,25){\bf \scriptsize{(d)}}
  \caption{Linear-log plots of the isotropy measures $C_{\rm iso}$ (a and b)
in the BAM framework and $F$ (c and d) in the AIM framework, versus the
scaled separation, with one-standard-deviation error bars at different heights.
Lines with black dash-dots indicate $r_{\rm iso}L_w^{-1}$; and the black dashed
line indicate $r_{\rm ani}L_w^{-1}$. Measurements at $z_{\rm bot}$ (a), (c) are
shown in blue; whereas  those at $z_{\rm top}$ (b), (d) are shown in
blue.\label{fig:F_C}}
\end{figure*}

\begin{eqnarray}\label{eq:xybam}
    x_{\rm BAM} &=& C_{1c} x_{1c} + C_{2c} x_{2c} + C_{3c} x_{3c}, \nonumber  \\
    y_{\rm BAM} &=& C_{1c} y_{1c} + C_{2c} y_{2c} + C_{3c} y_{3c}; \label{eq:bamcoord} \\
    C_{1c} = \lambda_1&-&\lambda_2,\,C_{2c} = 2(\lambda_2-\lambda_3), \, C_{3c} = 1+3\lambda_3. \nonumber
\end{eqnarray}

In Fig.~\ref{fig:avg_pred_paths}, the averaged relaxation-to-isotropy
trajectories in the $(-I_2, I_3)$ plane (AIM) and in the $(x_{\rm BAM}, y_{\rm
BAM})$ plane (BAM) from measurements along with their counterparts in the
quadratic Rotta model~\cite{Brugger2018} are shown. To obtain these plots, we
average over initial starting points in the AIM and BAM maps for the following
different stability conditions: stable (green), unstable (red), neutral (blue).
Data from measurements at different heights are also presented: dashes with a
circle for $z_{\rm bot}$ and dashed-dotted lines for $z_{\rm top}$. For all
these stability conditions, the averaged relaxation-to-isotropy path occurs
along the plain-strain limit, corresponding to $I_3 = 0$ ($x=0$ in the AIM
framework and the central dotted line in the BAM framework). That is, once
trajectories intersect the plane-strain limit, they are forced to approach
local isotropy along this line.

These measurements differ substantially from the prediction of the quadratic
Rotta model~\cite{Sarkar1990}, which is given as
\begin{eqnarray}\label{eq:quad_rotta}
     \frac{\dd I_2}{\dd \tau} &=& -2(B_1-2)I_2 + 2 B_2 I_3, \nonumber \\
     \frac{\dd I_3}{\dd \tau} &=& -3(B_1-2)I_3 + \frac{1}{2} B_2 I_2^2,
\end{eqnarray}
where $B1 = 3.4,\, B2 = 3(B1-2)$. We integrate these Rotta-model ordinary
differential equations (ODEs) numerically by a fourth-order Runge-Kutta scheme.
Their solutions predicts axisymmetric expansion for initial conditions with
$I_3 > 0$; these solutions lead to the trajectories that are shown in
Figs.\ref{fig:avg_pred_paths} (b) and (d).  Some of these trajectories lie
outside the AIM domain; they are not physically realisable. It is equally
significant that, the return to isotropy, by this Rotta scheme in the BAM
representation, favors the prolate direction when the starting trajectory is at
$x_{BAM}>0.5$.  This return-to-isotropy trajectory, which is obtained in the
Rotta scheme, is {\it opposite} to that in the RSL measurements, in which this
trajectories is attracted to the plain-strain line.   

One plausible explanation is that production terms, which are neglected in the
derivation of the Rotta model, must still be active. These production terms do
not favor axisymmetric expansion or a return to isotropy along the prolate
side in Fig.~\ref{fig:avg_pred_paths} (a), (c). Therefore, we conclude that
common models of scalewise return to isotropy require updating when we employ
then in the RSL.

\begin{table*}
\centering
\caption{ Ensemble-averaged (e.a.) correlation coefficients and the test
statistic $t$ (see text) of $L_T$ with $r_{\rm ani}$, $r_{\rm iso}$, $r_{\rm
ani}L_T^{-1}$, $r_{\rm iso}L_T^{-1}$, at $z_{\rm top}$ and $z_{\rm bot}$ and
for the different stability conditions, stable (s) and unstable (u).
\label{table:corr_coeff}}
\small
\begin{adjustbox}{}
 \begin{tabular}{c c | c c |c c |c c |c c }
 \hline\hline\\ [-1em]
Height & Filter  & $\rho(r_{\rm ani},L_{T})$ & $t(r_{\rm ani},L_{T})$ & $\rho(r_{\rm ani}L_{w}^{-1},L_{T})$ & $t(r_{\rm ani}L_{w}^{-1},L_{T})$ & $\rho(r_{\rm iso},L_{T})$ & $t(r_{\rm iso},L_{T})$ &  $\rho(r_{\rm iso}{L_{w}^{-1}},L_{T})$ & $t(r_{\rm iso}{L_{w}^{-1}},L_{T})$ \\ \\  [-1em]
 \hline
$z_{\rm bot}$  &    e.a.    & $0.160$     & $6.75$   &	$0.035$       & $1.44$     & $0.110$       & $4.60$  & $-0.011$   &$0.43$\\
            &s       & $0.146$     & $4.08$   &	$0.033$       & $0.91$     & $0.187$       & $5.28$  & $ 0.046$   &$1.27$\\
            &u       & $0.098$     & $2.66$   &	$0.079$       & $2.13$     & $0.016$       & $0.43$  & $ 0.001$   &$0.02$\\
\hline
$z_{\rm top}$  &    e.a.    & $0.167$     & $7.05$   &	$0.016$       & $0.68$     & $0.019$       & $0.80$  & $-0.014$   &$0.57$\\
            &s       & $0.183$     & $5.73$   &	$0.008$       & $0.24$     & $0.048$       & $1.49$  & $ 0.001$   &$0.03$\\
            &u       & $0.078$     & $2.03$   &	$0.105$       & $2.74$     & $0.056$       & $1.46$  & $-0.044$   &$1.14$\\

 \hline\hline
 \end{tabular}
\end{adjustbox}
\end{table*}

In Figs.\ref{fig:F_C}, linear-log plots of the isotropy measures 
$C_{\rm iso}$ (Figs.\ref{fig:F_C} a and b) and $F$ (Figs.\ref{fig:F_C} c and d) are shown in the BAM and the AIM frameworks,
respectively,  versus the scaled separation. Data for different heights, blue
for $z_{\rm bot}$ and red for $z_{\rm top}$, are shown along with the lines
$r_{\rm iso}L_w^{-1}$ (dashed dots) and $r_{\rm ani}L_w^{-1}$ (dashed) defined
in Eqs. ~(\ref{eq:L_iso_def}) and (\ref{eq:L_ani_def}). Relaxation to small-scale
quasi-isotropy, commences at $r_{\rm ani}/L_w$, whence the anisotropy measure
$F$ increases with decreasing scales, close to the $F\simeq 1+9I_2$ limit, as a
consequence of the relaxation along the plain-strain limit. This relaxation
concludes scales $r_{\rm iso}/L_w$, given by:
\begin{gather}
r_{\rm iso}/L_w = \max_{r/L_w}\{C_{\rm iso} (r/L_w) \geq 0.9 \max {\{C_{\rm iso}\}} \} \label{eq:L_iso_def}; \\
r_{\rm ani}/L_w = \min_{r/L_w} \Big\{C_{\rm iso}(r/L_w) < 1.1 \min\{C_{\rm iso}\} \Big\} \label{eq:L_ani_def}.
\end{gather}

We find that $r_{\rm iso}/L_w \simeq l^{z_{\rm top}}/L_w$ for $z_{\rm top}$ and
$r_{\rm iso}/L_w \simeq l^{z_{\rm bot}}/L_w$ for $z_{\rm bot}$. In the BAM
measure (Fig.~\ref{fig:F_C}), at the smallest scales resolved, $C_{\rm iso}$
saturates to a value $~ 0.8$ indicating that signatures of anisotropy are
still retained in the small-scale eddies. At the same time, the largest scales
above the canopy top are less anisotropic, in the AIM measure, compared to
higher up in the RSL. Hence, we conjecture that the interactions of the
residual wakes, which originate from vegetation elements, with anisotropic
attached eddies, play a role in the randomization of energies and the breakdown
of eddies in the cascade.

We also explore the dependence of these characteristic scales, $r_{\rm ani}$
and $r_{\rm iso}$, on the thermal integral length scale $L_T$ (Table
\ref{table:all_params}). Such studies have been carried out in experiments
above urban canopies~\cite{Liu2017} and also in the roughness and inertial
sublayers~\cite{Brugger2018}. In particular, we characterize the correlations
between $L_T$ and $r_{\rm ani}, \, r_{\rm iso}, \, r_{\rm ani}L_T^{-1}, \,
r_{\rm iso}L_T^{-1}$ at $z_{\rm top}$ and $z_{\rm bot}$, for the different
stability conditions given in Table~\ref{table:corr_coeff}.

The $t$-test in Eq. (~\ref{eq:t(A,B)_def}) is performed at a $5\%$ significance
level (i.e., $t_{\alpha/2,n-2} \simeq 1.96$.) with variables $A$ and $B$ set to
one of the variables $L_T$ and $r_{\rm ani}, \, r_{\rm iso}, \, r_{\rm
ani}L_T^{-1}, \, r_{\rm iso}L_T^{-1}$. The resulting correlation coefficients
and their significance at $z_{\rm top}$ and $z_{\rm bot}$ and for the different
stability conditions are presented in Table~\ref{table:corr_coeff}. These
correlation coefficients are small, $<0.2$~\cite{Brugger2018}; however, the
following points can be made:
\begin{itemize}
    \item For unstable stratification, and at both $z_{\rm top}$ and $z_{\rm
bot}$, there is significant correlation between $r_{\rm ani}$ and $L_T$,
irrespective of the normalization with $L_w$.
    \item For stable stratification and unfiltered ensemble averages,  $r_{\rm
ani}$ has a significant correlation with $L_T$; this is lost on normalization
with $L_w$.
    \item For unstable stratification, $r_{\rm iso}$ is uncorrelated with
$L_T$,  irrespective of the normalization.
    \item For stable stratification and unfiltered, ensemble-averages, just
above the canopy top (at $z_{\rm bot}$), $r_{\rm iso}$ is correlated with
$L_T$, but, on normalisation with $L_w$, the correlation coefficients are
insignificant; higher up in the roughness sublayer, we do not find conclusive
evidence for such an interpretation.
\end{itemize}

We normalise the separation $r$ by $L_w$  because it is the most restricted
length scale. Our significance tests also show that much of the dependence of
$r_{\rm ani}$ on $L_T$, at both $z_{\rm bot}$ and $z_{\rm top}$ (i.e.,
throughout the RSL), and of $r_{\rm iso}$, at $z_{\rm bot}$, arises from the
dependence of $L_w$ on $\zeta$. In general, these results are similar to the
results of measurements in the ASL and CSL \cite{Brugger2018}. For unstable
stratification, however, we do not find conclusive evidence in support of such
dependence.

\subsection{\label{sec:Result_Multifractality_Analysis}Multifractality Analysis}

The results for the Hurst exponents $h(q)$ (obtained from the MFDFA) are
presented first, followed by the results for Hurst surfaces $h(q,s)$ (obtained
from MMA) based on stability. We seek to find the cause of multifractality of
the temperature and velocity series. For the analysis below, only those
$30$-minute periods that satisfy the tests mentioned in
Sec.~\ref{sec:Postprocessing} are used. 

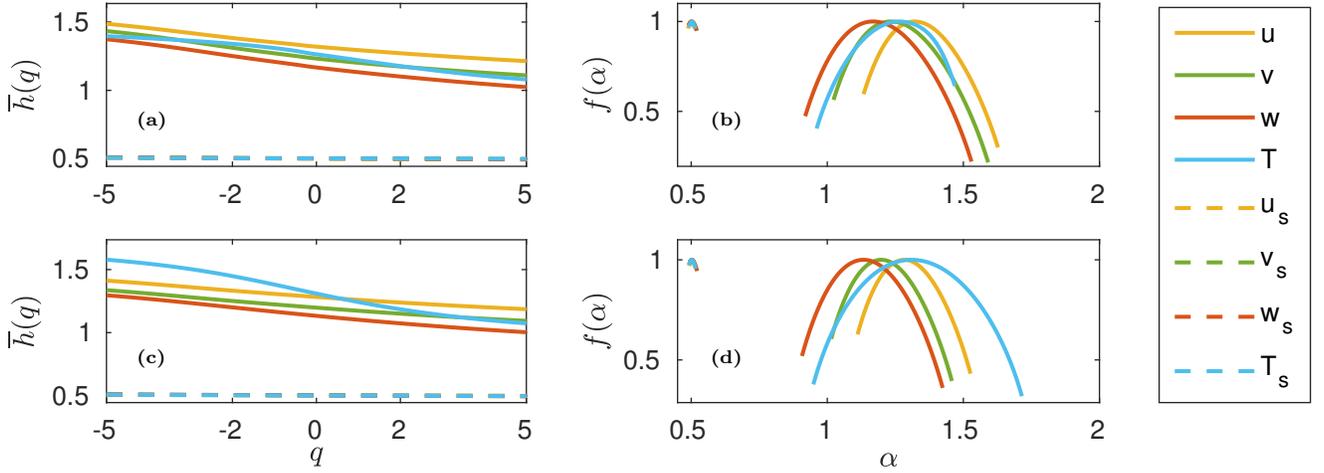
\begin{figure*}[ht]
\includegraphics[]{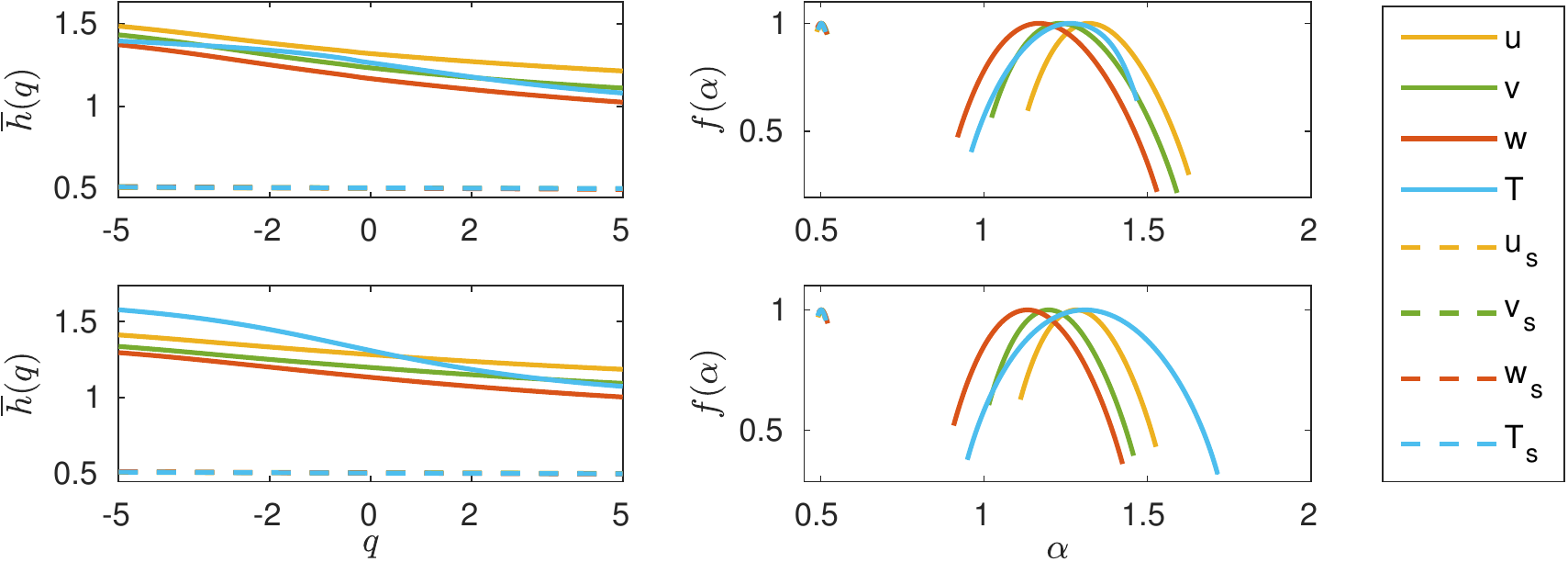}
\put(-445,130){\bf \scriptsize{(a)}} \put(-228,130){\bf \scriptsize{(b)}}
\put(-445, 40){\bf \scriptsize{(c)}} \put(-228, 40){\bf \scriptsize{(d)}}
\caption{Plots versus $q$ of the Hurst exponents $\overline{h}_{u},\,
\overline{h}_{v},\, \overline{h}_{w},$ and $\overline{h}_{T}$ and the
corresponding  the $f(\alpha)$ spectra in (b) and (d), for the original (full
lines) and shuffled (dashed lines) series from the MFDFA at $z_{\rm bot}$ (a)
and (b) and $z_{\rm top}$ (c) and (d). The shuffled series are monofractal with
Hurst exponent $\overline{h}(q)=0.5 \pm 0.01$; correspondingly the multifractal
spectrum has a width $\delta \alpha = \alpha_{\rm max} - \alpha_{\rm min}$,
which is an order of magnitude smaller than its counterpart for the original
series. \label{fig:mfdfa1}}
\end{figure*}

In Figs.~\ref{fig:mfdfa1} (a) and (c) we plot versus $q$ the Hurst exponents
$\overline{h}_{u},\, \overline{h}_{v},\, \overline{h}_{w},$ and
$\overline{h}_{T}$ and the corresponding  $f(\alpha)$ spectra in
Figs.~\ref{fig:mfdfa1} (b) and (d), for the original (full lines) and shuffled
(dashed lines) series from the MFDFA at $z_{\rm bot}$  in
Figs.~\ref{fig:mfdfa1} (a) and (b) and $z_{\rm top}$ in Figs.~\ref{fig:mfdfa1}
(c) and (d). 

The shuffled series are monofractal, with a $q$-independent Hurst exponent
$=0.5 \pm 0.01$, so the multifractal spectrum has a width $\delta \alpha =
\alpha_{\rm max} - \alpha_{\rm min}$, which is an order of magnitude smaller
than its counterpart for $f(\alpha)$ spectra that we obtain from the original
series. The corresponding Hurst surfaces for all the velocity components and
temperature (not shown) are constant surfaces with $h(q,s)=0.5 \pm 0.01$ and
contain no extra information than the corresponding Hurst exponents $h(q)$. 

Multifractality can be traced to (1) long-range correlations or (2) a broad
probability distribution function~\cite{Kantelhardt2002}. Random shuffling of a
time series eliminates the long-range correlations, but it preserves the
probability distribution. Thus, we can conclude that multifractality in the
velocity and the temperature time series arises because of long-range
correlations, which scale differently for the different velocity components and
temperature.  The cause of  multifractality in our studies is different from
that in Refs. \cite{Cava2009, Cava2012},  where amplitude variations, studied
by a telegraphic approximation~\cite{Sreenivasan2006}, leads to multifractal
fluctuations.

At this point, it is instructive to recall the relation of the Hurst exponents
$h(q)$, with the conventional, order-$p$, multiscaling exponents that are
defined via order-$p$ moments of velocity increments. We illustrate this for
the velocity component $w$; our discussion follows that of Meneveau and
Sreenivasan (see especially pages 64-67 of Ref.~\cite{Meneveau1987} and  Eq.
(5.6)).\\ We begin with order-$p$ velocity structure function $S^w_p (r) \equiv
\langle |[w({\bf r}_0+{\bf r}) - w({\bf r})_0]|^p \rangle$, where the angular
brackets denote an average over the statistically steady state and the origin
$\bf{r}_0$; for the separation $r = |{\bf{r}}|$ in the inertial range of scales
(much larger than the dissipation scale and much smaller than the length scale
at which energy is injected into the system), $S_p^w \sim r^{\xi_p^w}$; this
defines the order-$p$ multiscaling exponents $\xi_p^w$. (In homogeneous,
isotropic turbulence, these exponents are defined often by using the
longitudinal component of the velocity ~\cite{Pandit2009}). As discussed
elsewhere~\cite{Meneveau1987}, these exponents are related to the Hurst
exponents as follows:
\begin{eqnarray} \label{eq:zetap}
\xi_p &=& (p/3-1) D_{p/3}+1; \\
\xi_p &=& \tau(q) +1 = qh(q); p/3 = q;
\end{eqnarray}
here, $D_{p/3}$ is the generalized dimensions and $h(q)$ is the Hurst exponent
calculated before and the superscript indicating the corresponding component
has been dropped. From Eqs. ({2.13,2.14}) in  Ref.~\cite{Meneveau1987} and Eq.
(\ref{eq:falphadef}) above, it is evident that $D_q = \tau(q)/(q-1)$.

\begin{figure}
    \centering
    \includegraphics[width =0.45\columnwidth]{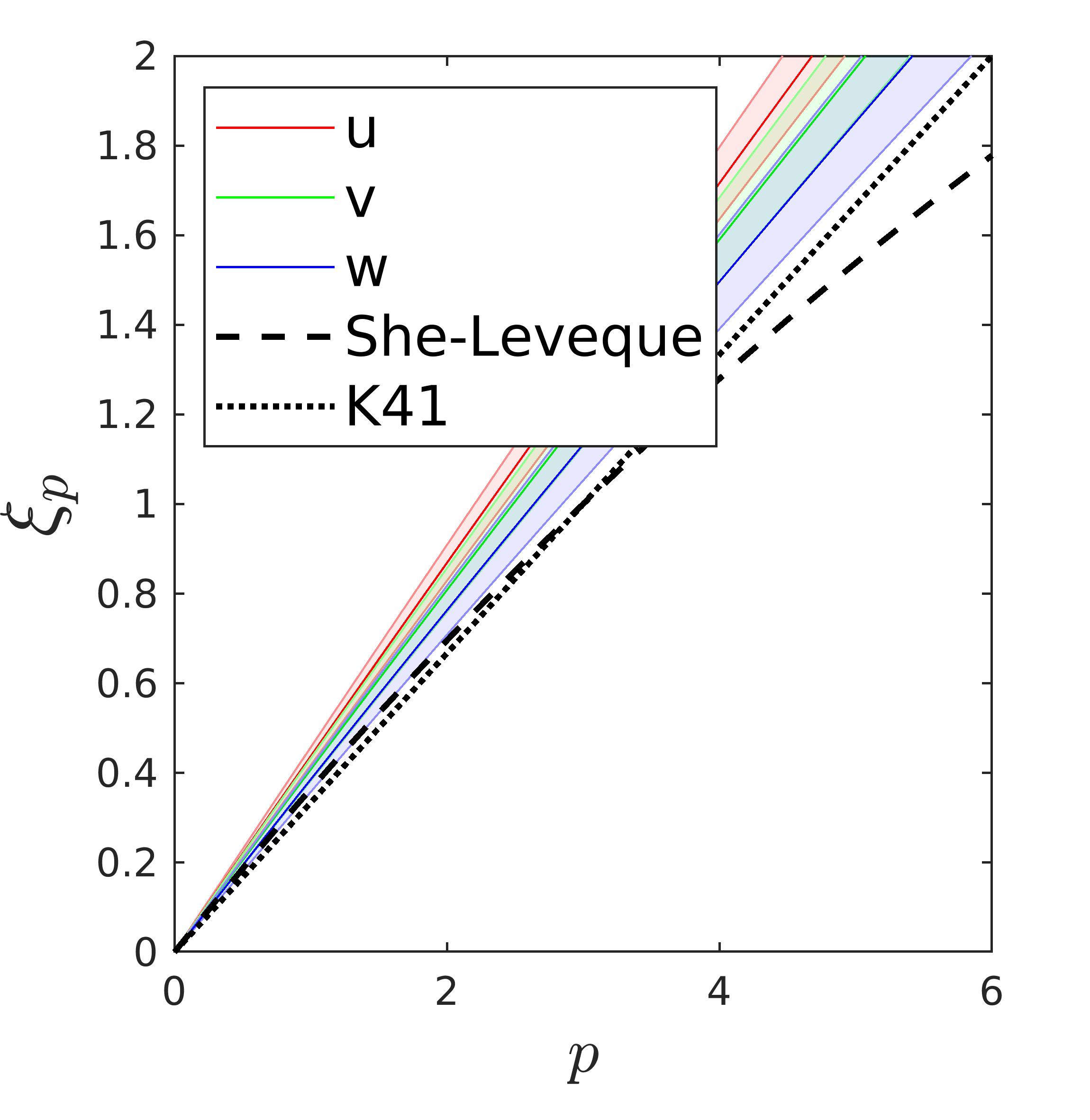} 
    \put(-25,25){{\makebox[0.01\textwidth][r]{\bf \scriptsize{(a)}}}}
    \includegraphics[width =0.46\columnwidth]{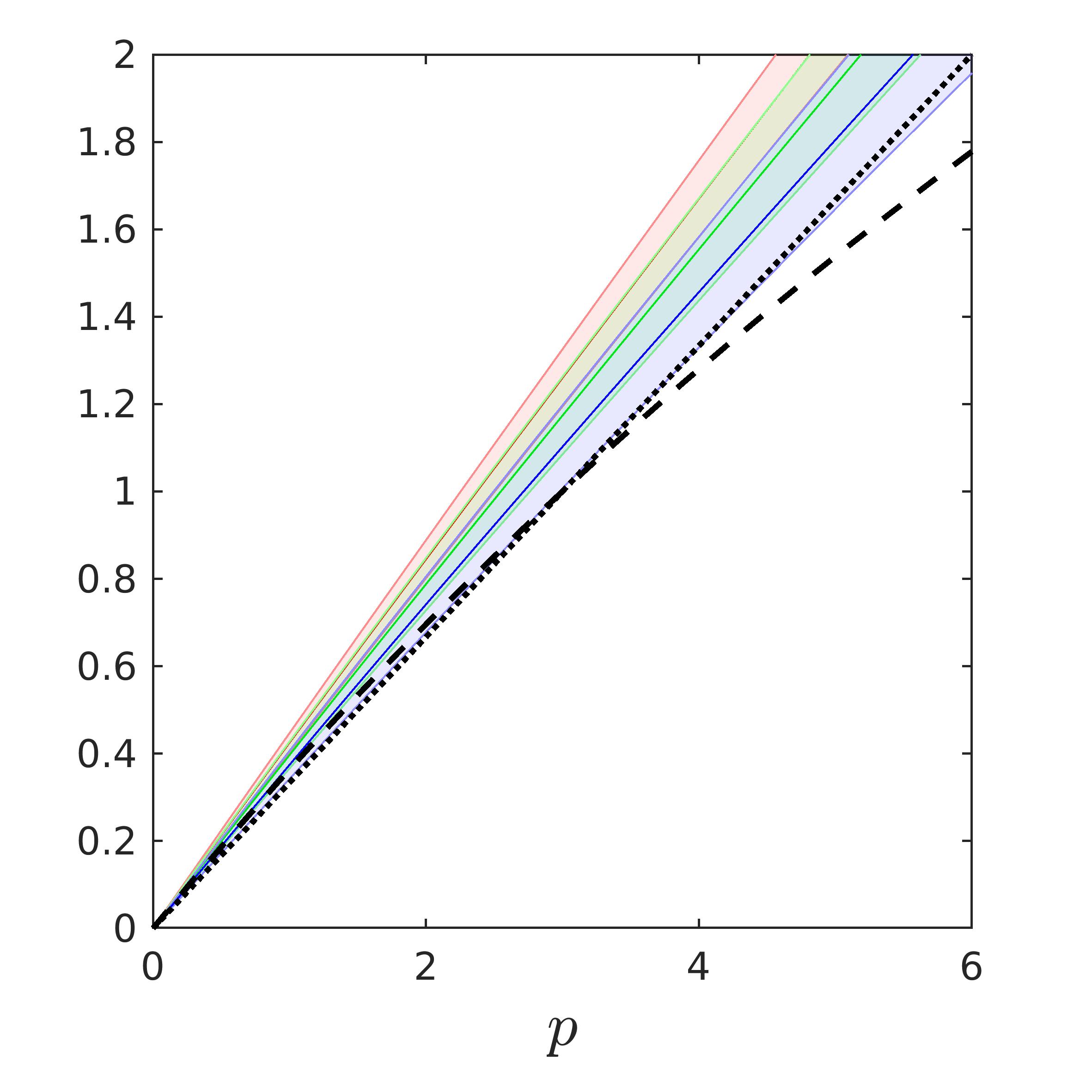}
    \put(-25,25){{\makebox[0.01\textwidth][r]{\bf \scriptsize{(b)}}}}
    \caption{Plots vs order $p$ of multiscaling exponents $\xi^{u_i}_p$ for
different velocity components $u$(red), $v$(green), $w$(blue) at different
heights, $z_{\rm bot}$ (a) and $z_{\rm top}$ (b), obtained from our
multifractal analysis. \label{fig:Xi}}
\end{figure}

\begin{figure}
    \centering
    \includegraphics[width =0.45\columnwidth]{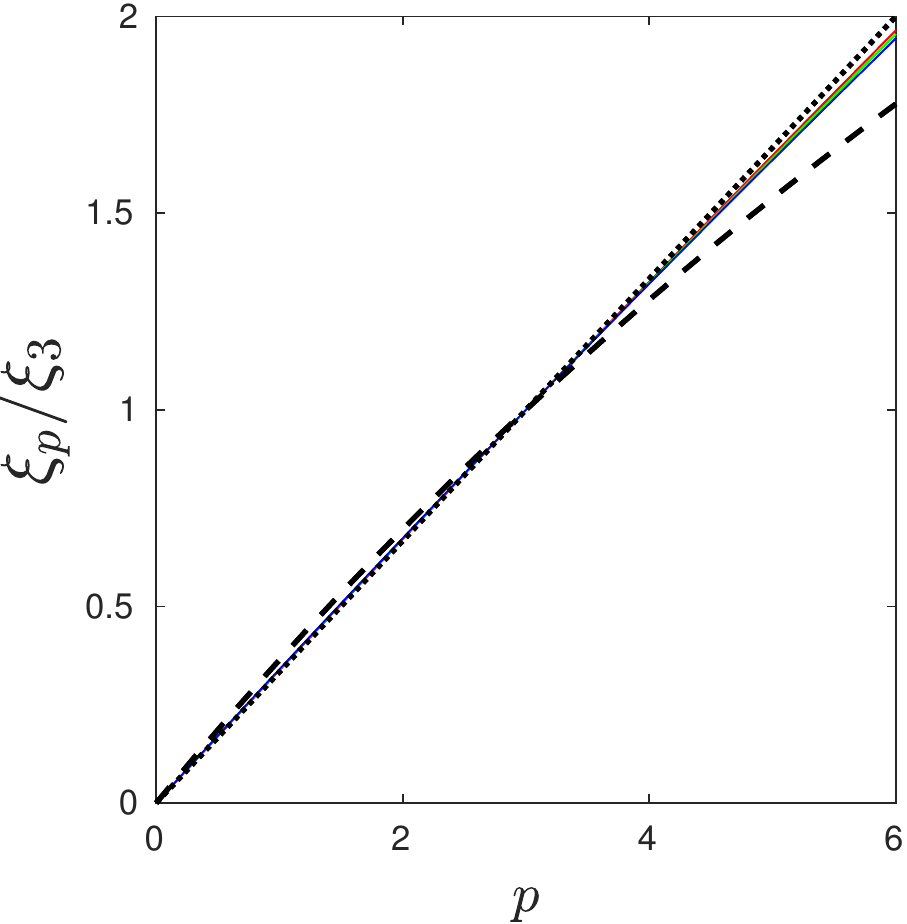}
    \put(-20,25){{\makebox[0.01\textwidth][r]{\bf \scriptsize{(a)}}}}
    \includegraphics[width =0.4\columnwidth]{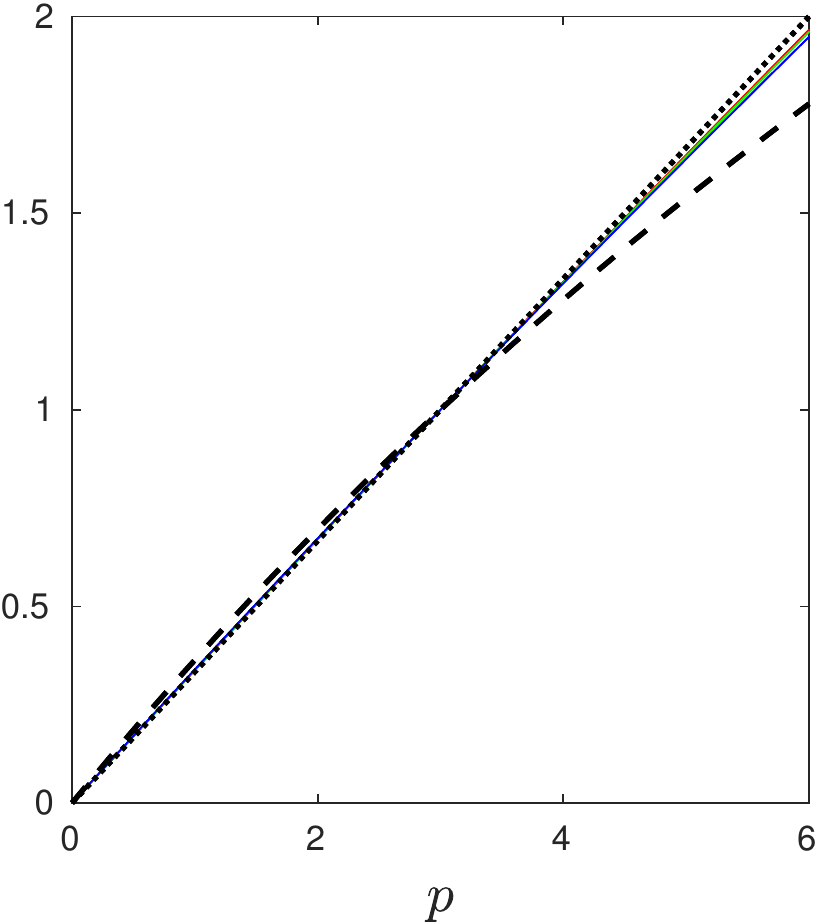}
    \put(-20,25){{\makebox[0.01\textwidth][r]{\bf \scriptsize{(b)}}}}
    \caption{Plots vs order $p$ of ESS-normalized multiscaling exponents
$\xi^{u_i}_p/\xi^{u_i}_3$, for different velocity components $u$(red),
$v$(green), $w$(blue) at different heights, $z_{\rm bot}$ (a) and $z_{\rm top}$
(b), obtained from the proposed multifractal analysis here. \label{fig:ESS_Xi}}
\end{figure}

In Fig.~\ref{fig:Xi}  we plot these order-$p$ exponents, for all the three
components $u$, $v$, and $w$ versus $p$ (for $0 \leq p \leq 6$); we also include
a plot of the K41 prediction $\xi_p^w = p/3$ and the She-Leveque (SL)
parameterization~\cite{SheLeveque,Pandit2009}. The SL model is chosen because it
has no tunable intermittency parameters. The error bars are large, so it is
difficult to use this plot to make definitive statements about multiscaling;
however, there is a systematic difference between the exponents for $u$, $v$, and
$w$~\cite{Katul2001,Katul2009};  $h_{w}(q)$ is consistently smaller than
$h_{u}(q)$ and $h_{v}(q)$. We also see that
$\delta\alpha_u<\delta\alpha_v<\delta\alpha_w$. Thus, we can conclude that the
largest (smallest) is found in intermittency in the wall-normal (longitudinal)
velocity components; this is consistent with the observations reported in the
atmospheric surface layer~\cite{Katul2001,Katul2009}. The anisotropy that we
observe in these scaling exponents is, however, less pronounced here than in
other studies over pine forests \cite{Katul2009}; we observe $L_u/L_w \simeq 6$
compared to $L_u/L_w>10$ in Ref.~\cite{Katul2009}. We find $\alpha_0 \simeq
1.12$, consistent with values observed in Ref.~\cite{Meneveau1987}.

The extended-self-similarity (ESS)
procedure~\cite{BenziESS,Pandit2009,Chakraborty2009}, in which the exponent
ratios, say $\xi_p^w/\xi_3^w$, are determined from the slopes of log-log plots
of $S_p^w$ versus $S_3^w$ has also been used frequently to measure the scaling
of these structure functions and thus to characterize multifractality
~\cite{Katul2001,Katul2009}. 

We have carried out the analog of this procedure by taking ratios of our Hurst exponents: 
\begin{equation}
    \xi_p^w/\xi_3^w = qh_w(q)/h_w(1),\, p = 3q.
\end{equation}

From these we obtain the ESS plots of the exponent ratios, which we display in
Fig.~\ref{fig:ESS_Xi}, for $0 \leq p \leq 6$. As we expect from the general
experience with ESS~\cite{Pandit2009,SheLeveque}, these exponent ratios do
indicate mutifractality; but the difference between the ratios for the
different components, $u, v,$ and $w$, is not very striking, for low orders ($0
\leq p \leq 6$); higher-order exponents require better statistics than can be
obtained from the field data here. In Appendix.\ref{sec:appendix_Sf}, we also
present the result of traditional structure function analysis. From one point
measurements, it is not possible to perform the SO(3) decomposition
~\cite{Arad1999, Biferale05, KurienSreenivasan}, which can be used
alternatively to calculate the anisotropic scaling of structure functions.
The nature and degree of multifractality of temperature fluctuations is quite
different from that of velocity fluctuations: immediately above the canopy top,
temperature fluctuations are insensitive to fluctuations of small magnitudes;
thus, we see a multifractal spectrum (Fig.~\ref{fig:mfdfa1} (b)) with a long
left tail \cite{Ihlen2012}; this is not evident higher up in the RSL, where
temperature fluctuations are more multifractal (Fig.~\ref{fig:mfdfa1} (b), (d))
compared to those above the canopy.

\begin{figure*}[ht]
\centering
\includegraphics[width=0.23\textwidth]{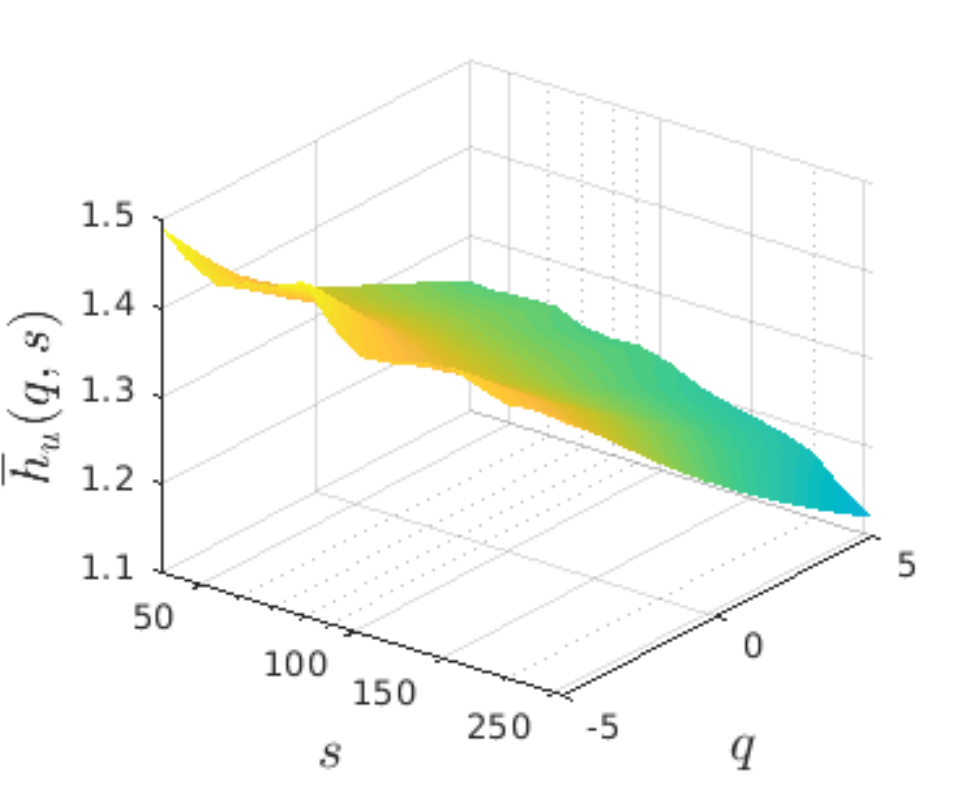}\put(-120,80){\bf
\scriptsize{(a)}}
\includegraphics[width=0.23\textwidth]{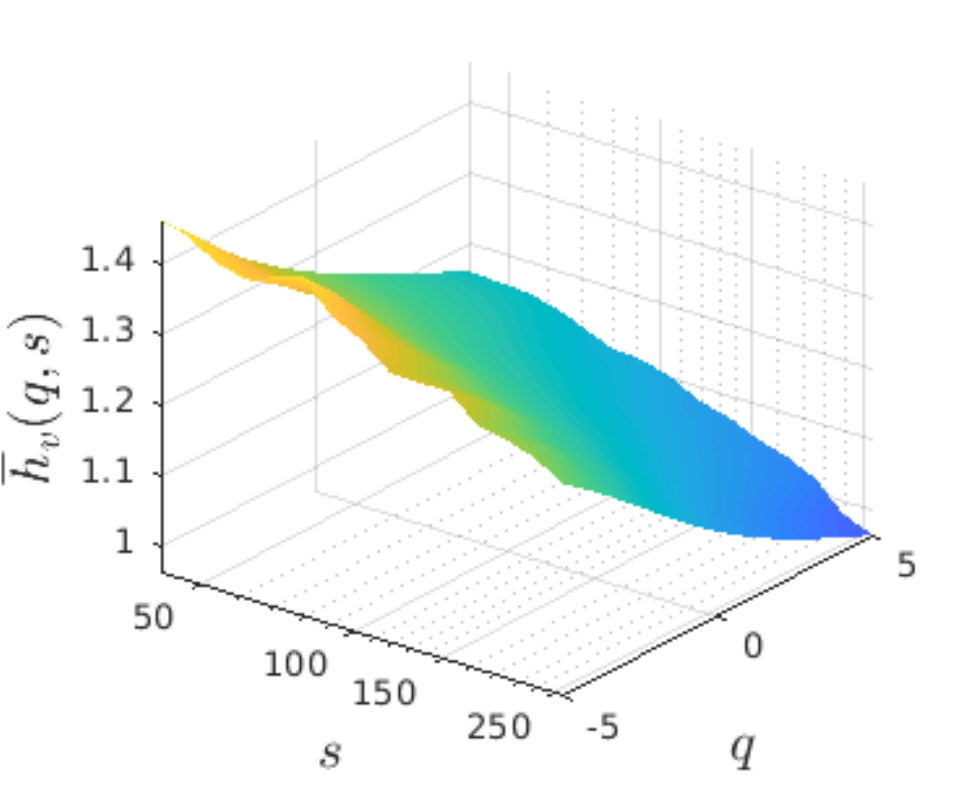}\put(-120,80){\bf
\scriptsize{(b)}}
\includegraphics[width=0.23\textwidth]{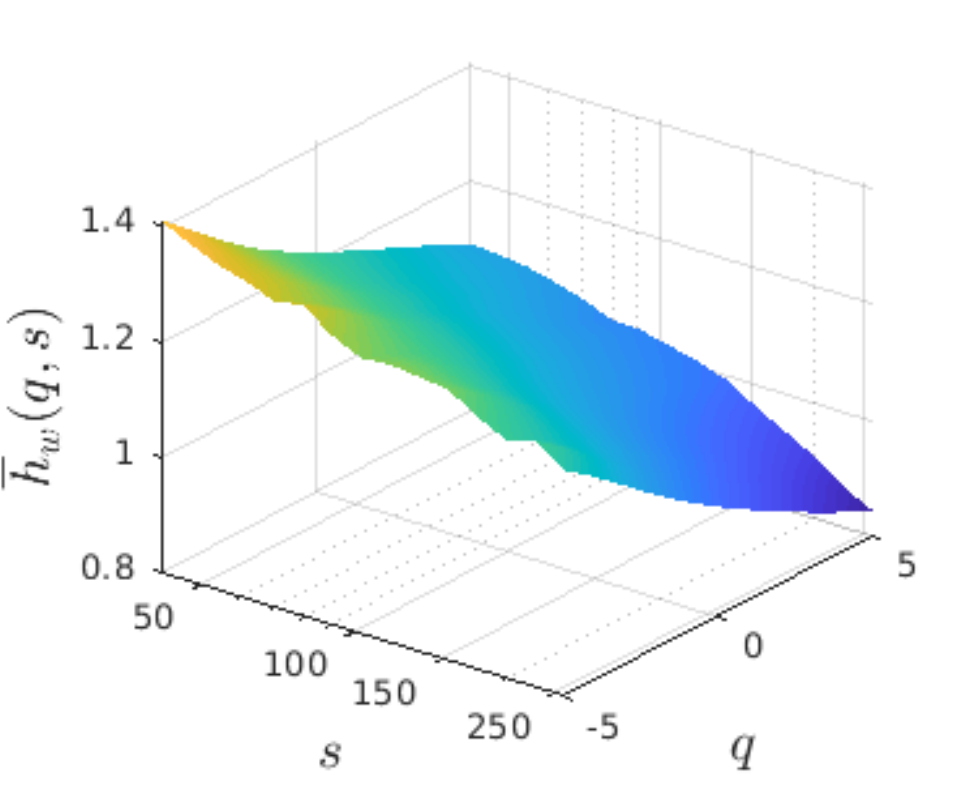}\put(-120,80){\bf
\scriptsize{(c)}}
\includegraphics[width=0.26\textwidth]{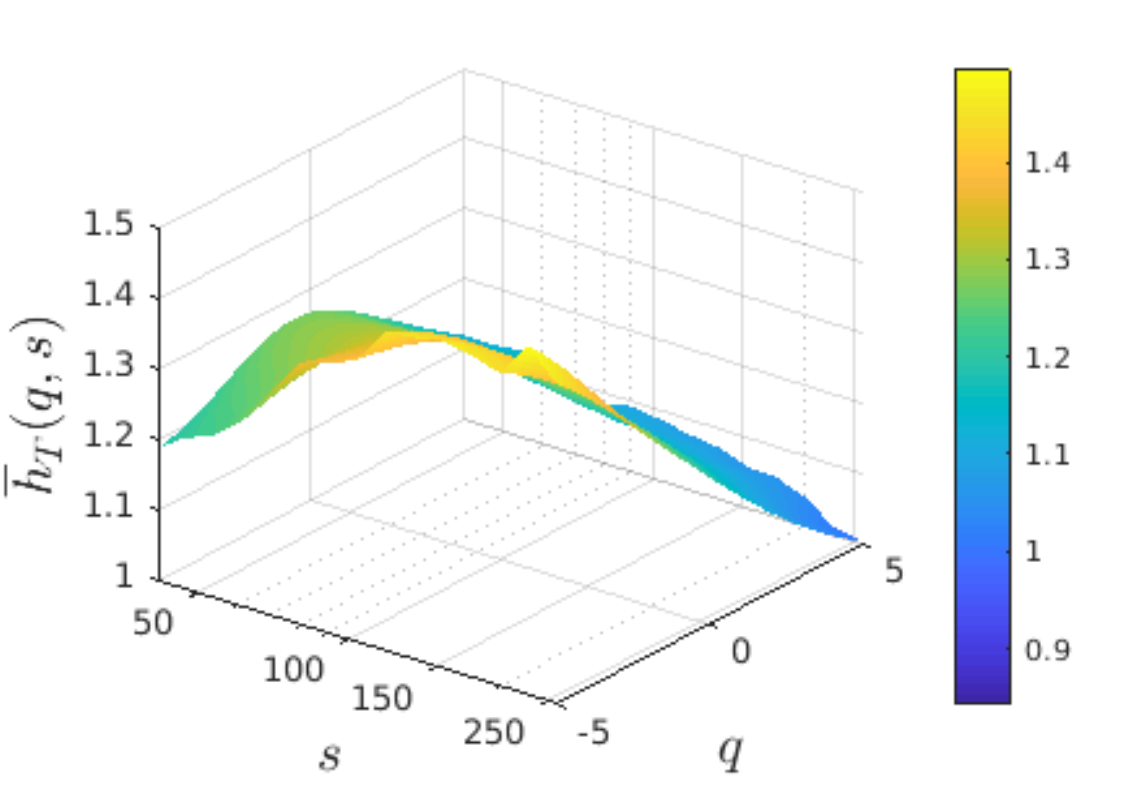}\put(-120,80){\bf
\scriptsize{(d)}}\\
\includegraphics[width=0.23\textwidth]{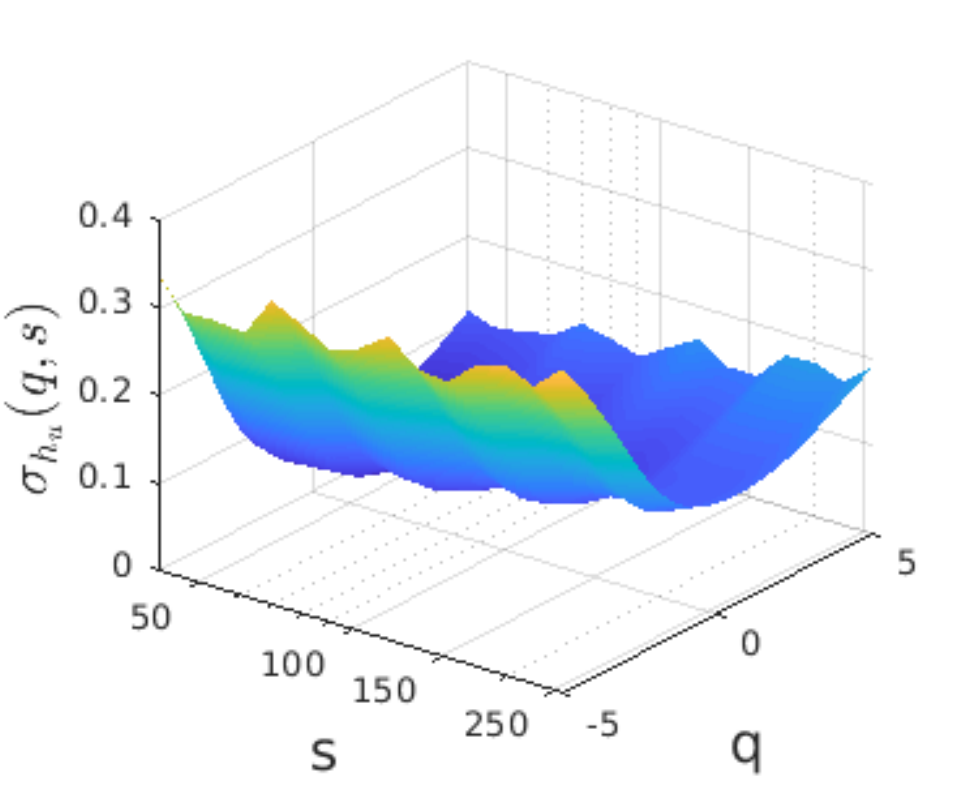}\put(-120,70){\bf
\scriptsize{(e)}}
\includegraphics[width=0.23\textwidth]{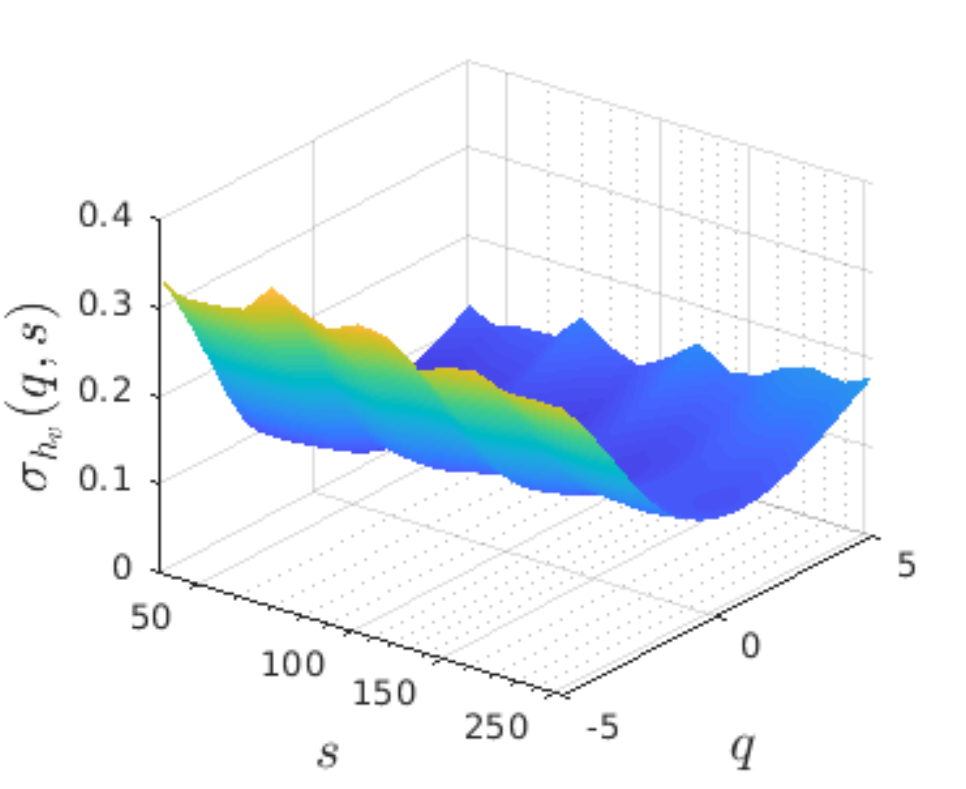}\put(-120,70){\bf
\scriptsize{(f)}}
\includegraphics[width=0.23\textwidth]{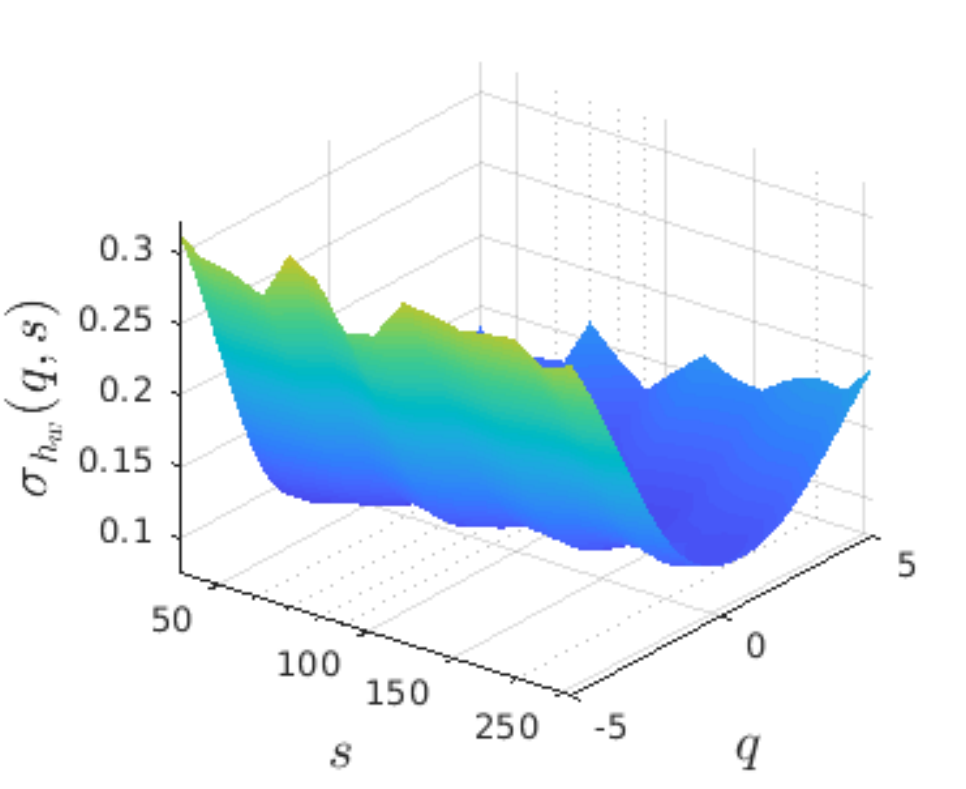}\put(-120,70){\bf
\scriptsize{(g)}}
\includegraphics[width=0.26\textwidth]{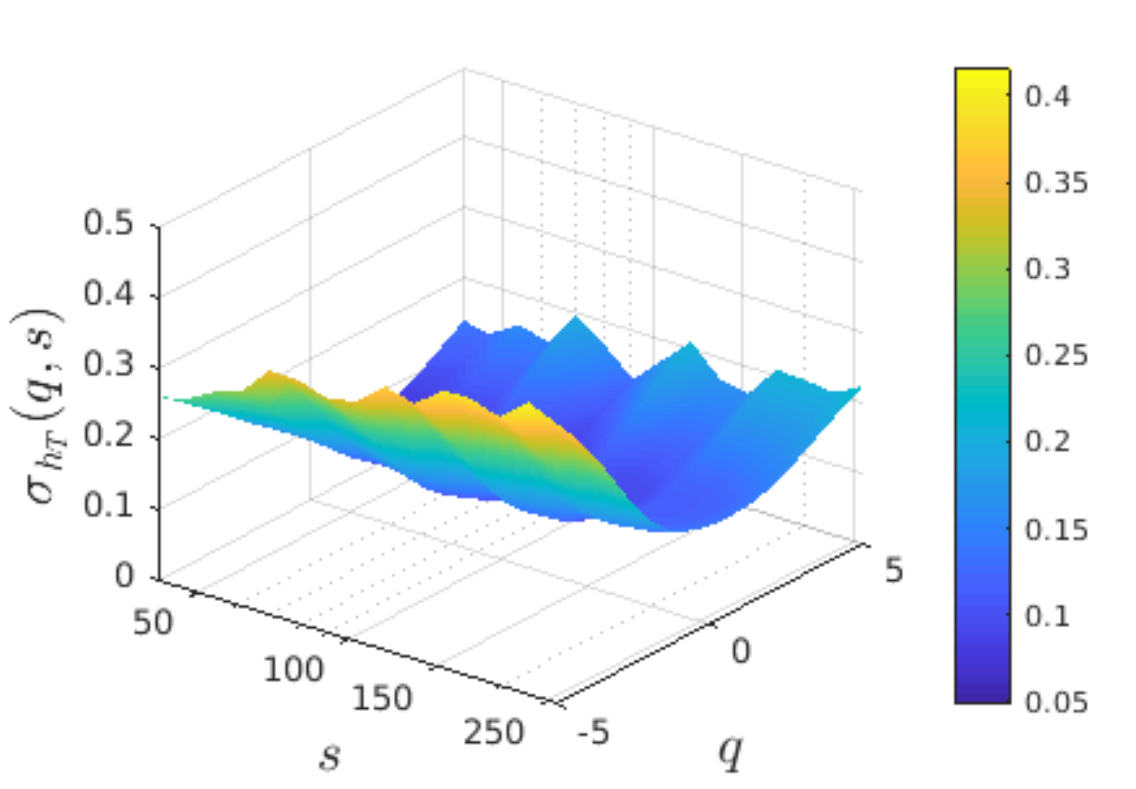}\put(-120,70){\bf
\scriptsize{(h)}}
\caption{Ensemble-averaged Hurst-surface plots versus $q$ and $s$ for (a) the
longitudinal and (b) the transverse components of the velocity and (c) of the
temperature; the surfaces showing the standard deviations of these surface
plots are given, respectively, in (d), (e), and (f), for stable stratification
and at $z_{\rm bot}$. Hurst-surface plots for different stability conditions
and at different heights are qualitatively similar. \label{fig:mHSstdHS}}
\end{figure*}

To distinguish between different Hurst surfaces, the distance $d_{\theta \phi}$
between the surfaces $h_{\theta}$ and $h_{\phi}$ for the variables $\theta$ and
$\phi$ is used and is defined as (see, e.g.,
Refs.~\cite{Gieraltowski2012},~\cite{Zeng2016}):
\begin{align}
h_{\phi^S}(q,s) &= h_{\phi}(q,s) + [\langle h_{\theta}(q,s) \rangle -\langle h_{\phi}(q,s) \rangle ]; \nonumber \\ 
d_{\theta \phi} &= \{\langle[h_{\theta}(q,s)-h_{\phi^S}(q,s)]^2\rangle\}^{1/2}[\langle h_{\theta}(q,s) \rangle ]^{-1} .\label{hd}
\end{align}

The variables $\theta$ and $\phi$ can be temperature or any one of the velocity
components $u, \, v,$ or $w$.\\ In Fig.~\ref{fig:mHSstdHS} representative
plots are given for  $q$ and $s$ of $\overline{h}_u(q,s)$,
$\overline{h}_v(q,s)$, $\overline{h}_w(q,s)$, $\overline{h}_T(q,s)$ (averaged
Hurst surfaces) and their standard deviations
($\sigma_{h_u},\sigma_{h_v},\sigma_{h_w},\sigma_{h_T}$), corresponding,
respectively, to streamwise, spanwise and wall-normal velocity components, and
the temperature, on the scale $s$ and for the order $q$.
$\overline{h}_{u_i}(q,s),\,i=1,2,3$ show weak scale $s$ dependence. Both MMA
and MFDFA show interesting features for temperature-fluctuation scaling
exponents, $h_T(q)$ and $\overline{h}_{u_i}(q,s)$ respectively; we refer the
reader to Ref.~\cite{Cava2009} for other explorations.

\subsubsection{Anisotropy in Hurst Surfaces}

\begin{figure*}[t]
\includegraphics[width = 0.5\textwidth]{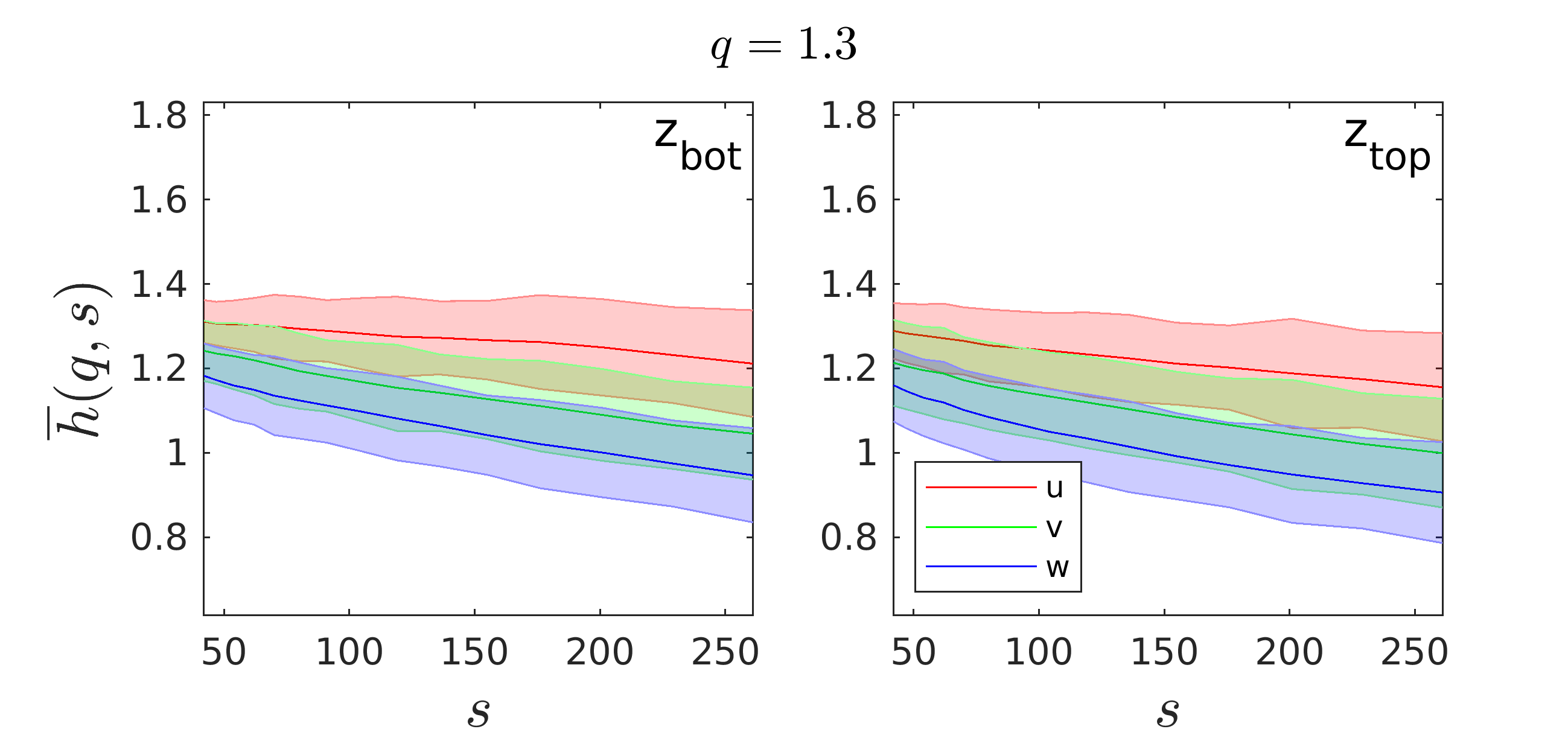}
\put(-210,90){\bf \scriptsize{(a)}} \put(-100,90){\bf \scriptsize{(b)}}
\includegraphics[width = 0.5\textwidth]{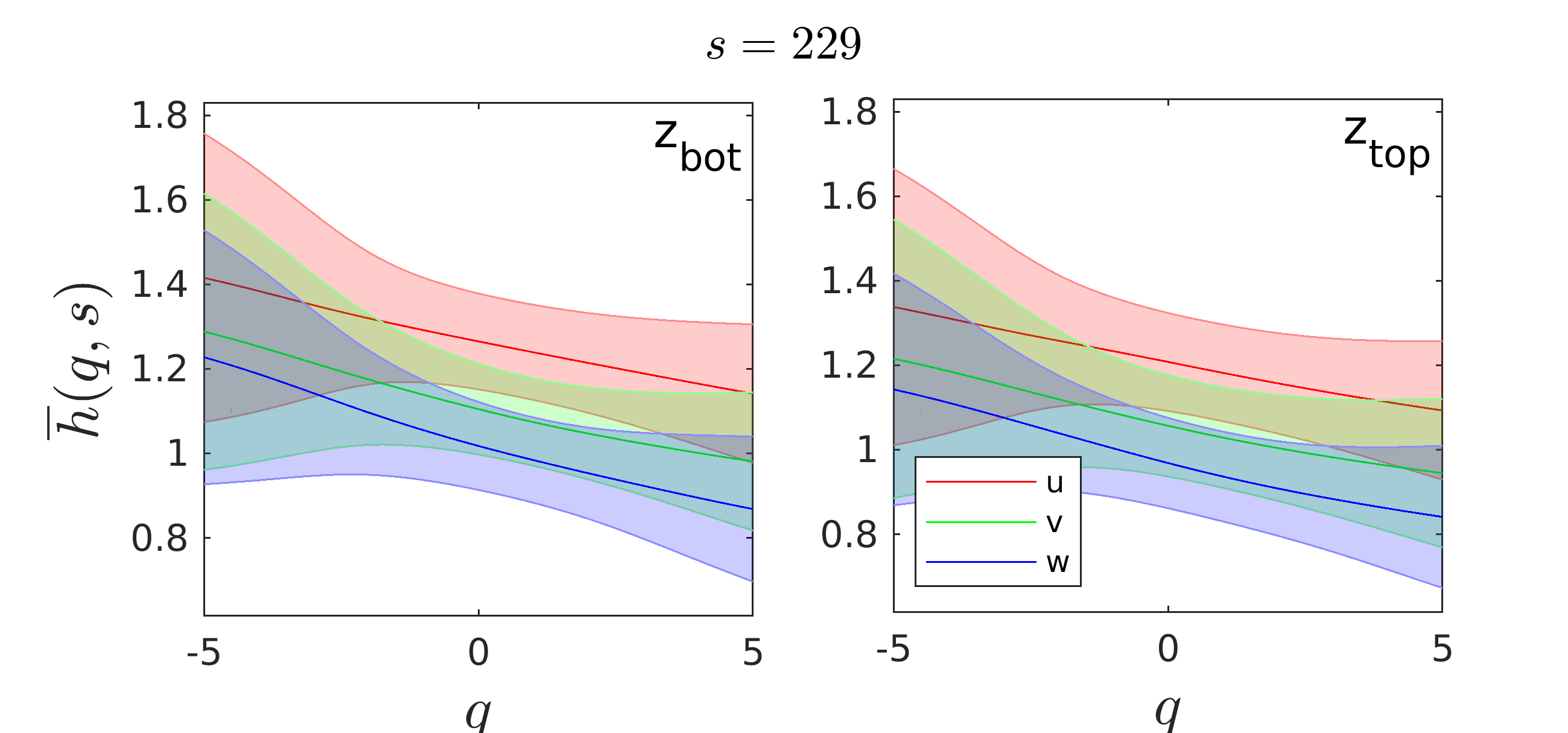}
\put(-210,90){\bf \scriptsize{(c)}} \put(-100,90){\bf \scriptsize{(d)}}
\caption{Cross-sections of averaged Hurst surfaces (with one-standard-deviation
error bars \cite{shadedEb}) corresponding to different velocity components
($\overline{h}_u(q,s)$ (red), $\overline{h}_v(q,s)$ (green)
$\overline{h}_w(q,s)$ (blue)) for constant $q=1.3$ (a) and (b) and
cross-sections for constant scale $s=229$ (c) and (d). (a) and (c) correspond
to measurements at $z_{\rm bot}$ and (c) and (d) correspond to measurements at
$z_{\rm top}$. These representative plots, correspond to measurements under
stable stratification ($\zeta > 0.01 $) and highlights the anisotropy in the
scaling of velocity fluctuations.} \label{fig:van_sq_stable}
\end{figure*}

In Figs.~\ref{fig:van_sq_stable} (a) and (b) we show, as a function of $s$,
the cross-section of constant $q=1.3$ of averaged Hurst surfaces (with
one-standard-deviation error bars); and in Figs.~\ref{fig:van_sq_stable} (c)
and (d) we show, as a function of $q$, the cross-section of constant $s=229$ of
averaged Hurst surfaces (with one-standard-deviation error bars). We also show
plots of $\overline{h}_u(q,s)$ (red), $\overline{h}_v(q,s)$ (green), and
$\overline{h}_w(q,s)$ (blue) at different heights $z_{\rm bot}$(a) and (c) and
$z_{\rm top}$ (b) and (d) for the case of stable stratification. These
variations of the Hurst surfaces can be best visualized by looking at the
animations that we provide in Ref.~\cite{vid:stable_anisotropy_23}.

\begin{figure*}[]
    \includegraphics[width=0.5\textwidth]{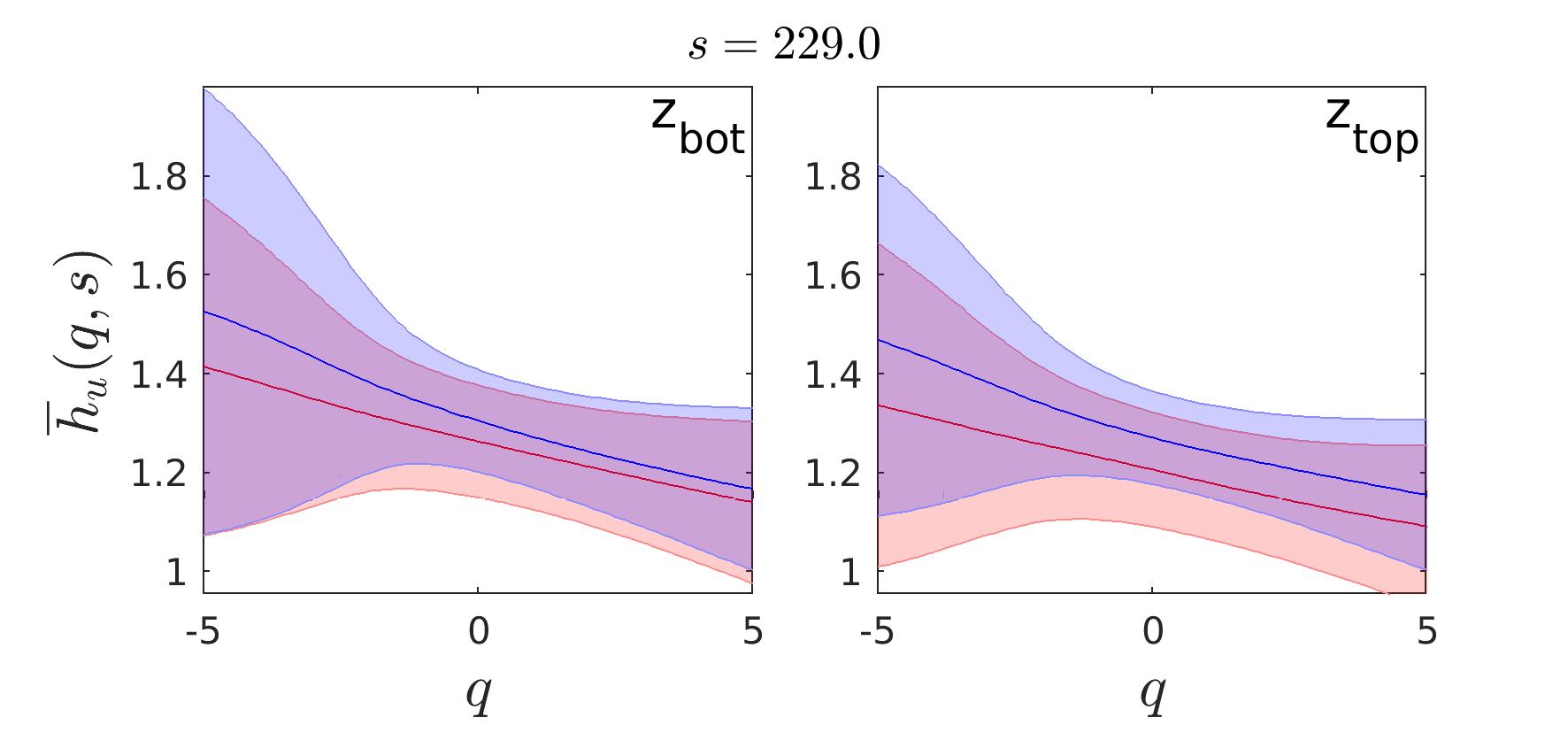}
    \put(-210,90){\bf \scriptsize{(a)}} \put(-100,90){\bf \scriptsize{(b)}}
    \includegraphics[width=0.5\textwidth]{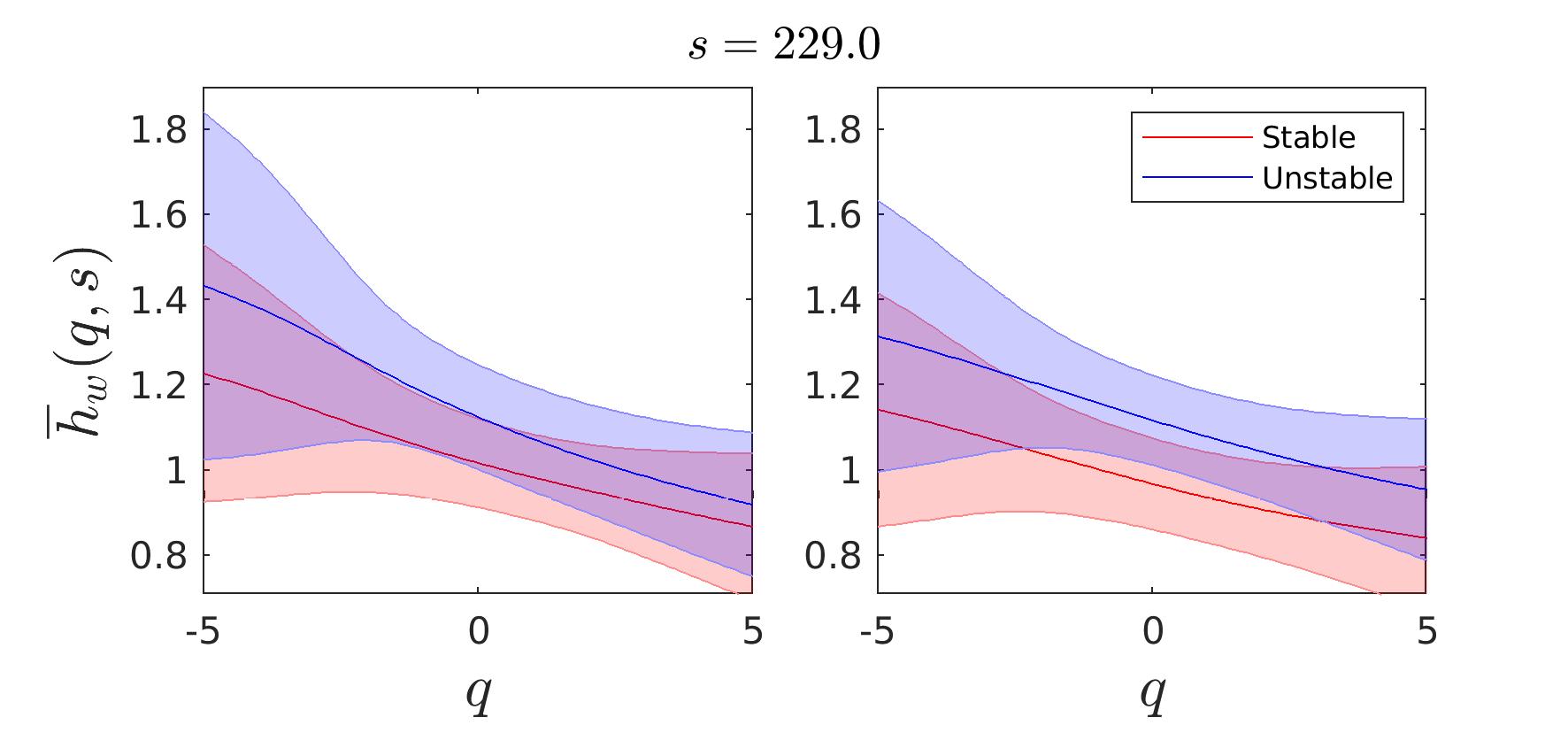}
    \put(-210,90){\bf \scriptsize{(c)}} \put(-100,90){\bf \scriptsize{(d)}}
\caption{Cross-sections of Hurst surfaces, $\overline{h}_u$ (a) and (b) and
$\overline{h}_w$ (c) and (d) for scale $s = 229$ with one-standard-deviation
error bars. Hurst surfaces corresponding to stable stratification ($\zeta >
0.01) $ is shown in red, whereas those corresponding to unstable stratification
($\zeta < -0.01 $) are shown in blue; measurements at $z_{\rm bot}$ are shown
in (a) and (c) whereas those at $z_{\rm top}$ are shown in (b) and (d). These
are representative plots that illustrate the effect of stability on the Hurst
surfaces. \label{fig:van_stability}}
\end{figure*}

\subsubsection{Effect of Shear on Anisotropy}

\begin{table}[ht]
 \caption{Distance between Hurst surfaces (see text) of different velocity
components along with their mean value.}
\centering
\begin{tabular}{c c c c c c c}
 \hline\hline \\ [-1em]
 $z$      & $d_{vu}$ & $d_{wu}$ & $d_{wv}$ & $\overline{h}_u$ &  $\overline{h}_v$  &  $\overline{h}_w$ \\ [0.5ex]
 \hline
$z_{\rm bot}$ & $0.026$ &	 $0.043$ &	 $0.017$  & $1.33$ &	$1.25$ &	$1.19$ \\ \hline \\  [-1em]
$z_{\rm top}$ & $0.018$ &	 $0.039$ &	 $0.022$  & $1.29$ &	$1.21$ &	$1.14$ \\ [1ex]   \hline
\end{tabular}
\label{table:d_shear_intercomp}
\end{table}

We now calculate the anisotropy in the scaling of the fluctuations of the
different components of the velocity by comparing their Hurst-surface plots.
For the calculations we present in this section, we have averaged over all
periods, independent of stratification condition. Thus, any differences may be
attributed to shear.

Table(\ref{table:d_shear_intercomp}) summarizes the distances between the Hurst
surface of the streamwise, transverse and wall-normal velocity components and
the mean values of the corresponding components. We observe that
$d_{wu}>d_{vu}>d_{wv}$, for both $z_{\rm top}$ and $z_{\rm bot}$, but this
anisotropy in the scaling of velocity fluctuations is especially evident near
the canopy surface, where such distances are larger. We also calculate the
componentwise distances $d_{u_i}=d(u_{i,z_{\rm top}},u_{i,z_{\rm bot}}), i
=1,2,3$: $d_{u} = 0.012, d_v = 0.020, d_w = 0.018$.

\subsubsection{Effect of Stability on Anisotropy}

\begin{table}
\centering
\caption{Distance between Hurst surfaces (see text) of corresponding velocity
components for stable and unstable stratification.} 
 \begin{tabular}{c c c c}
 \hline\hline\\ [-1em]
 Height & $d^{\rm su}_{uu}$ & $d^{\rm su}_{vv}$ & $d^{\rm su}_{ww}$ \\ 
 \hline
$z_{\rm bot}$    & $0.019$ & $0.028$ & $0.031$ \\
$z_{\rm top}$    & $0.019$ & $0.028$ & $0.021$ \\ [1ex] 
 \hline\hline
 \end{tabular}
\label{table:DHS_stabilitywise1}
\end{table}

\begin{table}[ht]
\caption{Distance between Hurst surfaces (see text) of different velocity
components along with their mean values for stable and unstable
stratification.} 
\centering
 \begin{tabular}{c c | c c c c c c}
 \hline\hline
 $z$ & Stability & $d_{vu}$ & $d_{wu}$ & $d_{wv}$ & $\overline{h}_u$ &  $\overline{h}_v$  &  $\overline{h}_w$ \\ [0.5ex]
 \hline
 $z_{\rm bot}$  & s  & $0.033$ &	 $0.043$ &	 $0.012$ & $1.31$ & $1.21$ & $1.13$ \\ 
                & u  & $0.020$ &	 $0.042$ &	 $0.024$ & $1.35$ & $1.30$ & $1.24$ \\ \hline \\  [-1em]
$z_{\rm top}$   & s  & $0.024$ &	 $0.041$ &	 $0.018$ & $1.27$ & $1.16$ & $1.09$ \\
                & u  & $0.012$ &	 $0.037$ &	 $0.028$ & $1.32$ & $1.27$ & $1.22$ \\ [1ex] 
\hline
\end{tabular}
\label{table:DHS_stabilitywise2}
\end{table}
The effect of stability is now quantified on each of the Hurst surfaces, for
the longitudinal, transverse and wall-normal velocity components in
Table~\ref{table:DHS_stabilitywise1}. Fig.~\ref{fig:van_stability} (a) and (b)
show the cross-sections of Hurst surfaces of the longitudinal velocity
component, $\overline{h}_u$  at $z_{\rm bot}$ and $z_{\rm top}$, respectively,
with one-standard-deviation error bars, whereas Fig.~\ref{fig:van_stability}
(c) and (d) show the cross-sections of Hurst surfaces of the wall-normal
velocity component, $\overline{h}_w$ at $z_{\rm bot}$ and $z_{\rm top}$,
respectively,  with one-standard-deviation error bars. These cross-sections
have been shown at a constant representative scale $s = 229$. The
cross-sections of Hurst surfaces corresponding to stable stratification ($\zeta
> 0.01 $) is shown in red, whereas those corresponding to unstable
stratification ($\zeta < -0.01 $) are shown in blue. The correlations of
multifractal fluctuations is very similar for the longitudinal and the
transverse velocity components for stable and unstable stratification, compared
to those for the wall-normal components. However, because of the significant
overlap between the Hurst surfaces, we cannot draw any conclusive evidence.

Table~\ref{table:DHS_stabilitywise2} shows the effect of stability on
inter-component Hurst surface distances $d_{u_iu_j}$.

The correlation of multifractal fluctuations for the wall-normal velocity
differ significantly, for both stable and unstable stratification. With
decreasing stability ($\zeta$), only $d_{vu}$ decreases, whereas $d_{wv}$
increases, albeit marginally, throughout the RSL. Furthermore, all fluctuations
show larger correlation for unstable stratification than for stable
stratification.

\begin{figure*}
    \includegraphics[width=0.29\textwidth]{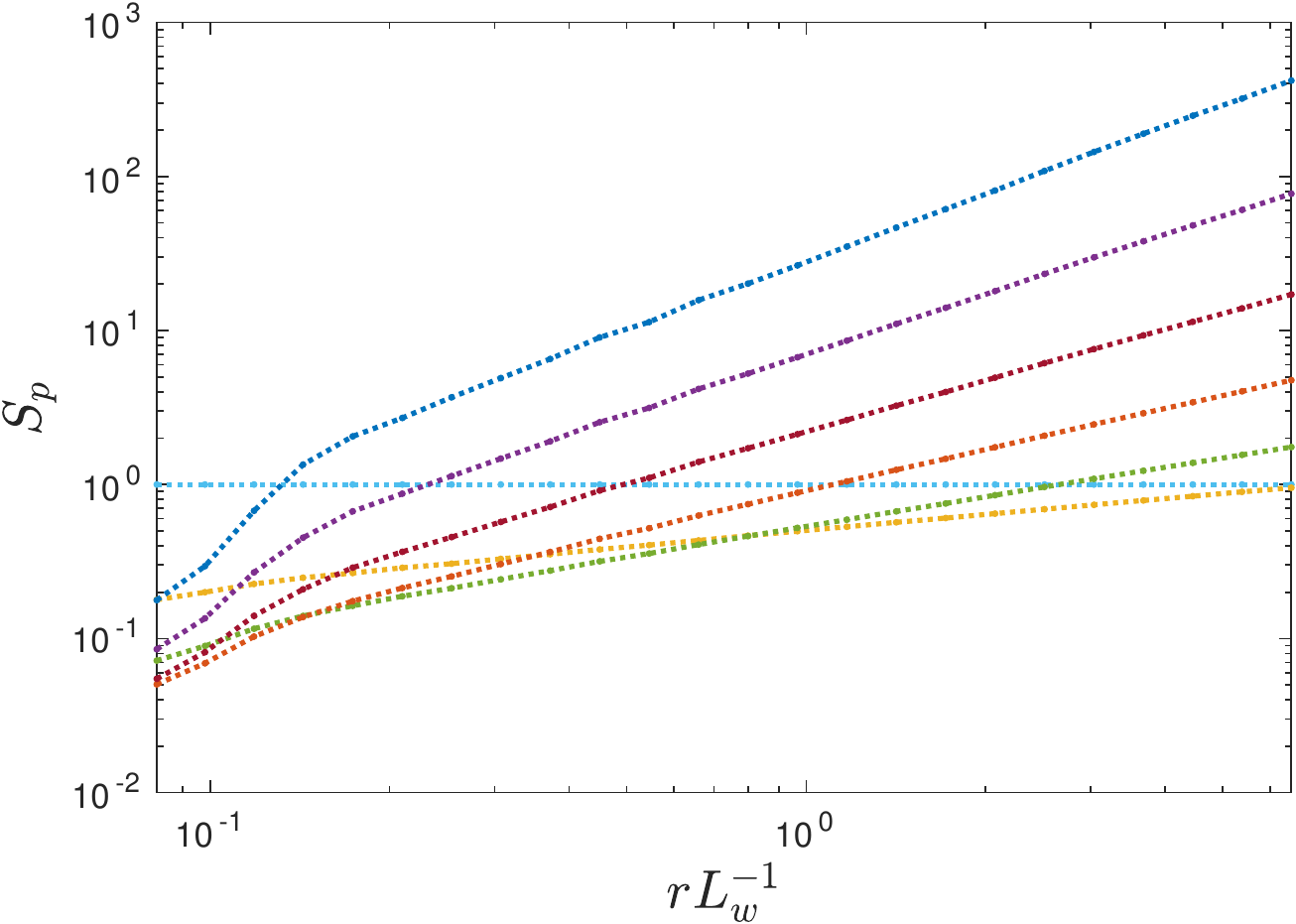}
    \put(-90, 95){{\makebox[0.01\textwidth][r]{\bf \scriptsize{(a)}} }}\hspace{1em}
    \includegraphics[width=0.29\textwidth]{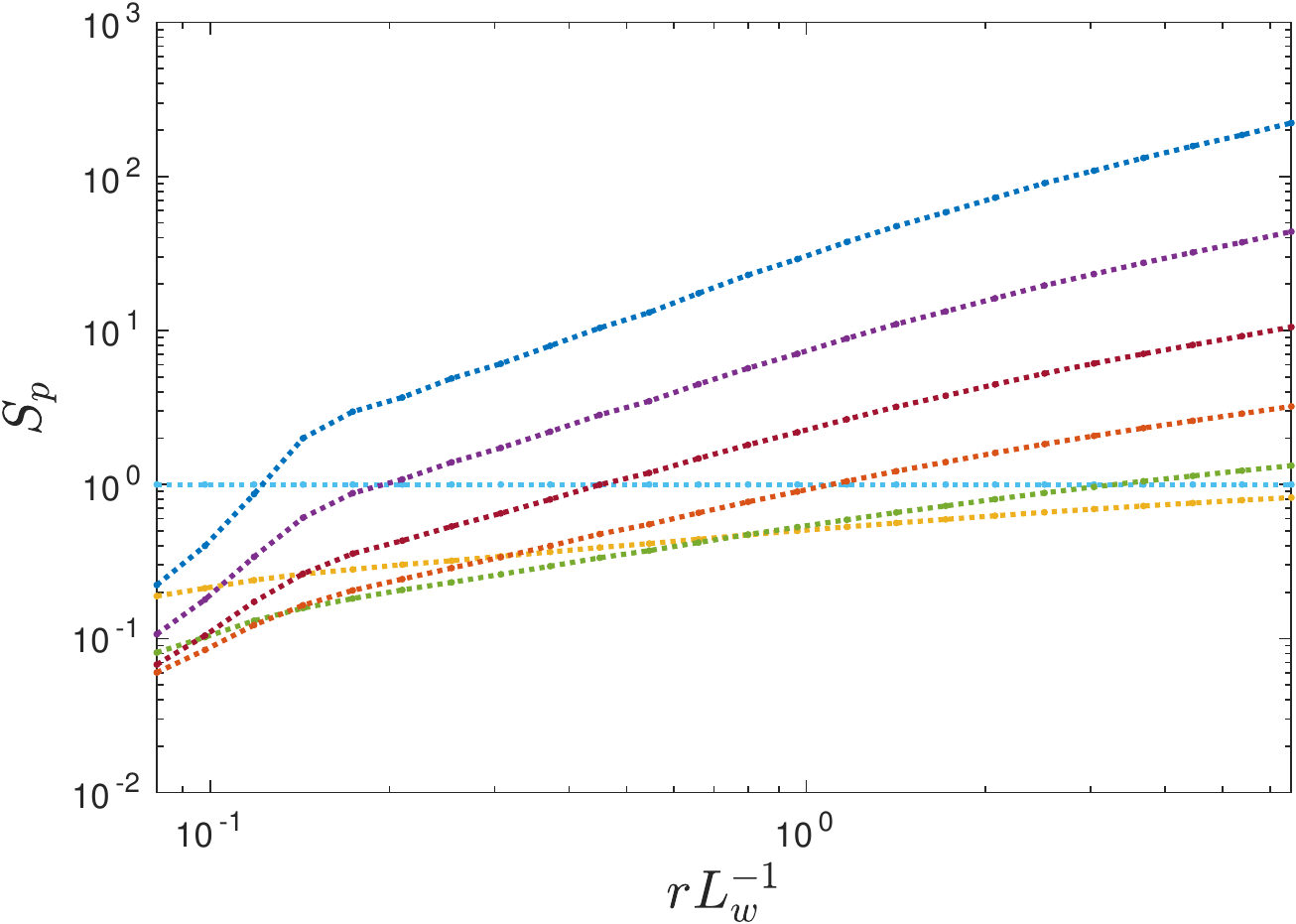}
    \put(-90, 95){{\makebox[0.01\textwidth][r]{\bf \scriptsize{(b)}} }}\hspace{1em}
    \includegraphics[width=0.29\textwidth]{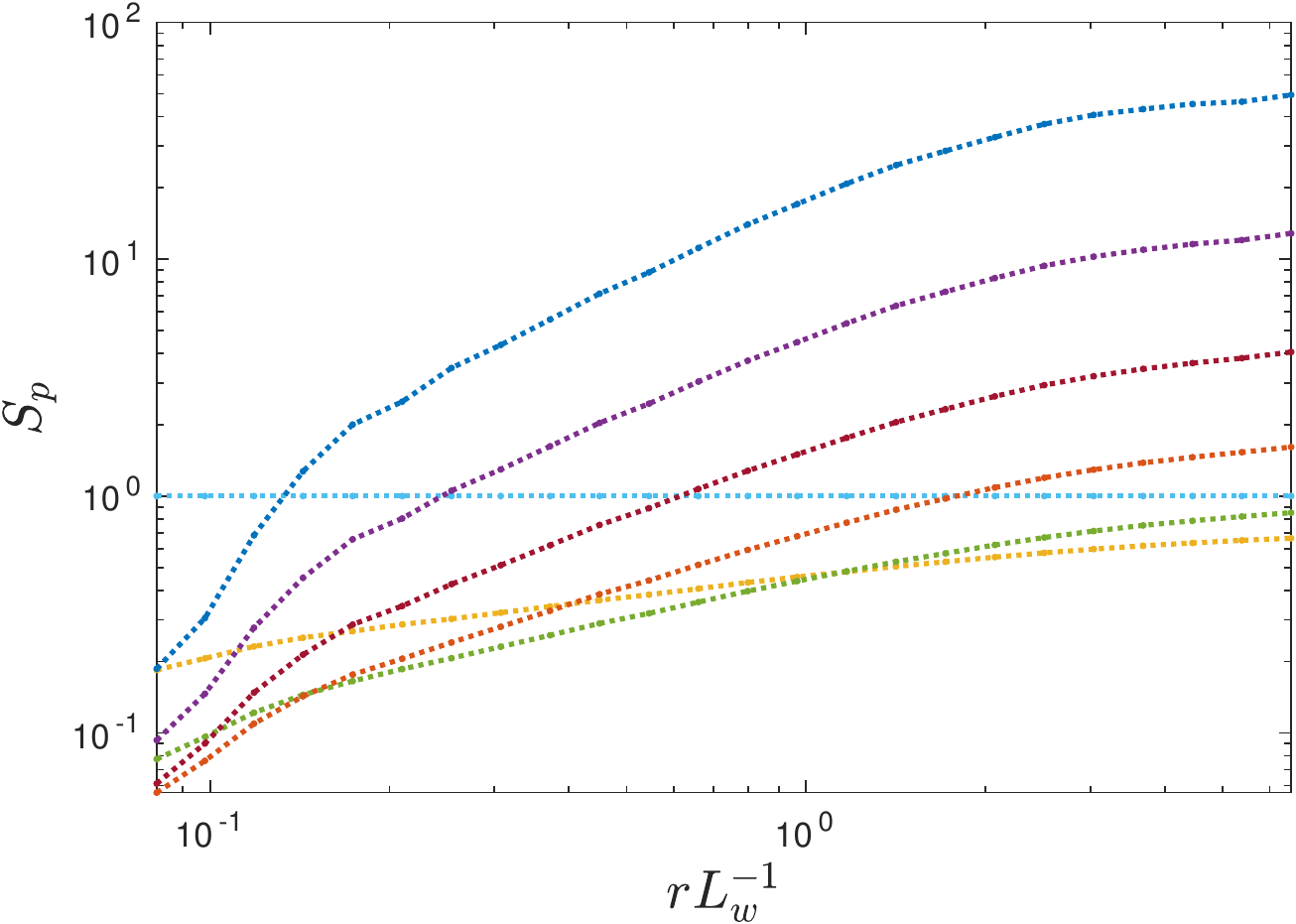}
    \put(-90, 95){{\makebox[0.01\textwidth][r]{\bf \scriptsize{(c)}} }}
\hfill

    \includegraphics[width=0.29\textwidth]{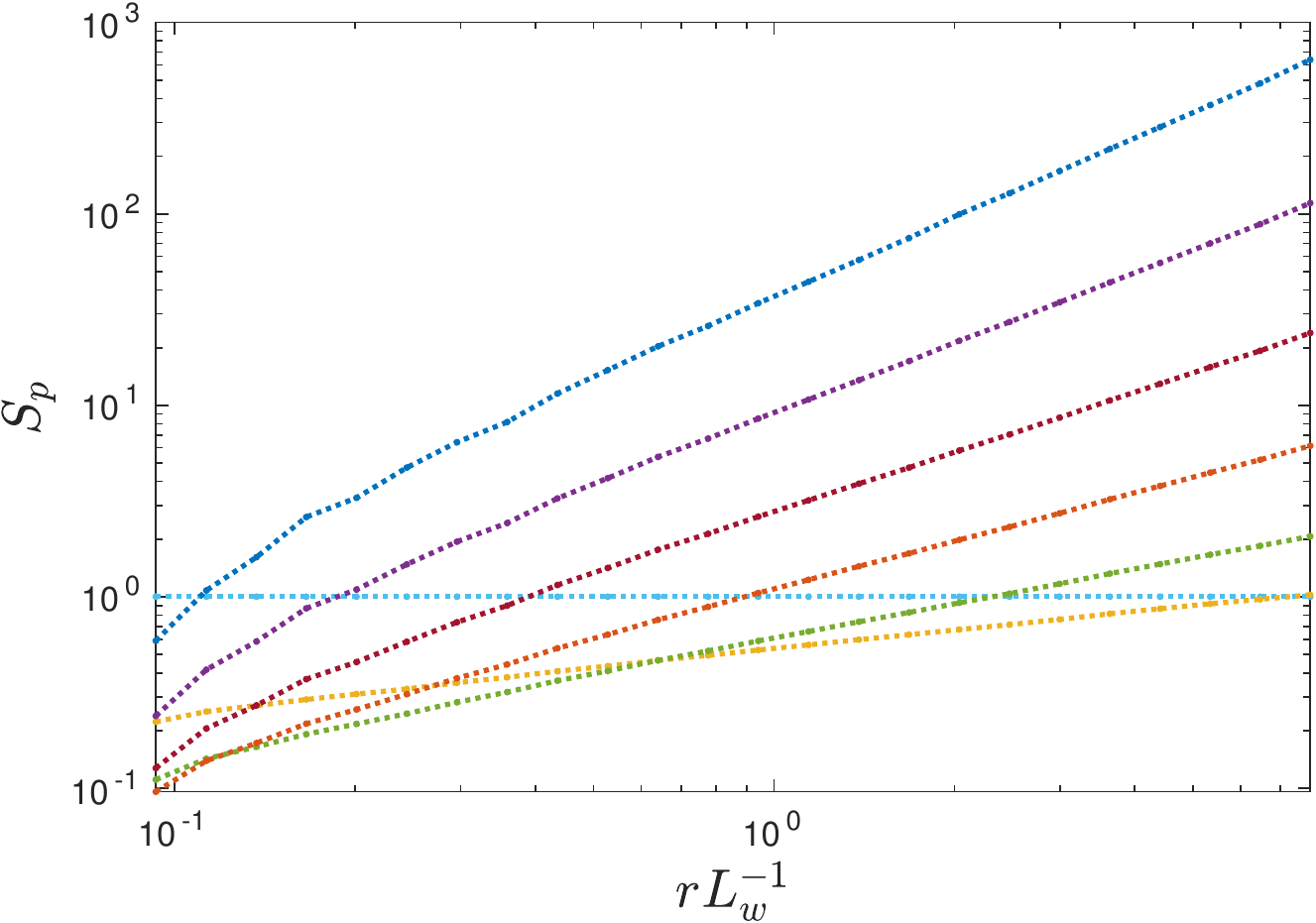}
    \put(-90, 95){{\makebox[0.01\textwidth][r]{\bf \scriptsize{(d)}} }}\hspace{1em}
    \includegraphics[width=0.29\textwidth]{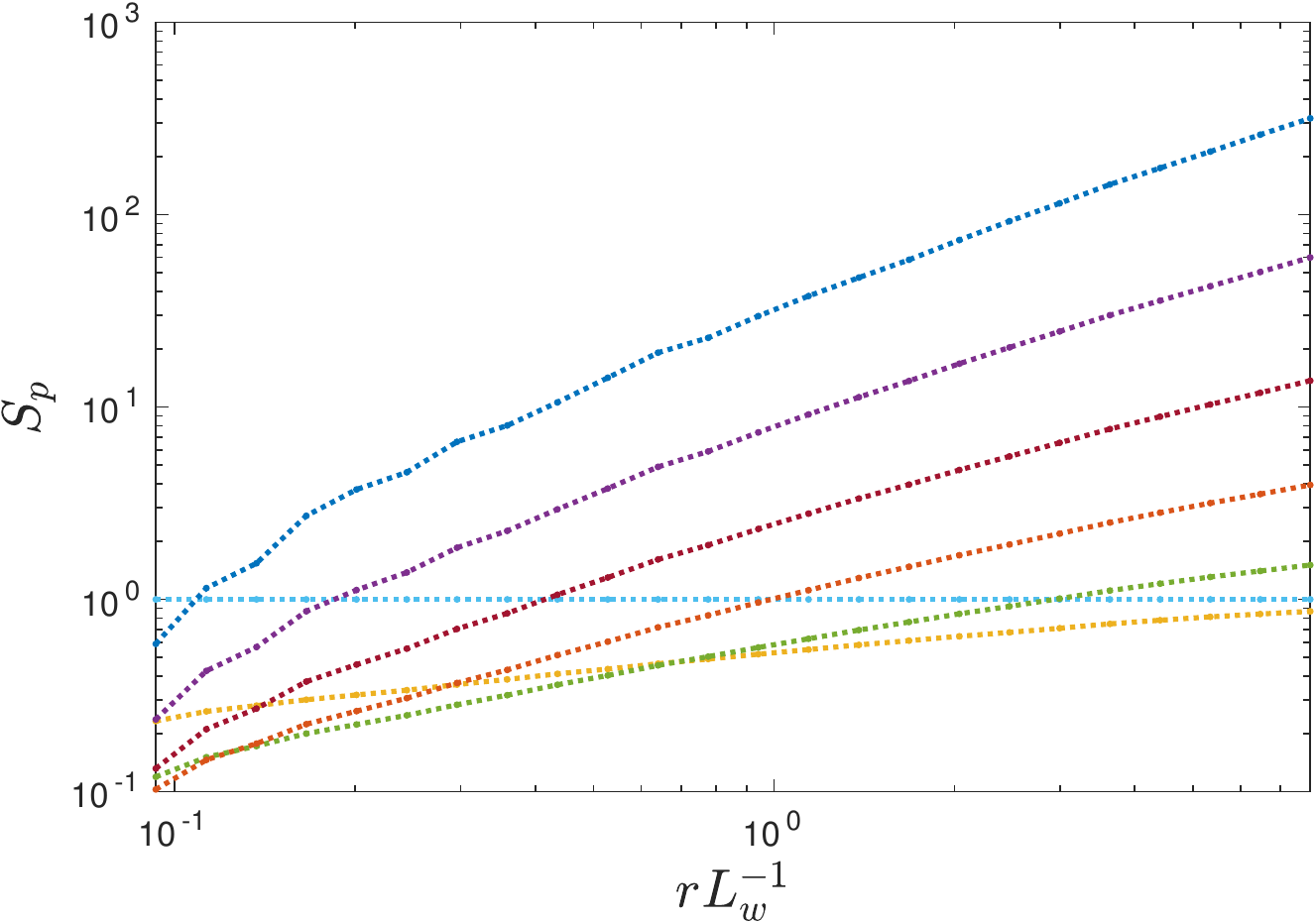}
    \put(-90, 95){{\makebox[0.01\textwidth][r]{\bf \scriptsize{(e)}} }}\hspace{1em}
    \includegraphics[width=0.29\textwidth]{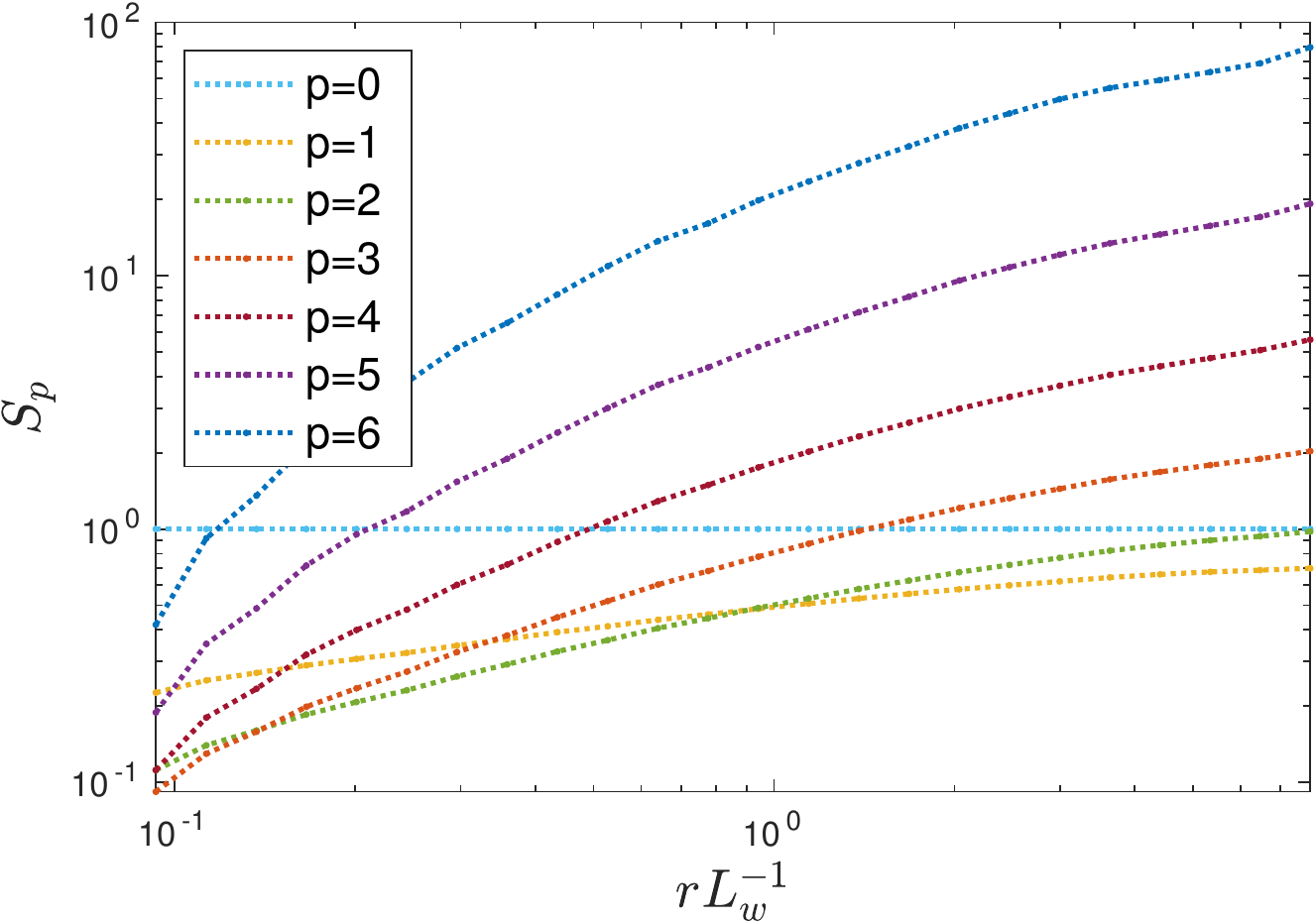}
    \put(-90, 95){{\makebox[0.01\textwidth][r]{\bf \scriptsize{(f)}} }}
\caption{Log-log plots versus the scaled separation $rL^{-1}_w$, of
ensemble-averaged structure functions $S^{u_i}_p$, corresponding to the
different velocity components: $u$ (a), (d), $v$ (b),(e), and $w$ (c),(f) for
orders $p=0,1,2,3,4,5,6$ (in cyan, yellow, green, orange, red, purple, and blue
respectively) at $z_{\rm bot}$ (a),(b),(c) and $z_{\rm bot}$ (d),(e),(f).
\label{fig:ap_sf} }
\end{figure*}


\begin{figure*}
    \includegraphics[width=0.29\textwidth]{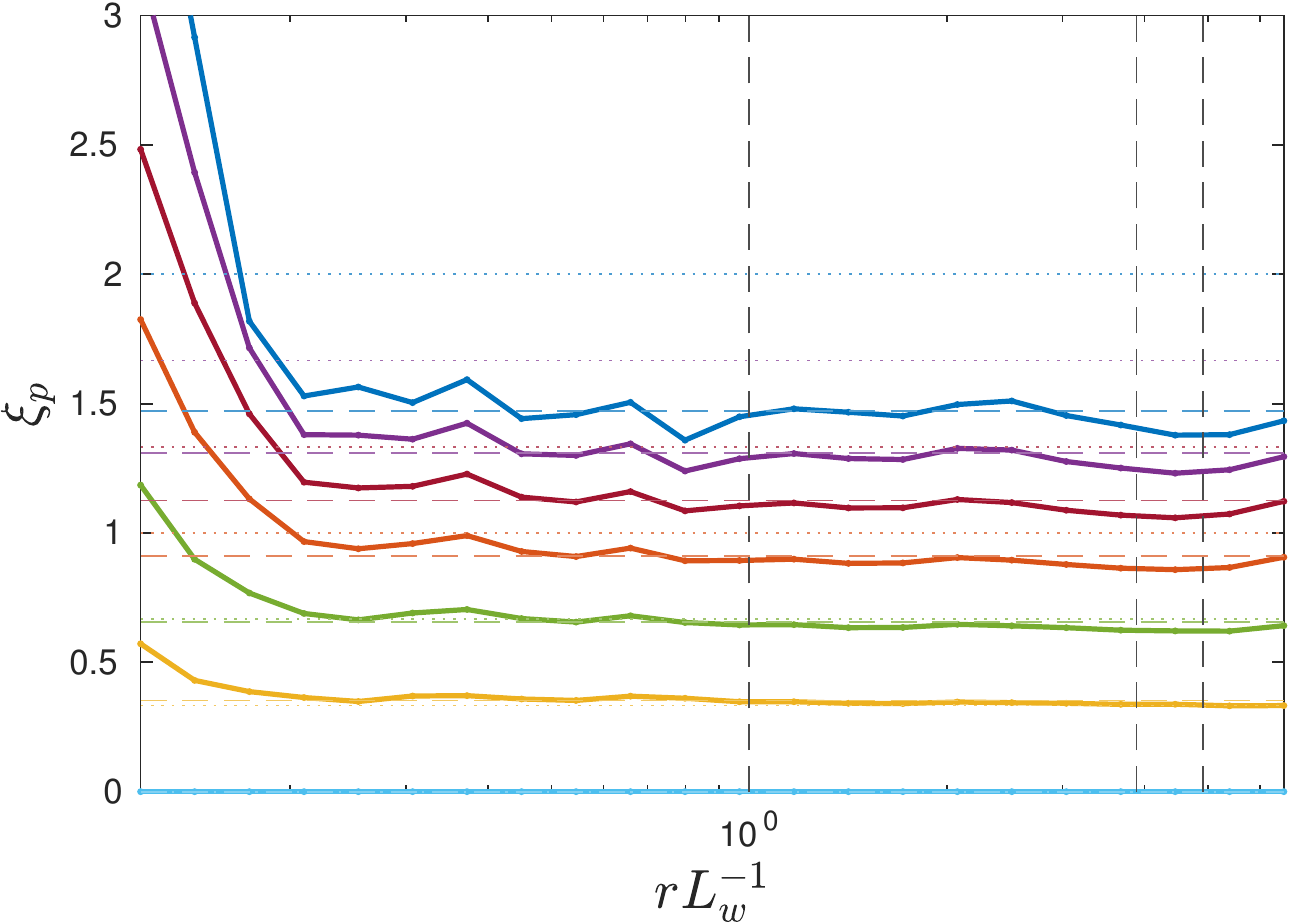}
    \put(-100, 85){{\makebox[0.01\textwidth][r]{\bf \scriptsize{(a)}} }}\hspace{1em}
    \includegraphics[width=0.29\textwidth]{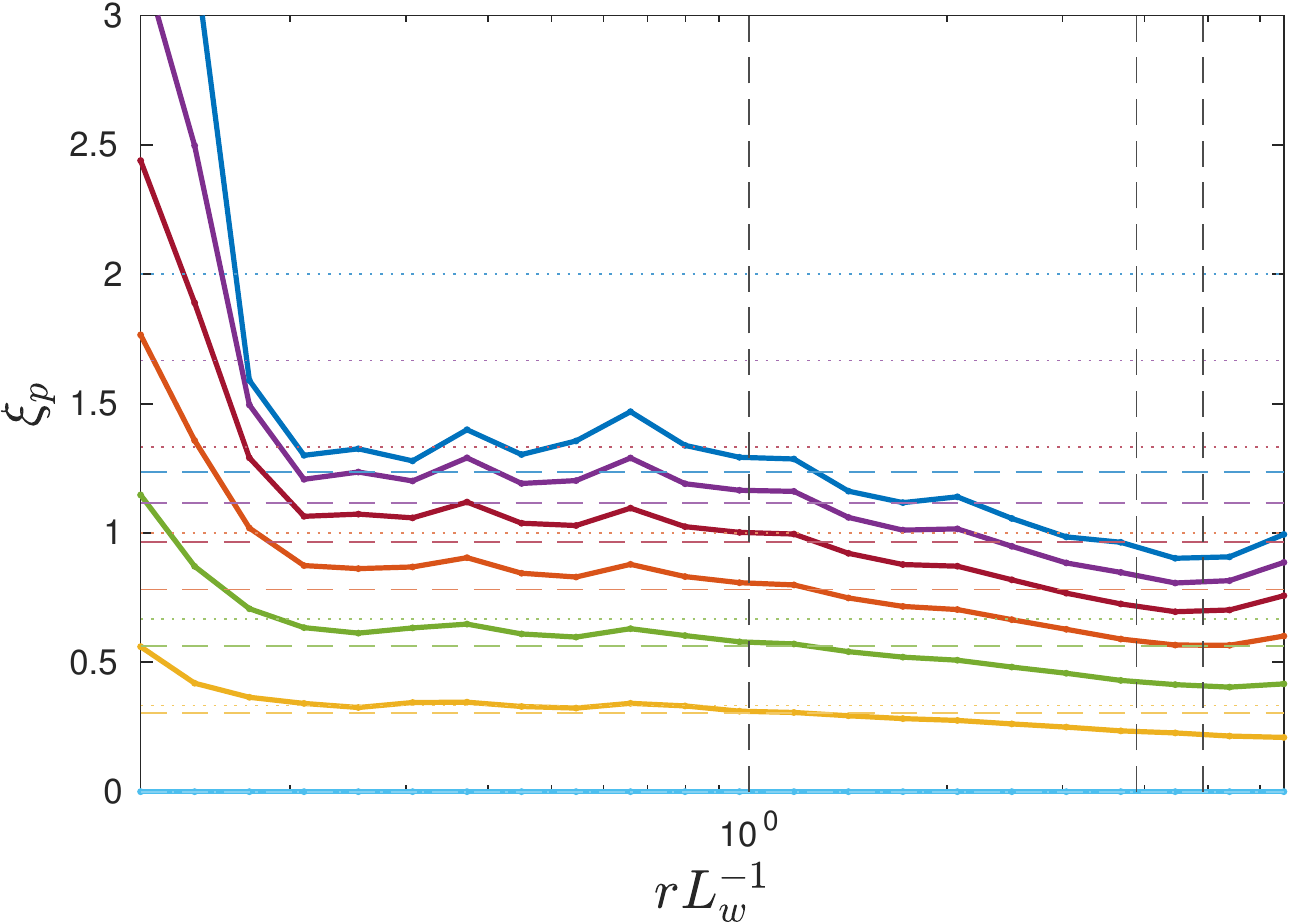}
    \put(-100, 85){{\makebox[0.01\textwidth][r]{\bf \scriptsize{(b)}} }}\hspace{1em}
    \includegraphics[width=0.29\textwidth]{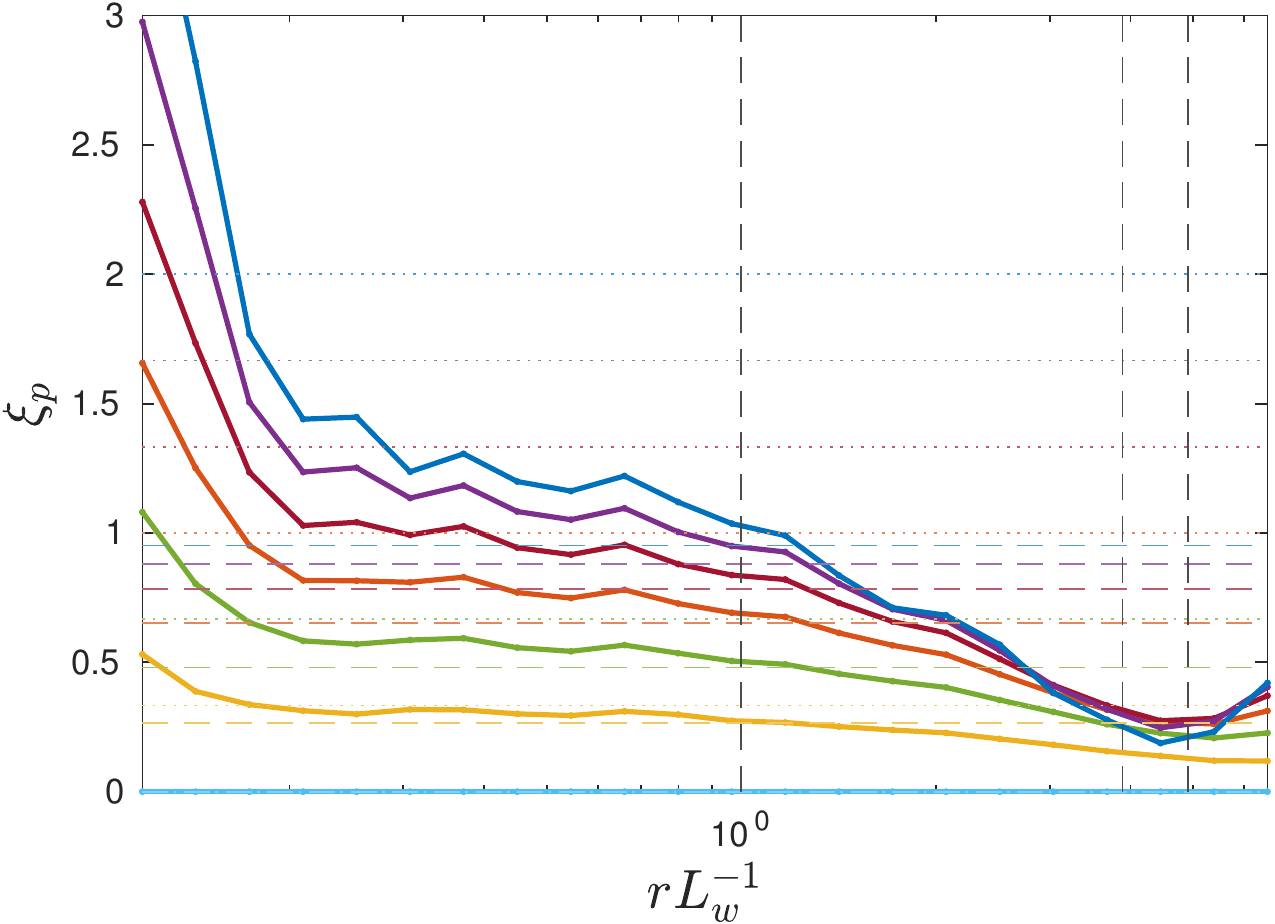}
    \put(-100, 85){{\makebox[0.01\textwidth][r]{\bf \scriptsize{(c)}} }}\hfill

    \includegraphics[width=0.29\textwidth]{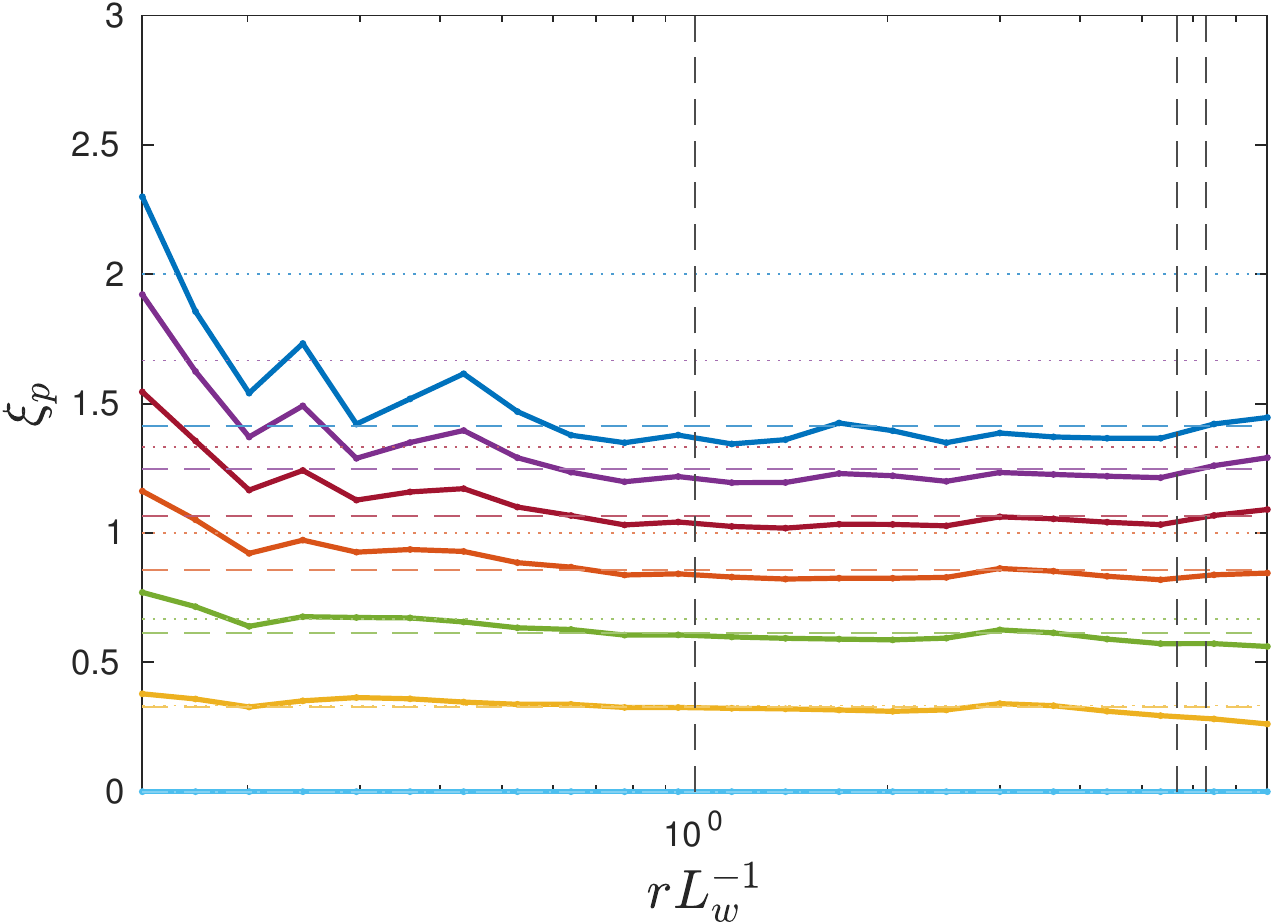}
    \put(-100,85){{\makebox[0.01\textwidth][r]{\bf \scriptsize{(d)}} }}\hspace{1em}
    \includegraphics[width=0.29\textwidth]{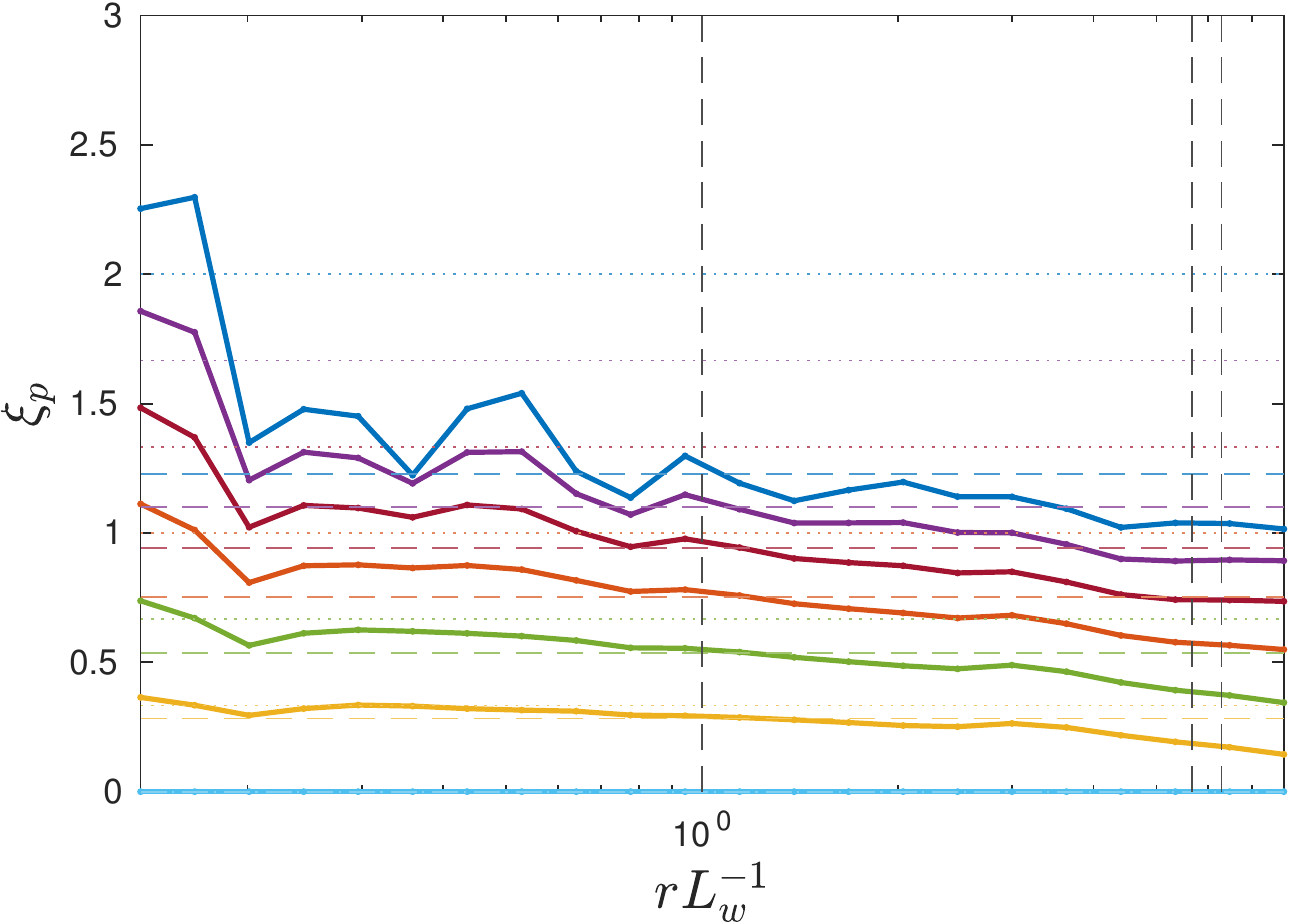}
    \put(-100,85){{\makebox[0.01\textwidth][r]{\bf \scriptsize{(e)}} }}\hspace{1em}
    \includegraphics[width=0.29\textwidth]{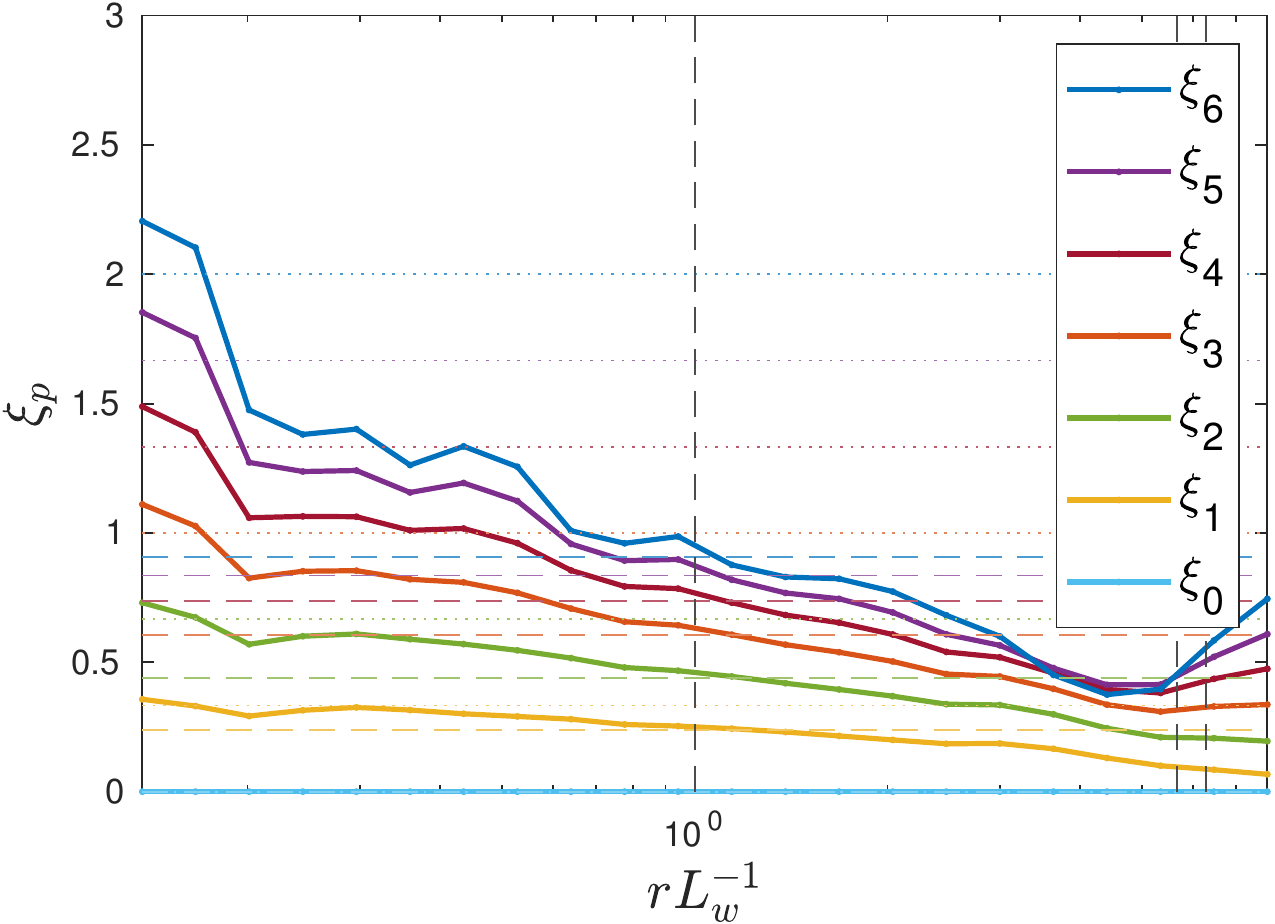}
    \put(-100,85){{\makebox[0.01\textwidth][r]{\bf \scriptsize{(f)}} }}
\caption{Semi-log plots versus the scaled separation $rL^{-1}_w$, of
local-slopes of ensemble-averaged structure functions $S^i_p$ corresponding to
the different velocity components: $u$ (a),(d), $v$ (b),(e), and $w$ (c),(f),
for orders $p=0,1,2,3,4,5,6$ (in cyan, yellow, green, orange, red, purple, and
blue respectively)  at $z_{\rm bot}$ (a),(b),(c) and $z_{\rm bot}$ (d),(e),(f).
Horizontal dashed lines indicate the linear-regression fit (see text) and
horizontal dotted line shows K41 values ($p/3$); vertical-dashed lines indicate
normalized (by $L_w$) integral scales $L_w,\, L_v,\, L_u$ (from left to right).
\label{fig:ap_locsc} }
\end{figure*}

\begin{figure*}
\includegraphics[width=0.26\textwidth]{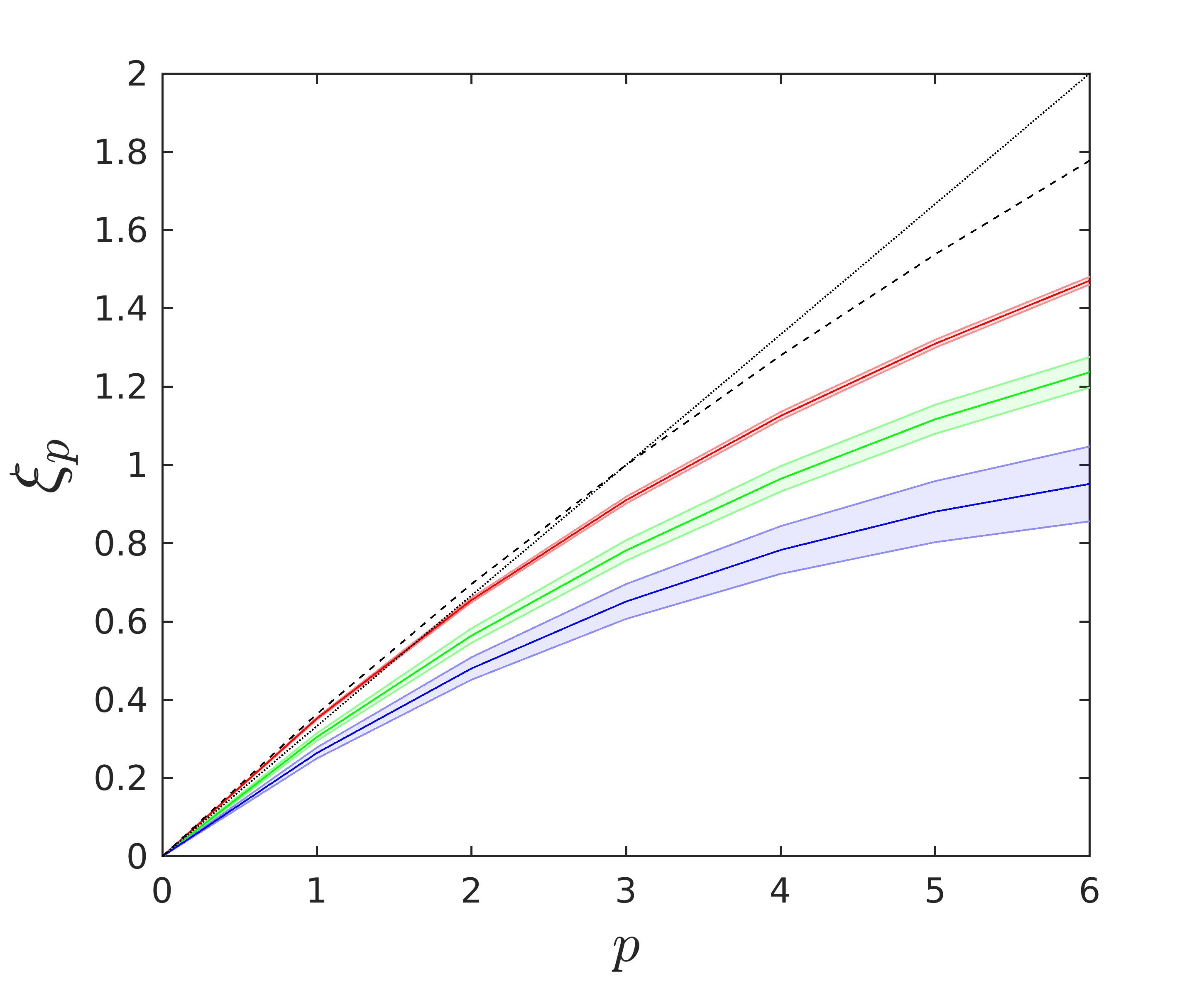}
\put(-90,65){{\makebox[0.01\textwidth][r]{\bf \scriptsize{(a)}} }}
\hspace{-1.5em}
\includegraphics[width=0.26\textwidth]{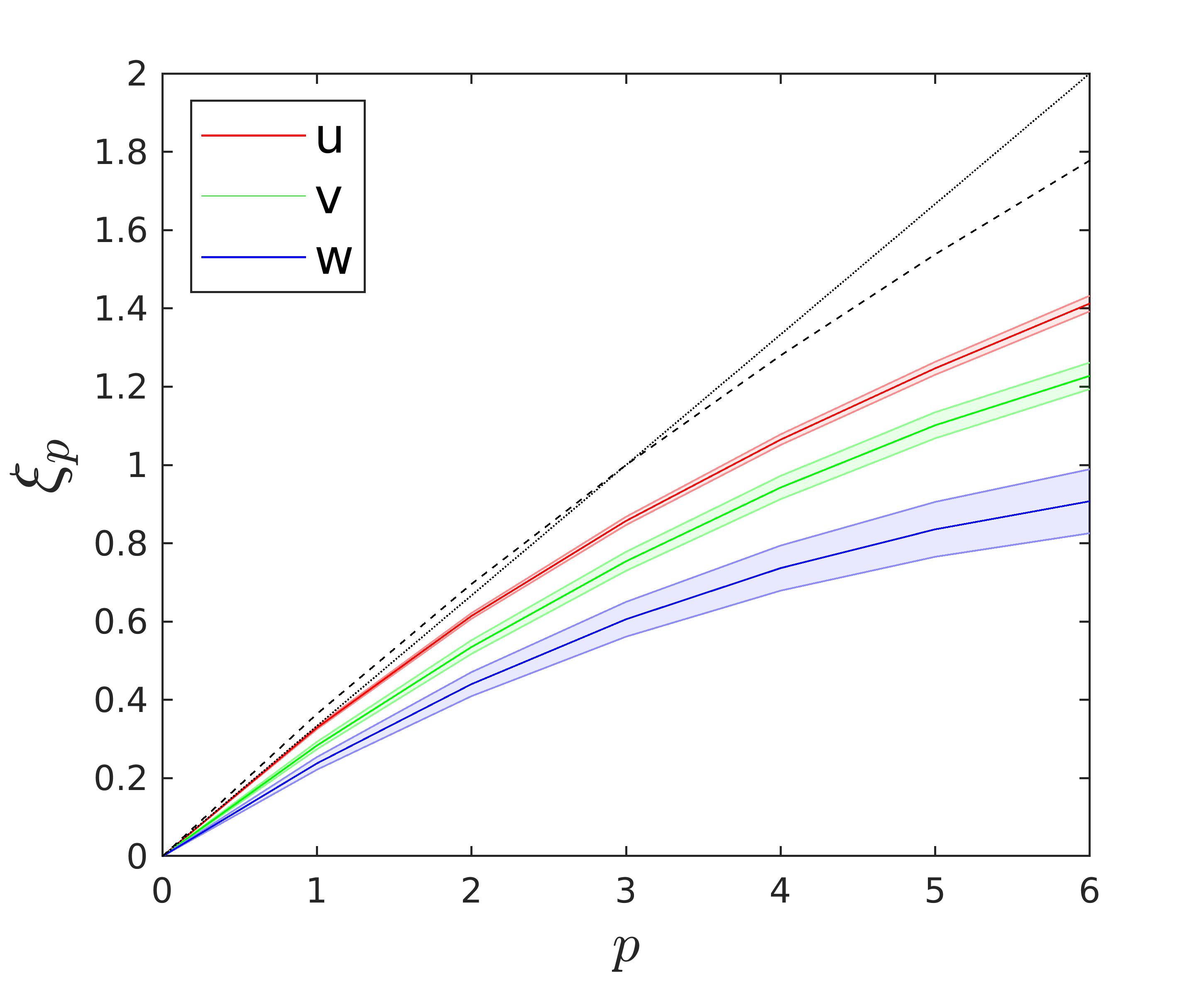}
\put(-90,65){{\makebox[0.01\textwidth][r]{\bf \scriptsize{(b)}} }}
\hspace{-1.5em}
\includegraphics[width=0.26\textwidth]{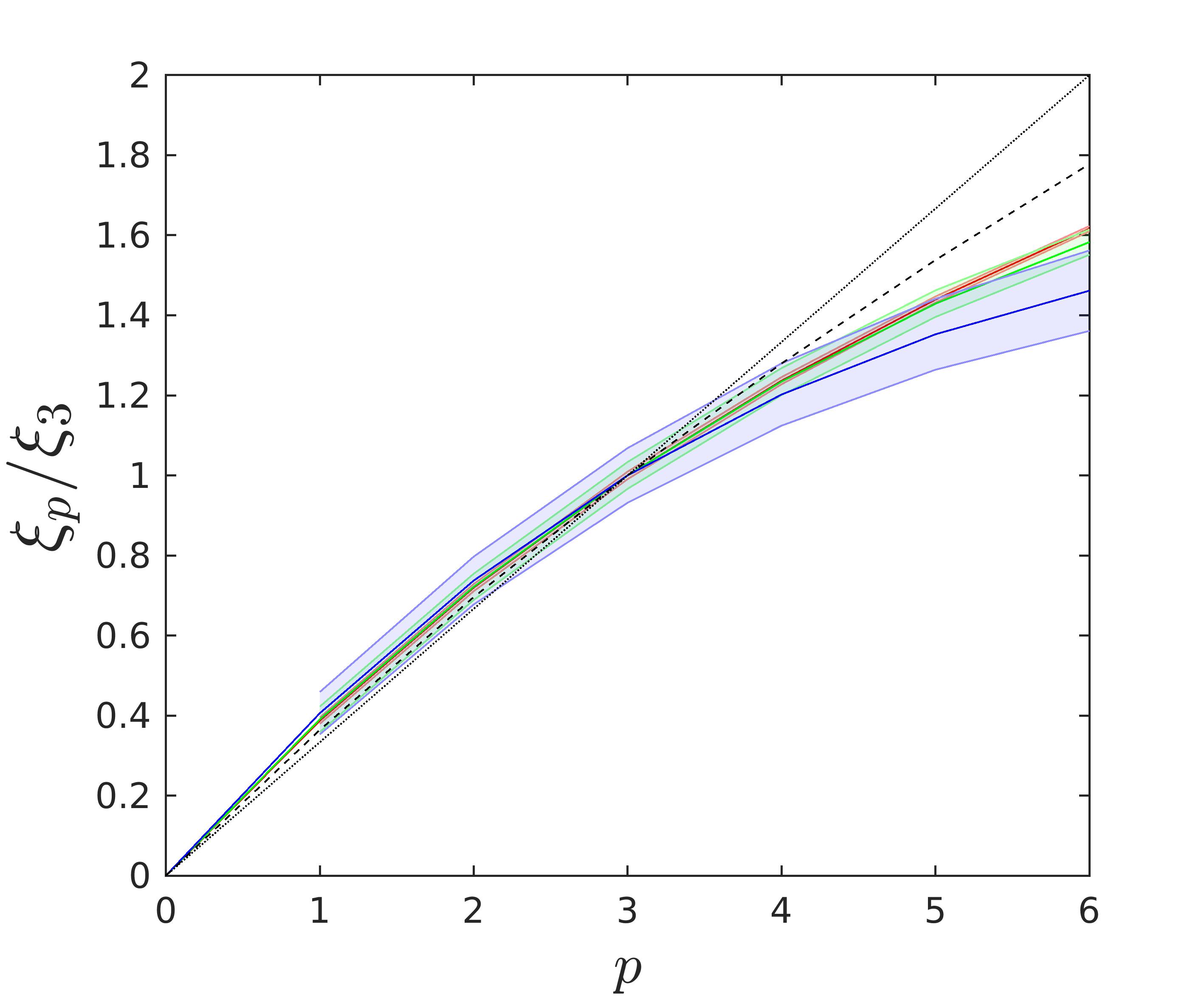}
\put(-90,65){{\makebox[0.01\textwidth][r]{\bf \scriptsize{(c)}} }}
\hspace{-1.5em}
\includegraphics[width=0.26\textwidth]{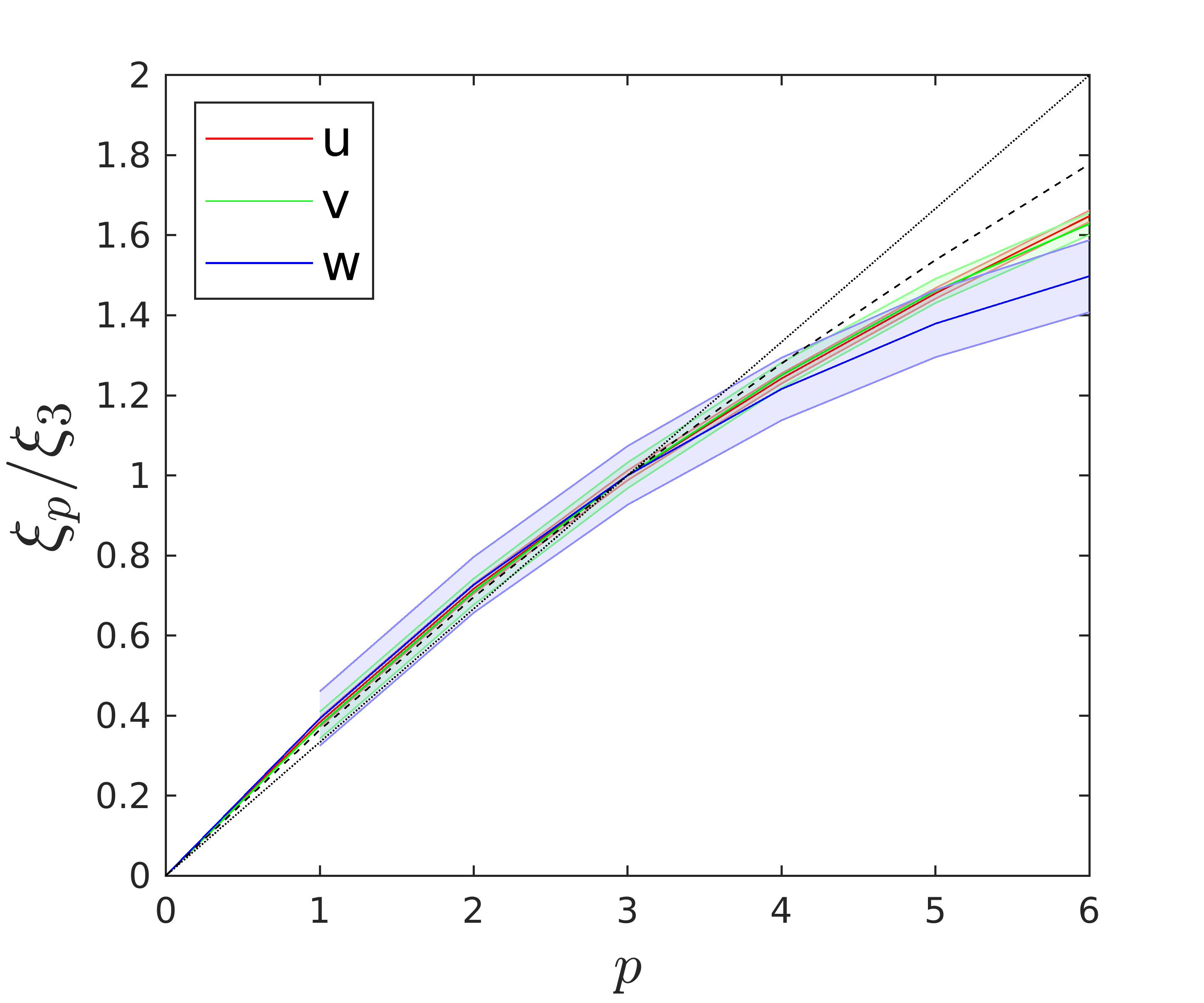}
\put(-90,65){{\makebox[0.01\textwidth][r]{\bf \scriptsize{(d)}} }}
\caption{Plots versus order $p$, of multiscaling exponents $\xi^{u_i}_p$
(a),(b) and ESS-normalized multiscaling exponents $\xi^{u_i}_p/\xi^{u_i}_3$
(c),(d) calculated from ensemble-averaged structure functions $S^{u_i}_p$ using
linear regression, for $u$ (red), $v$ (green) $w$ (blue) versus order $p$ along
with $95\%$ confidence intervals (shown by shaded region) from our linear
regression (see text) at $z_{\rm bot}$ (a),(c) and $z_{\rm bot}$ (b),(d).
\label{fig:ap_zeta} }
\end{figure*}

\section{\label{Conclusions}Discussions and Conclusions}

We have  carried out a detailed study of the statistical properties of
turbulent velocity and air-temperature in the roughness sublayer at different
heights above the Hyyti\"al\"a forest. The foci of our study have been (a)
anisotropy of turbulence at large scales, (b) its relaxation to isotropy at
small scales, and (c) its effect on the multiscaling exponents. This relaxation
is of interest given that turbulence in the RSL exhibits multifractality in
velocity and air-temperature fluctuations. 
 
The return to isotropy commences and concludes at scales, $r_{\rm ani}$ and
$r_{\rm iso}$, respectively, that increase the higher up we go in RSL, because
surface effects are weaker than they are just above the canopy. At the same
time, the largest scales above the canopy top are less anisotropic, in the AIM
measure, compared to higher up in the RSL. This leads us to the conjecture that
the interactions of the residual wakes, which originate from vegetation
elements, with anisotropic attached eddies, play a role in the randomization of
energies and the breakdown of eddies in the cascade. 

The scale $r_{\rm ani}$ is smaller for stable than for unstable stratification
at both heights. Both the AIM and BAM frameworks indicate that relaxation to
small-scale quasi-isotropy occurs (Fig.~\ref{fig:F_C}); however, this return is
not concluded in the BAM measure (Fig.~\ref{fig:F_C}), and at the smallest
scales resolved, $C_{\rm iso}$ saturates to a value $~ 0.8$ indicating that
signatures of anisotropy are still retained in the small-scale eddies. We see
that the return-to-isotropy trajectories occur along the plain-strain limit, at
all heights; and this similarity is most pronounced for stable stratification.
By contrast, this is quite different from the predictions of the quadratic
Rotta model, which predicts relaxation along the axisymmetric expansion
limit. 

The ensemble-averaged correlation coefficients indicate that normalized scales,
at which the return to isotropy commences and concludes, are generally seen to
be dependent on thermal stratification only through $L_w$. Whereas this is seen
to be true for stable stratification, we cannot draw such a conclusion for
unstable stratification, because we find $r_{\rm ani}$ and $r_{\rm iso}$ are
dependent on and independent of, respectively, the thermal integral scale,
irrespective of the normalization with $L_w$. 

We find that temperature and velocity fluctuations in the inertial range
display multifractality because of long-range correlations. The multifractal
exponents are slightly different (although the error bars are quite large) for
different velocity components; the longitudinal component being the least
intermittent. The nature of multifractality differs primarily in two aspects
when compared to \cite{Katul2009}; the reduced anisotropy in multifractal
exponents and the role of amplitude variability (albeit calculated by different
algorithms).  The multifractality of the temperature fluctuations is more than
that of velocity fluctuations (Fig.~\ref{fig:mfdfa1}).  Immediately above the
canopy top, temperature fluctuations are insensitive to fluctuations of small
magnitudes; multifractality of temperature increases with height in the RSL. In
contrast, the multifractality of the velocity components is found to be similar
above the canopy top and higher up in the RSL. 

We have also looked for signatures of anisotropy in the Hurst exponents $h(q)$
and the multiscaling exponents $\xi_p$, for different components of the
velocity field, by using our single-point measurements. We find that the error
bars for $h(q)$ and $\xi_p$ are large. Therefore, we have extended the scaling
range by using the ESS procedure~\cite{BenziESS, Pandit2009}. This ESS
procedure yields multiscaling exponent ratios $\xi_p/\xi_3$ with smaller error
bars than those we find for $h(q)$ and $\xi_p$. Unfortunately, the signatures
of anisotropy in $\xi_p/\xi_3$ are much less evident than in $\xi_p$. An
earlier study~\cite{Katul2009}, with slightly different experimental
conditions, has reported multiscaling exponent ratios $\xi_p/\xi_3$ that do
retain signatures of anisotropy. Improved measurements in future should allow
us to use $\xi_p$ or $h(q)$, without ESS, to uncover the signatures of
anisotropy in these multiscaling exponents.  

As we expect from field experiments, it is clearly more challenging to obtain
$h(q)$ and $\xi_p$, with small error bars, from measurements over forests than
from laboratory experiments.  We note, in passing, that the universal values of
$\xi_p$, for statistically homogeneous and isotropic fluid turbulence, can be
obtained from measurements in atmospheric boundary layers by using the $SO(3)$
decomposition discussed in Refs.~\cite{Arad1999, Biferale05,
KurienSreenivasan}; this is not our purpose here; but we observe that the ESS
procedure, which we use here, yields exponent ratios that are not very
sensitive to the anisotropy of the flow and are, therefore, very close to the
universal values of these ratios for statistically homogeneous and isotropic
turbulence.

 We also compare the results of our MFDFA with the conventional
structure-functions analysis. We find that both  multiscaling exponents $\xi_p$
and ESS multiscaling ratios $\xi_p/\xi_3$, obtained from structure functions,
show stronger deviations from K41 compared to their MFDFA counterparts. By
contrast, the anisotropy  in $\xi_p/\xi_3$ are comparable in MFDFA  and the ESS
procedure.  

We believe that in the future, the methods developed here shall be used for the
following: (a) different classes of stability, for e.g., near-neutral, mildly
unstable, unstable, dynamic-convective, and free-convective, as well as stable
- moderately stable, very stable with turbulence becoming patchy; and (b)
different roughness configurations (say snow-no-snow, or winter vs summer -
where the foliage density is quite different). It is envisaged that the methods
introduced here can be used for a comprehensive characterization of anisotropy,
return to isotropy, and multifractality of turbulence in the roughness sublayer
at different heights and many forest types to begin unfolding connections
between distribution of foliage and turbulence in the RSL.  

\begin{acknowledgments}
We thank the AtMath Collaboration at the University of Helsinki, ICOS by
University of Helsinki and the DST, CSIR and KVPY (India) for support. GK
acknowledges partial support from the US National Science Foundation
(NSF-AGS-1644382, NSF-IOS-1754893, and NSF-AGS-2028633).
\end{acknowledgments}

\appendix

\section{Structure function scaling exponents\label{sec:appendix_Sf}}

In Fig.~\ref{fig:ap_sf}, we show ensemble-averaged structure functions
$S^{u_i}_p,\,i=1,2,3$, for $u$ (a), (d), $v$ (b), (e), and $w$ (c), (f) for
orders $p=0,1,2,3,4,5,6$ (in cyan, yellow, green, orange, red, purple, and blue
respectively), in the inertial range. From $S^{u_i}_p$, we calculate the local
scaling exponents (as local slopes of the structure function) shown in
Fig.~\ref{fig:ap_locsc}. Vertical dashed lines(from left to right) indicate the
integral scales, $L_w,\,L_v,\,L_u$. We calculate the universal multiscaling
exponents $\xi^i_p,\, i =1,2,3$ from Fig.~\ref{fig:ap_sf} by using
linear-regression over the range of $rL_w^{-1}$ that  corresponds to the trough
in Fig.~\ref{fig:ap_locsc}. Horizontal dashed lines in Fig.~\ref{fig:ap_locsc}
indicate the linear-regression fit ($\xi^{u_i}_p$) and vertical-dashed lines
indicate normalised integral scales $L_w,\, L_v$, and $L_u$ (from left to
right). In Fig.~\ref{fig:ap_zeta}, we plot the multiscaling exponents
$\xi^{u_i}_p$ (a),(b) and ESS-normalised multiscaling exponents $\xi^{u_i}_p$
(c),(d) versus order-$p$. We see that the multiscaling exponents  obtained from
the traditional structure function analysis, indicate larger intermittency
compared to MFDFA; by contrast the anisotropy from both the calculations are
comparable.

\bibliography{References}

\begin{thebibliography}{68}%
\makeatletter
\providecommand \@ifxundefined [1]{%
 \@ifx{#1\undefined}
}%
\providecommand \@ifnum [1]{%
 \ifnum #1\expandafter \@firstoftwo
 \else \expandafter \@secondoftwo
 \fi
}%
\providecommand \@ifx [1]{%
 \ifx #1\expandafter \@firstoftwo
 \else \expandafter \@secondoftwo
 \fi
}%
\providecommand \natexlab [1]{#1}%
\providecommand \enquote  [1]{``#1''}%
\providecommand \bibnamefont  [1]{#1}%
\providecommand \bibfnamefont [1]{#1}%
\providecommand \citenamefont [1]{#1}%
\providecommand \href@noop [0]{\@secondoftwo}%
\providecommand \href [0]{\begingroup \@sanitize@url \@href}%
\providecommand \@href[1]{\@@startlink{#1}\@@href}%
\providecommand \@@href[1]{\endgroup#1\@@endlink}%
\providecommand \@sanitize@url [0]{\catcode `\\12\catcode `\$12\catcode
  `\&12\catcode `\#12\catcode `\^12\catcode `\_12\catcode `\%12\relax}%
\providecommand \@@startlink[1]{}%
\providecommand \@@endlink[0]{}%
\providecommand \url  [0]{\begingroup\@sanitize@url \@url }%
\providecommand \@url [1]{\endgroup\@href {#1}{\urlprefix }}%
\providecommand \urlprefix  [0]{URL }%
\providecommand \Eprint [0]{\href }%
\providecommand \doibase [0]{http://dx.doi.org/}%
\providecommand \selectlanguage [0]{\@gobble}%
\providecommand \bibinfo  [0]{\@secondoftwo}%
\providecommand \bibfield  [0]{\@secondoftwo}%
\providecommand \translation [1]{[#1]}%
\providecommand \BibitemOpen [0]{}%
\providecommand \bibitemStop [0]{}%
\providecommand \bibitemNoStop [0]{.\EOS\space}%
\providecommand \EOS [0]{\spacefactor3000\relax}%
\providecommand \BibitemShut  [1]{\csname bibitem#1\endcsname}%
\let\auto@bib@innerbib\@empty
\bibitem [{\citenamefont {Yuan}\ and\ \citenamefont
  {Aghaei~Jouybari}(2018)}]{PhysRevFluidsSurfTopo}%
  \BibitemOpen
  \bibfield  {author} {\bibinfo {author} {\bibfnamefont {J.}~\bibnamefont
  {Yuan}}\ and\ \bibinfo {author} {\bibfnamefont {M.}~\bibnamefont
  {Aghaei~Jouybari}},\ }\href {\doibase 10.1103/PhysRevFluids.3.114603}
  {\bibfield  {journal} {\bibinfo  {journal} {Phys. Rev. Fluids}\ }\textbf
  {\bibinfo {volume} {3}},\ \bibinfo {pages} {114603} (\bibinfo {year}
  {2018})}\BibitemShut {NoStop}%
\bibitem [{\citenamefont {Chen}\ \emph {et~al.}(2019)\citenamefont {Chen},
  \citenamefont {Chamecki},\ and\ \citenamefont {Katul}}]{chen2019effects}%
  \BibitemOpen
  \bibfield  {author} {\bibinfo {author} {\bibfnamefont {B.}~\bibnamefont
  {Chen}}, \bibinfo {author} {\bibfnamefont {M.}~\bibnamefont {Chamecki}}, \
  and\ \bibinfo {author} {\bibfnamefont {G.~G.}\ \bibnamefont {Katul}},\
  }\href@noop {} {\bibfield  {journal} {\bibinfo  {journal} {Quarterly Journal
  of the Royal Meteorological Society}\ }\textbf {\bibinfo {volume} {145}},\
  \bibinfo {pages} {2101} (\bibinfo {year} {2019})}\BibitemShut {NoStop}%
\bibitem [{\citenamefont {Poggi}\ and\ \citenamefont
  {Katul}(2008)}]{poggi2008turbulent}%
  \BibitemOpen
  \bibfield  {author} {\bibinfo {author} {\bibfnamefont {D.}~\bibnamefont
  {Poggi}}\ and\ \bibinfo {author} {\bibfnamefont {G.~G.}\ \bibnamefont
  {Katul}},\ }\href@noop {} {\bibfield  {journal} {\bibinfo  {journal}
  {Boundary-layer meteorology}\ }\textbf {\bibinfo {volume} {129}},\ \bibinfo
  {pages} {25} (\bibinfo {year} {2008})}\BibitemShut {NoStop}%
\bibitem [{\citenamefont {Raupach}\ and\ \citenamefont
  {Thom}(1981)}]{TurbPlantsI}%
  \BibitemOpen
  \bibfield  {author} {\bibinfo {author} {\bibfnamefont {M.~R.}\ \bibnamefont
  {Raupach}}\ and\ \bibinfo {author} {\bibfnamefont {A.~S.}\ \bibnamefont
  {Thom}},\ }\href {\doibase 10.1146/annurev.fl.13.010181.000525} {\bibfield
  {journal} {\bibinfo  {journal} {Annual Review of Fluid Mechanics}\ }\textbf
  {\bibinfo {volume} {13}},\ \bibinfo {pages} {97} (\bibinfo {year} {1981})},\
  \Eprint
  {http://arxiv.org/abs/https://doi.org/10.1146/annurev.fl.13.010181.000525}
  {https://doi.org/10.1146/annurev.fl.13.010181.000525} \BibitemShut {NoStop}%
\bibitem [{\citenamefont {Katul}\ \emph {et~al.}(2011)\citenamefont {Katul},
  \citenamefont {Konings},\ and\ \citenamefont {Porporato}}]{GK2011}%
  \BibitemOpen
  \bibfield  {author} {\bibinfo {author} {\bibfnamefont {G.~G.}\ \bibnamefont
  {Katul}}, \bibinfo {author} {\bibfnamefont {A.~G.}\ \bibnamefont {Konings}},
  \ and\ \bibinfo {author} {\bibfnamefont {A.}~\bibnamefont {Porporato}},\
  }\href {\doibase 10.1103/PhysRevLett.107.268502} {\bibfield  {journal}
  {\bibinfo  {journal} {Phys. Rev. Lett.}\ }\textbf {\bibinfo {volume} {107}},\
  \bibinfo {pages} {268502} (\bibinfo {year} {2011})}\BibitemShut {NoStop}%
\bibitem [{\citenamefont {Kurien}\ and\ \citenamefont
  {Sreenivasan}(2000)}]{KurienSreenivasan}%
  \BibitemOpen
  \bibfield  {author} {\bibinfo {author} {\bibfnamefont {S.}~\bibnamefont
  {Kurien}}\ and\ \bibinfo {author} {\bibfnamefont {K.~R.}\ \bibnamefont
  {Sreenivasan}},\ }\href {\doibase 10.1103/PhysRevE.62.2206} {\bibfield
  {journal} {\bibinfo  {journal} {Phys. Rev. E}\ }\textbf {\bibinfo {volume}
  {62}},\ \bibinfo {pages} {2206} (\bibinfo {year} {2000})}\BibitemShut
  {NoStop}%
\bibitem [{\citenamefont {Brugger}\ \emph {et~al.}(2018)\citenamefont
  {Brugger}, \citenamefont {Katul}, \citenamefont {De~Roo}, \citenamefont
  {Kr\"oniger}, \citenamefont {Rotenberg}, \citenamefont {Rohatyn},\ and\
  \citenamefont {Mauder}}]{Brugger2018}%
  \BibitemOpen
  \bibfield  {author} {\bibinfo {author} {\bibfnamefont {P.}~\bibnamefont
  {Brugger}}, \bibinfo {author} {\bibfnamefont {G.~G.}\ \bibnamefont {Katul}},
  \bibinfo {author} {\bibfnamefont {F.}~\bibnamefont {De~Roo}}, \bibinfo
  {author} {\bibfnamefont {K.}~\bibnamefont {Kr\"oniger}}, \bibinfo {author}
  {\bibfnamefont {E.}~\bibnamefont {Rotenberg}}, \bibinfo {author}
  {\bibfnamefont {S.}~\bibnamefont {Rohatyn}}, \ and\ \bibinfo {author}
  {\bibfnamefont {M.}~\bibnamefont {Mauder}},\ }\href {\doibase
  10.1103/PhysRevFluids.3.054608} {\bibfield  {journal} {\bibinfo  {journal}
  {Phys. Rev. Fluids}\ }\textbf {\bibinfo {volume} {3}},\ \bibinfo {pages}
  {054608} (\bibinfo {year} {2018})}\BibitemShut {NoStop}%
\bibitem [{\citenamefont {Ummels}\ \emph {et~al.}(2007)\citenamefont {Ummels},
  \citenamefont {Gibescu}, \citenamefont {Pelgrum}, \citenamefont {Kling},\
  and\ \citenamefont {Brand}}]{Ummels2007}%
  \BibitemOpen
  \bibfield  {author} {\bibinfo {author} {\bibfnamefont {B.~C.}\ \bibnamefont
  {Ummels}}, \bibinfo {author} {\bibfnamefont {M.}~\bibnamefont {Gibescu}},
  \bibinfo {author} {\bibfnamefont {E.}~\bibnamefont {Pelgrum}}, \bibinfo
  {author} {\bibfnamefont {W.~L.}\ \bibnamefont {Kling}}, \ and\ \bibinfo
  {author} {\bibfnamefont {A.~J.}\ \bibnamefont {Brand}},\ }\href {\doibase
  10.1109/TEC.2006.889616} {\bibfield  {journal} {\bibinfo  {journal} {IEEE
  Transactions on Energy Conversion}\ }\textbf {\bibinfo {volume} {22}},\
  \bibinfo {pages} {44} (\bibinfo {year} {2007})}\BibitemShut {NoStop}%
\bibitem [{\citenamefont {Carrasco}\ \emph {et~al.}(2006)\citenamefont
  {Carrasco}, \citenamefont {Franquelo}, \citenamefont {Bialasiewicz},
  \citenamefont {Galvan}, \citenamefont {PortilloGuisado}, \citenamefont
  {Prats}, \citenamefont {Leon},\ and\ \citenamefont
  {Moreno-Alfonso}}]{energyproductionii}%
  \BibitemOpen
  \bibfield  {author} {\bibinfo {author} {\bibfnamefont {J.~M.}\ \bibnamefont
  {Carrasco}}, \bibinfo {author} {\bibfnamefont {L.~G.}\ \bibnamefont
  {Franquelo}}, \bibinfo {author} {\bibfnamefont {J.~T.}\ \bibnamefont
  {Bialasiewicz}}, \bibinfo {author} {\bibfnamefont {E.}~\bibnamefont
  {Galvan}}, \bibinfo {author} {\bibfnamefont {R.~C.}\ \bibnamefont
  {PortilloGuisado}}, \bibinfo {author} {\bibfnamefont {M.~A.~M.}\ \bibnamefont
  {Prats}}, \bibinfo {author} {\bibfnamefont {J.~I.}\ \bibnamefont {Leon}}, \
  and\ \bibinfo {author} {\bibfnamefont {N.}~\bibnamefont {Moreno-Alfonso}},\
  }\href {\doibase 10.1109/TIE.2006.878356} {\bibfield  {journal} {\bibinfo
  {journal} {IEEE Transactions on Industrial Electronics}\ }\textbf {\bibinfo
  {volume} {53}} (\bibinfo {year} {2006}),\
  10.1109/TIE.2006.878356}\BibitemShut {NoStop}%
\bibitem [{\citenamefont {Demirci}\ and\ \citenamefont
  {Cuhadaroglu}(2000)}]{Demirci2000}%
  \BibitemOpen
  \bibfield  {author} {\bibinfo {author} {\bibfnamefont {E.}~\bibnamefont
  {Demirci}}\ and\ \bibinfo {author} {\bibfnamefont {B.}~\bibnamefont
  {Cuhadaroglu}},\ }\href {\doibase
  https://doi.org/10.1016/S0378-7788(99)00002-X} {\bibfield  {journal}
  {\bibinfo  {journal} {Energy and Buildings}\ }\textbf {\bibinfo {volume}
  {31}},\ \bibinfo {pages} {49 } (\bibinfo {year} {2000})}\BibitemShut
  {NoStop}%
\bibitem [{\citenamefont {Tesfuhuney}\ \emph {et~al.}(2013)\citenamefont
  {Tesfuhuney}, \citenamefont {Walker}, \citenamefont {Rensburg},\ and\
  \citenamefont {Everson}}]{Tesfuhuney2013}%
  \BibitemOpen
  \bibfield  {author} {\bibinfo {author} {\bibfnamefont {W.~A.}\ \bibnamefont
  {Tesfuhuney}}, \bibinfo {author} {\bibfnamefont {S.}~\bibnamefont {Walker}},
  \bibinfo {author} {\bibfnamefont {L.~D.~V.}\ \bibnamefont {Rensburg}}, \ and\
  \bibinfo {author} {\bibfnamefont {C.~S.}\ \bibnamefont {Everson}},\ }\href
  {\doibase 10.3390/atmos4040428} {\bibfield  {journal} {\bibinfo  {journal}
  {Atmosphere}\ }\textbf {\bibinfo {volume} {4}},\ \bibinfo {pages} {428}
  (\bibinfo {year} {2013})}\BibitemShut {NoStop}%
\bibitem [{\citenamefont {Cermak}(2003)}]{Cermak2003}%
  \BibitemOpen
  \bibfield  {author} {\bibinfo {author} {\bibfnamefont {J.~E.}\ \bibnamefont
  {Cermak}},\ }\href {\doibase https://doi.org/10.1016/S0167-6105(02)00396-3}
  {\bibfield  {journal} {\bibinfo  {journal} {Journal of Wind Engineering and
  Industrial Aerodynamics}\ }\textbf {\bibinfo {volume} {91}},\ \bibinfo
  {pages} {355 } (\bibinfo {year} {2003})}\BibitemShut {NoStop}%
\bibitem [{\citenamefont {Nathan}\ and\ \citenamefont
  {Katul}(2005)}]{nathan2005foliage}%
  \BibitemOpen
  \bibfield  {author} {\bibinfo {author} {\bibfnamefont {R.}~\bibnamefont
  {Nathan}}\ and\ \bibinfo {author} {\bibfnamefont {G.~G.}\ \bibnamefont
  {Katul}},\ }\href@noop {} {\bibfield  {journal} {\bibinfo  {journal}
  {Proceedings of the National Academy of Sciences}\ }\textbf {\bibinfo
  {volume} {102}},\ \bibinfo {pages} {8251} (\bibinfo {year}
  {2005})}\BibitemShut {NoStop}%
\bibitem [{\citenamefont {Sreenivasan}\ \emph {et~al.}(1979)\citenamefont
  {Sreenivasan}, \citenamefont {Antonia},\ and\ \citenamefont
  {Britz}}]{Sreenivasan1979}%
  \BibitemOpen
  \bibfield  {author} {\bibinfo {author} {\bibfnamefont {K.~R.}\ \bibnamefont
  {Sreenivasan}}, \bibinfo {author} {\bibfnamefont {R.~A.}\ \bibnamefont
  {Antonia}}, \ and\ \bibinfo {author} {\bibfnamefont {D.}~\bibnamefont
  {Britz}},\ }\href {\doibase 10.1017/S0022112079001270} {\bibfield  {journal}
  {\bibinfo  {journal} {Journal of Fluid Mechanics}\ }\textbf {\bibinfo
  {volume} {94}},\ \bibinfo {pages} {745–775} (\bibinfo {year}
  {1979})}\BibitemShut {NoStop}%
\bibitem [{\citenamefont {Warhaft}(2000)}]{Warhaft2000}%
  \BibitemOpen
  \bibfield  {author} {\bibinfo {author} {\bibfnamefont {Z.}~\bibnamefont
  {Warhaft}},\ }\href {\doibase 10.1146/annurev.fluid.32.1.203} {\bibfield
  {journal} {\bibinfo  {journal} {Annual Review of Fluid Mechanics}\ }\textbf
  {\bibinfo {volume} {32}},\ \bibinfo {pages} {203} (\bibinfo {year} {2000})},\
  \Eprint {http://arxiv.org/abs/https://doi.org/10.1146/annurev.fluid.32.1.203}
  {https://doi.org/10.1146/annurev.fluid.32.1.203} \BibitemShut {NoStop}%
\bibitem [{\citenamefont {Katul}\ \emph {et~al.}(2006)\citenamefont {Katul},
  \citenamefont {Porporato}, \citenamefont {Cava},\ and\ \citenamefont
  {Siqueira}}]{Katul2006}%
  \BibitemOpen
  \bibfield  {author} {\bibinfo {author} {\bibfnamefont {G.}~\bibnamefont
  {Katul}}, \bibinfo {author} {\bibfnamefont {A.}~\bibnamefont {Porporato}},
  \bibinfo {author} {\bibfnamefont {D.}~\bibnamefont {Cava}}, \ and\ \bibinfo
  {author} {\bibfnamefont {M.}~\bibnamefont {Siqueira}},\ }\href {\doibase
  https://doi.org/10.1016/j.physd.2006.02.004} {\bibfield  {journal} {\bibinfo
  {journal} {Physica D: Nonlinear Phenomena}\ }\textbf {\bibinfo {volume}
  {215}},\ \bibinfo {pages} {117 } (\bibinfo {year} {2006})}\BibitemShut
  {NoStop}%
\bibitem [{\citenamefont {Lumley}(1979)}]{LUMLEY1979123}%
  \BibitemOpen
  \bibfield  {author} {\bibinfo {author} {\bibfnamefont {J.~L.}\ \bibnamefont
  {Lumley}}\ }(\bibinfo  {publisher} {Elsevier},\ \bibinfo {year} {1979})\ pp.\
  \bibinfo {pages} {123 -- 176}\BibitemShut {NoStop}%
\bibitem [{\citenamefont {Banerjee}\ \emph {et~al.}(2007)\citenamefont
  {Banerjee}, \citenamefont {Krahl}, \citenamefont {Durst},\ and\ \citenamefont
  {Zenger}}]{Banerjee2007}%
  \BibitemOpen
  \bibfield  {author} {\bibinfo {author} {\bibfnamefont {S.}~\bibnamefont
  {Banerjee}}, \bibinfo {author} {\bibfnamefont {R.}~\bibnamefont {Krahl}},
  \bibinfo {author} {\bibfnamefont {F.}~\bibnamefont {Durst}}, \ and\ \bibinfo
  {author} {\bibfnamefont {C.}~\bibnamefont {Zenger}},\ }\href {\doibase
  10.1080/14685240701506896} {\bibfield  {journal} {\bibinfo  {journal}
  {Journal of Turbulence}\ }\textbf {\bibinfo {volume} {8}},\ \bibinfo {pages}
  {N32} (\bibinfo {year} {2007})},\ \Eprint
  {http://arxiv.org/abs/https://doi.org/10.1080/14685240701506896}
  {https://doi.org/10.1080/14685240701506896} \BibitemShut {NoStop}%
\bibitem [{\citenamefont {Foken}\ \emph {et~al.}(2012)\citenamefont {Foken},
  \citenamefont {Leuning}, \citenamefont {Oncley}, \citenamefont {Mauder},\
  and\ \citenamefont {Aubinet}}]{ECBook}%
  \BibitemOpen
  \bibfield  {author} {\bibinfo {author} {\bibfnamefont {T.}~\bibnamefont
  {Foken}}, \bibinfo {author} {\bibfnamefont {R.}~\bibnamefont {Leuning}},
  \bibinfo {author} {\bibfnamefont {S.~R.}\ \bibnamefont {Oncley}}, \bibinfo
  {author} {\bibfnamefont {M.}~\bibnamefont {Mauder}}, \ and\ \bibinfo {author}
  {\bibfnamefont {M.}~\bibnamefont {Aubinet}},\ }in\ \href {\doibase
  https://doi.org/10.1007/978-94-007-2351-1} {\emph {\bibinfo {booktitle} {Eddy
  Covariance A Practical Guide to Measurement and Data Analysis}}},\ \bibinfo
  {editor} {edited by\ \bibinfo {editor} {\bibfnamefont {D.~P.}\ \bibnamefont
  {Marc~Aubinet}, \bibfnamefont {Timo~Vesala}}}\ (\bibinfo  {publisher}
  {Springer Netherlands},\ \bibinfo {year} {2012})\ Chap.~\bibinfo {chapter}
  {4}, pp.\ \bibinfo {pages} {85--125}\BibitemShut {NoStop}%
\bibitem [{\citenamefont {{Kolmogorov}}(1941)}]{K41}%
  \BibitemOpen
  \bibfield  {author} {\bibinfo {author} {\bibfnamefont {A.}~\bibnamefont
  {{Kolmogorov}}},\ }\href
  {https://ui.adsabs.harvard.edu/abs/1941DoSSR..30..301K} {\bibfield  {journal}
  {\bibinfo  {journal} {Akademiia Nauk SSSR Doklady}\ }\textbf {\bibinfo
  {volume} {30}},\ \bibinfo {pages} {301} (\bibinfo {year} {1941})}\BibitemShut
  {NoStop}%
\bibitem [{\citenamefont {Frisch}(1995)}]{Frisch96}%
  \BibitemOpen
  \bibfield  {author} {\bibinfo {author} {\bibfnamefont {U.}~\bibnamefont
  {Frisch}},\ }\href {\doibase 10.1017/CBO9781139170666} {\emph {\bibinfo
  {title} {Turbulence: The Legacy of A. N. Kolmogorov}}}\ (\bibinfo
  {publisher} {Cambridge University Press},\ \bibinfo {year}
  {1995})\BibitemShut {NoStop}%
\bibitem [{\citenamefont {Ghil}\ \emph {et~al.}(1985)\citenamefont {Ghil},
  \citenamefont {Benzi},\ and\ \citenamefont {Parisi}}]{ParisiFrisch}%
  \BibitemOpen
  \bibinfo {editor} {\bibfnamefont {M.}~\bibnamefont {Ghil}}, \bibinfo {editor}
  {\bibfnamefont {R.}~\bibnamefont {Benzi}}, \ and\ \bibinfo {editor}
  {\bibfnamefont {G.}~\bibnamefont {Parisi}},\ eds.,\ \href@noop {} {\emph
  {\bibinfo {title} {Turbulence and Predictability in Geophysical Fluid
  Dynamics and Climate Dynamics}}}\ (\bibinfo  {publisher} {North-Holland Publ.
  Co., Amsterdam/New York},\ \bibinfo {year} {1985})\ p.\ \bibinfo {pages}
  {449}\BibitemShut {NoStop}%
\bibitem [{\citenamefont {Benzi}\ \emph {et~al.}(1984)\citenamefont {Benzi},
  \citenamefont {Paladin}, \citenamefont {Parisi},\ and\ \citenamefont
  {Vulpiani}}]{Benzi84}%
  \BibitemOpen
  \bibfield  {author} {\bibinfo {author} {\bibfnamefont {R.}~\bibnamefont
  {Benzi}}, \bibinfo {author} {\bibfnamefont {G.}~\bibnamefont {Paladin}},
  \bibinfo {author} {\bibfnamefont {G.}~\bibnamefont {Parisi}}, \ and\ \bibinfo
  {author} {\bibfnamefont {A.}~\bibnamefont {Vulpiani}},\ }\href {\doibase
  10.1088/0305-4470/17/18/021} {\bibfield  {journal} {\bibinfo  {journal}
  {Journal of Physics A: Mathematical and General}\ }\textbf {\bibinfo {volume}
  {17}},\ \bibinfo {pages} {3521} (\bibinfo {year} {1984})}\BibitemShut
  {NoStop}%
\bibitem [{\citenamefont {Halsey}\ \emph {et~al.}(1986)\citenamefont {Halsey},
  \citenamefont {Jensen}, \citenamefont {Kadanoff}, \citenamefont {Procaccia},\
  and\ \citenamefont {Shraiman}}]{Halsey86}%
  \BibitemOpen
  \bibfield  {author} {\bibinfo {author} {\bibfnamefont {T.~C.}\ \bibnamefont
  {Halsey}}, \bibinfo {author} {\bibfnamefont {M.~H.}\ \bibnamefont {Jensen}},
  \bibinfo {author} {\bibfnamefont {L.~P.}\ \bibnamefont {Kadanoff}}, \bibinfo
  {author} {\bibfnamefont {I.}~\bibnamefont {Procaccia}}, \ and\ \bibinfo
  {author} {\bibfnamefont {B.~I.}\ \bibnamefont {Shraiman}},\ }\href {\doibase
  10.1103/PhysRevA.33.1141} {\bibfield  {journal} {\bibinfo  {journal} {Phys.
  Rev. A}\ }\textbf {\bibinfo {volume} {33}},\ \bibinfo {pages} {1141}
  (\bibinfo {year} {1986})}\BibitemShut {NoStop}%
\bibitem [{\citenamefont {Argoul}\ \emph {et~al.}(1989)\citenamefont {Argoul},
  \citenamefont {Arneodo}, \citenamefont {Grasseau}, \citenamefont {Gagne},
  \citenamefont {Hopfinger},\ and\ \citenamefont {Frisch}}]{Argoul89}%
  \BibitemOpen
  \bibfield  {author} {\bibinfo {author} {\bibfnamefont {F.}~\bibnamefont
  {Argoul}}, \bibinfo {author} {\bibfnamefont {A.}~\bibnamefont {Arneodo}},
  \bibinfo {author} {\bibfnamefont {G.}~\bibnamefont {Grasseau}}, \bibinfo
  {author} {\bibfnamefont {Y.}~\bibnamefont {Gagne}}, \bibinfo {author}
  {\bibfnamefont {E.~J.}\ \bibnamefont {Hopfinger}}, \ and\ \bibinfo {author}
  {\bibfnamefont {U.}~\bibnamefont {Frisch}},\ }\href@noop {} {\bibfield
  {journal} {\bibinfo  {journal} {Nature}\ }\textbf {\bibinfo {volume} {338}},\
  \bibinfo {pages} {51} (\bibinfo {year} {1989})}\BibitemShut {NoStop}%
\bibitem [{\citenamefont {Meneveau}\ and\ \citenamefont
  {Sreenivasan}(1991)}]{MenSreeni91}%
  \BibitemOpen
  \bibfield  {author} {\bibinfo {author} {\bibfnamefont {C.}~\bibnamefont
  {Meneveau}}\ and\ \bibinfo {author} {\bibfnamefont {K.~R.}\ \bibnamefont
  {Sreenivasan}},\ }\href {\doibase 10.1017/S0022112091001830} {\bibfield
  {journal} {\bibinfo  {journal} {Journal of Fluid Mechanics}\ }\textbf
  {\bibinfo {volume} {224}},\ \bibinfo {pages} {429–484} (\bibinfo {year}
  {1991})}\BibitemShut {NoStop}%
\bibitem [{\citenamefont {Boffetta}\ \emph {et~al.}(2008)\citenamefont
  {Boffetta}, \citenamefont {Mazzino},\ and\ \citenamefont
  {Vulpiani}}]{Boffetta08}%
  \BibitemOpen
  \bibfield  {author} {\bibinfo {author} {\bibfnamefont {G.}~\bibnamefont
  {Boffetta}}, \bibinfo {author} {\bibfnamefont {A.}~\bibnamefont {Mazzino}}, \
  and\ \bibinfo {author} {\bibfnamefont {A.}~\bibnamefont {Vulpiani}},\ }\href
  {\doibase 10.1088/1751-8113/41/36/363001} {\bibfield  {journal} {\bibinfo
  {journal} {Journal of Physics A: Mathematical and Theoretical}\ }\textbf
  {\bibinfo {volume} {41}},\ \bibinfo {pages} {363001} (\bibinfo {year}
  {2008})}\BibitemShut {NoStop}%
\bibitem [{\citenamefont {Smalley}\ \emph {et~al.}(2002)\citenamefont
  {Smalley}, \citenamefont {Leonardi}, \citenamefont {Antonia}, \citenamefont
  {Djenidi},\ and\ \citenamefont {Orlandi}}]{Smalley2002}%
  \BibitemOpen
  \bibfield  {author} {\bibinfo {author} {\bibfnamefont {R.}~\bibnamefont
  {Smalley}}, \bibinfo {author} {\bibfnamefont {S.}~\bibnamefont {Leonardi}},
  \bibinfo {author} {\bibfnamefont {R.}~\bibnamefont {Antonia}}, \bibinfo
  {author} {\bibfnamefont {L.}~\bibnamefont {Djenidi}}, \ and\ \bibinfo
  {author} {\bibfnamefont {P.}~\bibnamefont {Orlandi}},\ }\href {\doibase
  10.1007/s00348-002-0466-z} {\bibfield  {journal} {\bibinfo  {journal}
  {Experiments in Fluids}\ }\textbf {\bibinfo {volume} {33}},\ \bibinfo {pages}
  {31} (\bibinfo {year} {2002})}\BibitemShut {NoStop}%
\bibitem [{\citenamefont {Liu}\ \emph {et~al.}(2017)\citenamefont {Liu},
  \citenamefont {Yuan}, \citenamefont {Mei}, \citenamefont {Sun}, \citenamefont
  {Liu},\ and\ \citenamefont {Wang}}]{Liu2017}%
  \BibitemOpen
  \bibfield  {author} {\bibinfo {author} {\bibfnamefont {H.}~\bibnamefont
  {Liu}}, \bibinfo {author} {\bibfnamefont {R.}~\bibnamefont {Yuan}}, \bibinfo
  {author} {\bibfnamefont {J.}~\bibnamefont {Mei}}, \bibinfo {author}
  {\bibfnamefont {J.}~\bibnamefont {Sun}}, \bibinfo {author} {\bibfnamefont
  {Q.}~\bibnamefont {Liu}}, \ and\ \bibinfo {author} {\bibfnamefont
  {Y.}~\bibnamefont {Wang}},\ }\href {\doibase 10.1007/s10546-017-0272-z}
  {\bibfield  {journal} {\bibinfo  {journal} {Boundary-Layer Meteorology}\
  }\textbf {\bibinfo {volume} {165}},\ \bibinfo {pages} {277} (\bibinfo {year}
  {2017})}\BibitemShut {NoStop}%
\bibitem [{\citenamefont {Sarkar}\ and\ \citenamefont
  {Speziale}(1990)}]{Sarkar1990}%
  \BibitemOpen
  \bibfield  {author} {\bibinfo {author} {\bibfnamefont {S.}~\bibnamefont
  {Sarkar}}\ and\ \bibinfo {author} {\bibfnamefont {C.~G.}\ \bibnamefont
  {Speziale}},\ }\href {\doibase 10.1063/1.857694} {\bibfield  {journal}
  {\bibinfo  {journal} {Physics of Fluids A: Fluid Dynamics}\ }\textbf
  {\bibinfo {volume} {2}},\ \bibinfo {pages} {84} (\bibinfo {year} {1990})},\
  \Eprint {http://arxiv.org/abs/https://doi.org/10.1063/1.857694}
  {https://doi.org/10.1063/1.857694} \BibitemShut {NoStop}%
\bibitem [{\citenamefont {Panda}\ \emph {et~al.}(2017)\citenamefont {Panda},
  \citenamefont {Warrior}, \citenamefont {Maity}, \citenamefont {Mitra},\ and\
  \citenamefont {Sasmal}}]{Panda2017}%
  \BibitemOpen
  \bibfield  {author} {\bibinfo {author} {\bibfnamefont {J.~P.}\ \bibnamefont
  {Panda}}, \bibinfo {author} {\bibfnamefont {H.~V.}\ \bibnamefont {Warrior}},
  \bibinfo {author} {\bibfnamefont {S.}~\bibnamefont {Maity}}, \bibinfo
  {author} {\bibfnamefont {A.}~\bibnamefont {Mitra}}, \ and\ \bibinfo {author}
  {\bibfnamefont {K.}~\bibnamefont {Sasmal}},\ }\href {\doibase
  10.1115/1.4035467} {\bibfield  {journal} {\bibinfo  {journal} {Journal of
  Fluids Engineering}\ }\textbf {\bibinfo {volume} {139}} (\bibinfo {year}
  {2017}),\ 10.1115/1.4035467},\ \bibinfo {note} {044503}\BibitemShut {NoStop}%
\bibitem [{\citenamefont {Schmitt}(2007)}]{Schmidt2007}%
  \BibitemOpen
  \bibfield  {author} {\bibinfo {author} {\bibfnamefont {F.~G.}\ \bibnamefont
  {Schmitt}},\ }\href {\doibase https://doi.org/10.1016/j.cnsns.2006.01.015}
  {\bibfield  {journal} {\bibinfo  {journal} {Communications in Nonlinear
  Science and Numerical Simulation}\ }\textbf {\bibinfo {volume} {12}},\
  \bibinfo {pages} {1251 } (\bibinfo {year} {2007})}\BibitemShut {NoStop}%
\bibitem [{\citenamefont {Lumley}\ and\ \citenamefont
  {Newman}(1977)}]{lumley1977}%
  \BibitemOpen
  \bibfield  {author} {\bibinfo {author} {\bibfnamefont {J.~L.}\ \bibnamefont
  {Lumley}}\ and\ \bibinfo {author} {\bibfnamefont {G.~R.}\ \bibnamefont
  {Newman}},\ }\href {\doibase 10.1017/S0022112077000585} {\bibfield  {journal}
  {\bibinfo  {journal} {Journal of Fluid Mechanics}\ }\textbf {\bibinfo
  {volume} {82}} (\bibinfo {year} {1977}),\
  10.1017/S0022112077000585}\BibitemShut {NoStop}%
\bibitem [{\citenamefont {Antonia}\ \emph {et~al.}(1991)\citenamefont
  {Antonia}, \citenamefont {Kim},\ and\ \citenamefont {Browne}}]{antonia1991}%
  \BibitemOpen
  \bibfield  {author} {\bibinfo {author} {\bibfnamefont {R.~A.}\ \bibnamefont
  {Antonia}}, \bibinfo {author} {\bibfnamefont {J.}~\bibnamefont {Kim}}, \ and\
  \bibinfo {author} {\bibfnamefont {L.~W.~B.}\ \bibnamefont {Browne}},\ }\href
  {\doibase 10.1017/S0022112091000526} {\bibfield  {journal} {\bibinfo
  {journal} {Journal of Fluid Mechanics}\ }\textbf {\bibinfo {volume} {233}}
  (\bibinfo {year} {1991}),\ 10.1017/S0022112091000526}\BibitemShut {NoStop}%
\bibitem [{\citenamefont {Krogstad}\ and\ \citenamefont
  {Torbergsen}(2000)}]{Krogstad2000}%
  \BibitemOpen
  \bibfield  {author} {\bibinfo {author} {\bibfnamefont {P.-A.}\ \bibnamefont
  {Krogstad}}\ and\ \bibinfo {author} {\bibfnamefont {L.~E.}\ \bibnamefont
  {Torbergsen}},\ }\href {\doibase 10.1023/A:1009996021533} {\bibfield
  {journal} {\bibinfo  {journal} {Flow, Turbulence and Combustion}\ }\textbf
  {\bibinfo {volume} {64}},\ \bibinfo {pages} {161} (\bibinfo {year}
  {2000})}\BibitemShut {NoStop}%
\bibitem [{\citenamefont {Katul}\ \emph {et~al.}(2001)\citenamefont {Katul},
  \citenamefont {Vidakovic},\ and\ \citenamefont {Albertson}}]{Katul2001}%
  \BibitemOpen
  \bibfield  {author} {\bibinfo {author} {\bibfnamefont {G.}~\bibnamefont
  {Katul}}, \bibinfo {author} {\bibfnamefont {B.}~\bibnamefont {Vidakovic}}, \
  and\ \bibinfo {author} {\bibfnamefont {J.}~\bibnamefont {Albertson}},\ }\href
  {\doibase 10.1063/1.1324706} {\bibfield  {journal} {\bibinfo  {journal}
  {Physics of Fluids}\ }\textbf {\bibinfo {volume} {13}},\ \bibinfo {pages}
  {241} (\bibinfo {year} {2001})},\ \Eprint
  {http://arxiv.org/abs/https://doi.org/10.1063/1.1324706}
  {https://doi.org/10.1063/1.1324706} \BibitemShut {NoStop}%
\bibitem [{\citenamefont {Katul}\ \emph {et~al.}(2009)\citenamefont {Katul},
  \citenamefont {Porporato},\ and\ \citenamefont {Poggi}}]{Katul2009}%
  \BibitemOpen
  \bibfield  {author} {\bibinfo {author} {\bibfnamefont {G.~G.}\ \bibnamefont
  {Katul}}, \bibinfo {author} {\bibfnamefont {A.}~\bibnamefont {Porporato}}, \
  and\ \bibinfo {author} {\bibfnamefont {D.}~\bibnamefont {Poggi}},\ }\href
  {\doibase 10.1063/1.3097005} {\bibfield  {journal} {\bibinfo  {journal}
  {Physics of Fluids}\ }\textbf {\bibinfo {volume} {21}},\ \bibinfo {pages}
  {035106} (\bibinfo {year} {2009})},\ \Eprint
  {http://arxiv.org/abs/https://doi.org/10.1063/1.3097005}
  {https://doi.org/10.1063/1.3097005} \BibitemShut {NoStop}%
\bibitem [{\citenamefont {Shi}\ \emph {et~al.}(2005)\citenamefont {Shi},
  \citenamefont {Vidakovic}, \citenamefont {Katul},\ and\ \citenamefont
  {Albertson}}]{shi2005assessing}%
  \BibitemOpen
  \bibfield  {author} {\bibinfo {author} {\bibfnamefont {B.}~\bibnamefont
  {Shi}}, \bibinfo {author} {\bibfnamefont {B.}~\bibnamefont {Vidakovic}},
  \bibinfo {author} {\bibfnamefont {G.~G.}\ \bibnamefont {Katul}}, \ and\
  \bibinfo {author} {\bibfnamefont {J.~D.}\ \bibnamefont {Albertson}},\
  }\href@noop {} {\bibfield  {journal} {\bibinfo  {journal} {Physics of
  Fluids}\ }\textbf {\bibinfo {volume} {17}},\ \bibinfo {pages} {055104}
  (\bibinfo {year} {2005})}\BibitemShut {NoStop}%
\bibitem [{\citenamefont {Arn\`eodo}\ \emph {et~al.}(2008)\citenamefont
  {Arn\`eodo}, \citenamefont {Benzi}, \citenamefont {Berg}, \citenamefont
  {Biferale}, \citenamefont {Bodenschatz}, \citenamefont {Busse}, \citenamefont
  {Calzavarini}, \citenamefont {Castaing}, \citenamefont {Cencini},
  \citenamefont {Chevillard}, \citenamefont {Fisher}, \citenamefont {Grauer},
  \citenamefont {Homann}, \citenamefont {Lamb}, \citenamefont {Lanotte},
  \citenamefont {L\'ev\`eque}, \citenamefont {L\"uthi}, \citenamefont {Mann},
  \citenamefont {Mordant}, \citenamefont {M\"uller}, \citenamefont {Ott},
  \citenamefont {Ouellette}, \citenamefont {Pinton}, \citenamefont {Pope},
  \citenamefont {Roux}, \citenamefont {Toschi}, \citenamefont {Xu},\ and\
  \citenamefont {Yeung}}]{Arneodo08}%
  \BibitemOpen
  \bibfield  {author} {\bibinfo {author} {\bibfnamefont {A.}~\bibnamefont
  {Arn\`eodo}}, \bibinfo {author} {\bibfnamefont {R.}~\bibnamefont {Benzi}},
  \bibinfo {author} {\bibfnamefont {J.}~\bibnamefont {Berg}}, \bibinfo {author}
  {\bibfnamefont {L.}~\bibnamefont {Biferale}}, \bibinfo {author}
  {\bibfnamefont {E.}~\bibnamefont {Bodenschatz}}, \bibinfo {author}
  {\bibfnamefont {A.}~\bibnamefont {Busse}}, \bibinfo {author} {\bibfnamefont
  {E.}~\bibnamefont {Calzavarini}}, \bibinfo {author} {\bibfnamefont
  {B.}~\bibnamefont {Castaing}}, \bibinfo {author} {\bibfnamefont
  {M.}~\bibnamefont {Cencini}}, \bibinfo {author} {\bibfnamefont
  {L.}~\bibnamefont {Chevillard}}, \bibinfo {author} {\bibfnamefont {R.~T.}\
  \bibnamefont {Fisher}}, \bibinfo {author} {\bibfnamefont {R.}~\bibnamefont
  {Grauer}}, \bibinfo {author} {\bibfnamefont {H.}~\bibnamefont {Homann}},
  \bibinfo {author} {\bibfnamefont {D.}~\bibnamefont {Lamb}}, \bibinfo {author}
  {\bibfnamefont {A.~S.}\ \bibnamefont {Lanotte}}, \bibinfo {author}
  {\bibfnamefont {E.}~\bibnamefont {L\'ev\`eque}}, \bibinfo {author}
  {\bibfnamefont {B.}~\bibnamefont {L\"uthi}}, \bibinfo {author} {\bibfnamefont
  {J.}~\bibnamefont {Mann}}, \bibinfo {author} {\bibfnamefont {N.}~\bibnamefont
  {Mordant}}, \bibinfo {author} {\bibfnamefont {W.-C.}\ \bibnamefont
  {M\"uller}}, \bibinfo {author} {\bibfnamefont {S.}~\bibnamefont {Ott}},
  \bibinfo {author} {\bibfnamefont {N.~T.}\ \bibnamefont {Ouellette}}, \bibinfo
  {author} {\bibfnamefont {J.-F.}\ \bibnamefont {Pinton}}, \bibinfo {author}
  {\bibfnamefont {S.~B.}\ \bibnamefont {Pope}}, \bibinfo {author}
  {\bibfnamefont {S.~G.}\ \bibnamefont {Roux}}, \bibinfo {author}
  {\bibfnamefont {F.}~\bibnamefont {Toschi}}, \bibinfo {author} {\bibfnamefont
  {H.}~\bibnamefont {Xu}}, \ and\ \bibinfo {author} {\bibfnamefont {P.~K.}\
  \bibnamefont {Yeung}} (\bibinfo {collaboration} {International Collaboration
  for Turbulence Research}),\ }\href {\doibase 10.1103/PhysRevLett.100.254504}
  {\bibfield  {journal} {\bibinfo  {journal} {Phys. Rev. Lett.}\ }\textbf
  {\bibinfo {volume} {100}},\ \bibinfo {pages} {254504} (\bibinfo {year}
  {2008})}\BibitemShut {NoStop}%
\bibitem [{\citenamefont {Ray}\ \emph {et~al.}(2008)\citenamefont {Ray},
  \citenamefont {Mitra},\ and\ \citenamefont {Pandit}}]{Ray08}%
  \BibitemOpen
  \bibfield  {author} {\bibinfo {author} {\bibfnamefont {S.~S.}\ \bibnamefont
  {Ray}}, \bibinfo {author} {\bibfnamefont {D.}~\bibnamefont {Mitra}}, \ and\
  \bibinfo {author} {\bibfnamefont {R.}~\bibnamefont {Pandit}},\ }\href
  {\doibase 10.1088/1367-2630/10/3/033003} {\bibfield  {journal} {\bibinfo
  {journal} {New Journal of Physics}\ }\textbf {\bibinfo {volume} {10}},\
  \bibinfo {pages} {033003} (\bibinfo {year} {2008})}\BibitemShut {NoStop}%
\bibitem [{\citenamefont {Pandit}\ \emph {et~al.}(2009)\citenamefont {Pandit},
  \citenamefont {Perlekar},\ and\ \citenamefont {Ray}}]{Pandit2009}%
  \BibitemOpen
  \bibfield  {author} {\bibinfo {author} {\bibfnamefont {R.}~\bibnamefont
  {Pandit}}, \bibinfo {author} {\bibfnamefont {P.}~\bibnamefont {Perlekar}}, \
  and\ \bibinfo {author} {\bibfnamefont {S.~S.}\ \bibnamefont {Ray}},\ }\href
  {\doibase 10.1007/s12043-009-0096-6} {\bibfield  {journal} {\bibinfo
  {journal} {Pramana - J Phys}\ }\textbf {\bibinfo {volume} {224}} (\bibinfo
  {year} {2009}),\ 10.1007/s12043-009-0096-6}\BibitemShut {NoStop}%
\bibitem [{\citenamefont {Pal}\ \emph {et~al.}(2016)\citenamefont {Pal},
  \citenamefont {Perlekar}, \citenamefont {Gupta},\ and\ \citenamefont
  {Pandit}}]{nairita2016}%
  \BibitemOpen
  \bibfield  {author} {\bibinfo {author} {\bibfnamefont {N.}~\bibnamefont
  {Pal}}, \bibinfo {author} {\bibfnamefont {P.}~\bibnamefont {Perlekar}},
  \bibinfo {author} {\bibfnamefont {A.}~\bibnamefont {Gupta}}, \ and\ \bibinfo
  {author} {\bibfnamefont {R.}~\bibnamefont {Pandit}},\ }\href {\doibase
  10.1103/PhysRevE.93.063115} {\bibfield  {journal} {\bibinfo  {journal} {Phys.
  Rev. E}\ }\textbf {\bibinfo {volume} {93}},\ \bibinfo {pages} {063115}
  (\bibinfo {year} {2016})}\BibitemShut {NoStop}%
\bibitem [{\citenamefont {Zeng}\ \emph {et~al.}(2016)\citenamefont {Zeng},
  \citenamefont {Zhang}, \citenamefont {Li},\ and\ \citenamefont
  {Meng}}]{Zeng2016}%
  \BibitemOpen
  \bibfield  {author} {\bibinfo {author} {\bibfnamefont {M.}~\bibnamefont
  {Zeng}}, \bibinfo {author} {\bibfnamefont {X.-N.}\ \bibnamefont {Zhang}},
  \bibinfo {author} {\bibfnamefont {J.-h.}\ \bibnamefont {Li}}, \ and\ \bibinfo
  {author} {\bibfnamefont {Q.-h.}\ \bibnamefont {Meng}},\ }\href {\doibase
  10.5506/APhysPolB.47.2205} {\bibfield  {journal} {\bibinfo  {journal} {ACTA
  Physica Polonica B}\ }\textbf {\bibinfo {volume} {47}},\ \bibinfo {pages}
  {2205} (\bibinfo {year} {2016})}\BibitemShut {NoStop}%
\bibitem [{\citenamefont {Buldyrev}\ \emph {et~al.}(1998)\citenamefont
  {Buldyrev}, \citenamefont {Dokholyan}, \citenamefont {Goldberger},
  \citenamefont {Havlin}, \citenamefont {Peng}, \citenamefont {Stanley},\ and\
  \citenamefont {Viswanathan}}]{DNAseq}%
  \BibitemOpen
  \bibfield  {author} {\bibinfo {author} {\bibfnamefont {S.}~\bibnamefont
  {Buldyrev}}, \bibinfo {author} {\bibfnamefont {N.}~\bibnamefont {Dokholyan}},
  \bibinfo {author} {\bibfnamefont {A.}~\bibnamefont {Goldberger}}, \bibinfo
  {author} {\bibfnamefont {S.}~\bibnamefont {Havlin}}, \bibinfo {author}
  {\bibfnamefont {C.-K.}\ \bibnamefont {Peng}}, \bibinfo {author}
  {\bibfnamefont {H.}~\bibnamefont {Stanley}}, \ and\ \bibinfo {author}
  {\bibfnamefont {G.}~\bibnamefont {Viswanathan}},\ }\href@noop {} {\bibfield
  {journal} {\bibinfo  {journal} {Physica A: Statistical Mechanics and its
  Applications}\ }\textbf {\bibinfo {volume} {249}},\ \bibinfo {pages} {430}
  (\bibinfo {year} {1998})}\BibitemShut {NoStop}%
\bibitem [{\citenamefont {J.~Gieraltowski}\ and\ \citenamefont
  {Baranowski}(2012)}]{Gieraltowski2012}%
  \BibitemOpen
  \bibfield  {author} {\bibinfo {author} {\bibfnamefont {J.~J.~Z.}\
  \bibnamefont {J.~Gieraltowski}}\ and\ \bibinfo {author} {\bibfnamefont
  {R.}~\bibnamefont {Baranowski}},\ }\href {\doibase
  10.1103/PhysRevE.85.021915} {\bibfield  {journal} {\bibinfo  {journal}
  {Physical Review E}\ }\textbf {\bibinfo {volume} {021915}},\ \bibinfo {pages}
  {1} (\bibinfo {year} {2012})}\BibitemShut {NoStop}%
\bibitem [{\citenamefont {Ivanov}\ \emph {et~al.}(1999)\citenamefont {Ivanov},
  \citenamefont {Amaral},\ and\ \citenamefont {Stanley}}]{Ivanov1999}%
  \BibitemOpen
  \bibfield  {author} {\bibinfo {author} {\bibfnamefont {P.~C.}\ \bibnamefont
  {Ivanov}}, \bibinfo {author} {\bibfnamefont {A.~N.}\ \bibnamefont {Amaral}},
  \ and\ \bibinfo {author} {\bibfnamefont {H.~E.}\ \bibnamefont {Stanley}},\
  }\href@noop {} {\bibfield  {journal} {\bibinfo  {journal} {Nature}\ }\textbf
  {\bibinfo {volume} {399}},\ \bibinfo {pages} {461} (\bibinfo {year}
  {1999})}\BibitemShut {NoStop}%
\bibitem [{\citenamefont {Ivanova}\ \emph {et~al.}(2000)\citenamefont
  {Ivanova}, \citenamefont {Ausloos}, \citenamefont {Clothiaux},\ and\
  \citenamefont {Ackerman}}]{cloudmfdfa}%
  \BibitemOpen
  \bibfield  {author} {\bibinfo {author} {\bibfnamefont {K.}~\bibnamefont
  {Ivanova}}, \bibinfo {author} {\bibfnamefont {M.}~\bibnamefont {Ausloos}},
  \bibinfo {author} {\bibfnamefont {E.~E.}\ \bibnamefont {Clothiaux}}, \ and\
  \bibinfo {author} {\bibfnamefont {T.~P.}\ \bibnamefont {Ackerman}},\ }\href
  {\doibase 10.1209/epl/i2000-00401-5} {\bibfield  {journal} {\bibinfo
  {journal} {Europhysics Letters ({EPL})}\ }\textbf {\bibinfo {volume} {52}},\
  \bibinfo {pages} {40} (\bibinfo {year} {2000})}\BibitemShut {NoStop}%
\bibitem [{\citenamefont {N.~Mantegna}\ and\ \citenamefont
  {Stanley}(1999)}]{ecophy1999}%
  \BibitemOpen
  \bibfield  {author} {\bibinfo {author} {\bibfnamefont {R.}~\bibnamefont
  {N.~Mantegna}}\ and\ \bibinfo {author} {\bibfnamefont {H.~E.}\ \bibnamefont
  {Stanley}},\ }\href@noop {} {\enquote {\bibinfo {title} {An introduction to
  econophysics: correlations and complexity in finance},}\ } (\bibinfo {year}
  {1999})\BibitemShut {NoStop}%
\bibitem [{\citenamefont {Amin}\ \emph {et~al.}(2018)\citenamefont {Amin},
  \citenamefont {Ray}, \citenamefont {Pal}, \citenamefont {Pandit},\ and\
  \citenamefont {Bid}}]{Amin}%
  \BibitemOpen
  \bibfield  {author} {\bibinfo {author} {\bibfnamefont {K.~R.}\ \bibnamefont
  {Amin}}, \bibinfo {author} {\bibfnamefont {S.~S.}\ \bibnamefont {Ray}},
  \bibinfo {author} {\bibfnamefont {N.}~\bibnamefont {Pal}}, \bibinfo {author}
  {\bibfnamefont {R.}~\bibnamefont {Pandit}}, \ and\ \bibinfo {author}
  {\bibfnamefont {A.}~\bibnamefont {Bid}},\ }\href {\doibase
  10.1038/s42005-017-0001-4} {\bibfield  {journal} {\bibinfo  {journal}
  {Communications Physics}\ }\textbf {\bibinfo {volume} {1}},\ \bibinfo {pages}
  {1} (\bibinfo {year} {2018})}\BibitemShut {NoStop}%
\bibitem [{\citenamefont {Sutradhar}\ \emph {et~al.}(2019)\citenamefont
  {Sutradhar}, \citenamefont {Mukerjee}, \citenamefont {Pandit},\ and\
  \citenamefont {Banerjee}}]{Sutradhar19}%
  \BibitemOpen
  \bibfield  {author} {\bibinfo {author} {\bibfnamefont {J.}~\bibnamefont
  {Sutradhar}}, \bibinfo {author} {\bibfnamefont {S.}~\bibnamefont {Mukerjee}},
  \bibinfo {author} {\bibfnamefont {R.}~\bibnamefont {Pandit}}, \ and\ \bibinfo
  {author} {\bibfnamefont {S.}~\bibnamefont {Banerjee}},\ }\href {\doibase
  10.1103/PhysRevB.99.224204} {\bibfield  {journal} {\bibinfo  {journal} {Phys.
  Rev. B}\ }\textbf {\bibinfo {volume} {99}},\ \bibinfo {pages} {224204}
  (\bibinfo {year} {2019})}\BibitemShut {NoStop}%
\bibitem [{\citenamefont {Ihlen}(2012)}]{Ihlen2012}%
  \BibitemOpen
  \bibfield  {author} {\bibinfo {author} {\bibfnamefont {E.~A.~F.}\
  \bibnamefont {Ihlen}},\ }\href {\doibase 10.3389/fphys.2012.00141} {\bibfield
   {journal} {\bibinfo  {journal} {Frontiers in Physiology}\ }\textbf {\bibinfo
  {volume} {3}},\ \bibinfo {pages} {1} (\bibinfo {year} {2012})}\BibitemShut
  {NoStop}%
\bibitem [{\citenamefont {Kantelhardt}\ \emph {et~al.}(2002)\citenamefont
  {Kantelhardt}, \citenamefont {Zschiegner}, \citenamefont {Koscielny-bunde},
  \citenamefont {Havlin}, \citenamefont {Bunde},\ and\ \citenamefont
  {Stanley}}]{Kantelhardt2002}%
  \BibitemOpen
  \bibfield  {author} {\bibinfo {author} {\bibfnamefont {J.~W.}\ \bibnamefont
  {Kantelhardt}}, \bibinfo {author} {\bibfnamefont {S.~A.}\ \bibnamefont
  {Zschiegner}}, \bibinfo {author} {\bibfnamefont {E.}~\bibnamefont
  {Koscielny-bunde}}, \bibinfo {author} {\bibfnamefont {S.}~\bibnamefont
  {Havlin}}, \bibinfo {author} {\bibfnamefont {A.}~\bibnamefont {Bunde}}, \
  and\ \bibinfo {author} {\bibfnamefont {H.~E.}\ \bibnamefont {Stanley}},\
  }\href@noop {} {\bibfield  {journal} {\bibinfo  {journal} {Physica A}\
  }\textbf {\bibinfo {volume} {316}},\ \bibinfo {pages} {87} (\bibinfo {year}
  {2002})}\BibitemShut {NoStop}%
\bibitem [{\citenamefont {Katul}\ \emph {et~al.}(2013)\citenamefont {Katul},
  \citenamefont {Porporato}, \citenamefont {Manes},\ and\ \citenamefont
  {Meneveau}}]{katul2013co}%
  \BibitemOpen
  \bibfield  {author} {\bibinfo {author} {\bibfnamefont {G.~G.}\ \bibnamefont
  {Katul}}, \bibinfo {author} {\bibfnamefont {A.}~\bibnamefont {Porporato}},
  \bibinfo {author} {\bibfnamefont {C.}~\bibnamefont {Manes}}, \ and\ \bibinfo
  {author} {\bibfnamefont {C.}~\bibnamefont {Meneveau}},\ }\href@noop {}
  {\bibfield  {journal} {\bibinfo  {journal} {Physics of Fluids}\ }\textbf
  {\bibinfo {volume} {25}},\ \bibinfo {pages} {091702} (\bibinfo {year}
  {2013})}\BibitemShut {NoStop}%
\bibitem [{\citenamefont {Katul}\ \emph {et~al.}(2014)\citenamefont {Katul},
  \citenamefont {Porporato}, \citenamefont {Shah},\ and\ \citenamefont
  {Bou-Zeid}}]{katul2014two}%
  \BibitemOpen
  \bibfield  {author} {\bibinfo {author} {\bibfnamefont {G.~G.}\ \bibnamefont
  {Katul}}, \bibinfo {author} {\bibfnamefont {A.}~\bibnamefont {Porporato}},
  \bibinfo {author} {\bibfnamefont {S.}~\bibnamefont {Shah}}, \ and\ \bibinfo
  {author} {\bibfnamefont {E.}~\bibnamefont {Bou-Zeid}},\ }\href@noop {}
  {\bibfield  {journal} {\bibinfo  {journal} {Physical Review E}\ }\textbf
  {\bibinfo {volume} {89}},\ \bibinfo {pages} {023007} (\bibinfo {year}
  {2014})}\BibitemShut {NoStop}%
\bibitem [{\citenamefont {Saddoughi}\ and\ \citenamefont
  {Veeravalli}(1994)}]{saddoughi_veeravalli_1994}%
  \BibitemOpen
  \bibfield  {author} {\bibinfo {author} {\bibfnamefont {S.~G.}\ \bibnamefont
  {Saddoughi}}\ and\ \bibinfo {author} {\bibfnamefont {S.~V.}\ \bibnamefont
  {Veeravalli}},\ }\href {\doibase 10.1017/S0022112094001370} {\bibfield
  {journal} {\bibinfo  {journal} {Journal of Fluid Mechanics}\ }\textbf
  {\bibinfo {volume} {268}},\ \bibinfo {pages} {333–372} (\bibinfo {year}
  {1994})}\BibitemShut {NoStop}%
\bibitem [{\citenamefont {Cava}\ and\ \citenamefont {Katul}(2012)}]{Cava2012}%
  \BibitemOpen
  \bibfield  {author} {\bibinfo {author} {\bibfnamefont {D.}~\bibnamefont
  {Cava}}\ and\ \bibinfo {author} {\bibfnamefont {G.~G.}\ \bibnamefont
  {Katul}},\ }\href {\doibase 10.1007/s10546-012-9737-2} {\bibfield  {journal}
  {\bibinfo  {journal} {Boundary-Layer Meteorology}\ }\textbf {\bibinfo
  {volume} {145}},\ \bibinfo {pages} {351} (\bibinfo {year}
  {2012})}\BibitemShut {NoStop}%
\bibitem [{\citenamefont {Davidson}\ \emph {et~al.}(2006)\citenamefont
  {Davidson}, \citenamefont {Nickels},\ and\ \citenamefont
  {Krogstad}}]{Davidson2006}%
  \BibitemOpen
  \bibfield  {author} {\bibinfo {author} {\bibfnamefont {P.~A.}\ \bibnamefont
  {Davidson}}, \bibinfo {author} {\bibfnamefont {T.~B.}\ \bibnamefont
  {Nickels}}, \ and\ \bibinfo {author} {\bibfnamefont {P.-A.}\ \bibnamefont
  {Krogstad}},\ }\href {\doibase 10.1017/S0022112005008001} {\bibfield
  {journal} {\bibinfo  {journal} {Journal of Fluid Mechanics}\ }\textbf
  {\bibinfo {volume} {550}},\ \bibinfo {pages} {51–60} (\bibinfo {year}
  {2006})}\BibitemShut {NoStop}%
\bibitem [{\citenamefont {Hsieh}\ and\ \citenamefont
  {Katul}(1997)}]{hsieh1997dissipation}%
  \BibitemOpen
  \bibfield  {author} {\bibinfo {author} {\bibfnamefont {C.-I.}\ \bibnamefont
  {Hsieh}}\ and\ \bibinfo {author} {\bibfnamefont {G.~G.}\ \bibnamefont
  {Katul}},\ }\href@noop {} {\bibfield  {journal} {\bibinfo  {journal} {Journal
  of Geophysical Research: Atmospheres}\ }\textbf {\bibinfo {volume} {102}},\
  \bibinfo {pages} {16391} (\bibinfo {year} {1997})}\BibitemShut {NoStop}%
\bibitem [{\citenamefont {Cava}\ and\ \citenamefont {Katul}(2009)}]{Cava2009}%
  \BibitemOpen
  \bibfield  {author} {\bibinfo {author} {\bibfnamefont {D.}~\bibnamefont
  {Cava}}\ and\ \bibinfo {author} {\bibfnamefont {G.~G.}\ \bibnamefont
  {Katul}},\ }\href {\doibase 10.1007/s10546-008-9342-6} {\bibfield  {journal}
  {\bibinfo  {journal} {Boundary-Layer Meteorology}\ }\textbf {\bibinfo
  {volume} {130}} (\bibinfo {year} {2009}),\
  10.1007/s10546-008-9342-6}\BibitemShut {NoStop}%
\bibitem [{\citenamefont {Sreenivasan}\ and\ \citenamefont
  {Bershadskii}(2006)}]{Sreenivasan2006}%
  \BibitemOpen
  \bibfield  {author} {\bibinfo {author} {\bibfnamefont {K.~R.}\ \bibnamefont
  {Sreenivasan}}\ and\ \bibinfo {author} {\bibfnamefont {A.}~\bibnamefont
  {Bershadskii}},\ }\href {\doibase 10.1007/s10955-006-9112-0} {\bibfield
  {journal} {\bibinfo  {journal} {Journal of Statistical Physics}\ }\textbf
  {\bibinfo {volume} {125}},\ \bibinfo {pages} {1141–1153} (\bibinfo {year}
  {2006})}\BibitemShut {NoStop}%
\bibitem [{\citenamefont {Meneveau}\ and\ \citenamefont
  {Sreenivasan}(1987)}]{Meneveau1987}%
  \BibitemOpen
  \bibfield  {author} {\bibinfo {author} {\bibfnamefont {C.}~\bibnamefont
  {Meneveau}}\ and\ \bibinfo {author} {\bibfnamefont {K.}~\bibnamefont
  {Sreenivasan}},\ }\href {\doibase
  https://doi.org/10.1016/0920-5632(87)90008-9} {\bibfield  {journal} {\bibinfo
   {journal} {Nuclear Physics B - Proceedings Supplements}\ }\textbf {\bibinfo
  {volume} {2}},\ \bibinfo {pages} {49 } (\bibinfo {year} {1987})}\BibitemShut
  {NoStop}%
\bibitem [{\citenamefont {She}\ and\ \citenamefont
  {Leveque}(1994)}]{SheLeveque}%
  \BibitemOpen
  \bibfield  {author} {\bibinfo {author} {\bibfnamefont {Z.-S.}\ \bibnamefont
  {She}}\ and\ \bibinfo {author} {\bibfnamefont {E.}~\bibnamefont {Leveque}},\
  }\href {\doibase 10.1103/PhysRevLett.72.336} {\bibfield  {journal} {\bibinfo
  {journal} {Phys. Rev. Lett.}\ }\textbf {\bibinfo {volume} {72}},\ \bibinfo
  {pages} {336} (\bibinfo {year} {1994})}\BibitemShut {NoStop}%
\bibitem [{\citenamefont {Benzi}\ \emph {et~al.}(1993)\citenamefont {Benzi},
  \citenamefont {Ciliberto}, \citenamefont {Tripiccione}, \citenamefont
  {Baudet}, \citenamefont {Massaioli},\ and\ \citenamefont {Succi}}]{BenziESS}%
  \BibitemOpen
  \bibfield  {author} {\bibinfo {author} {\bibfnamefont {R.}~\bibnamefont
  {Benzi}}, \bibinfo {author} {\bibfnamefont {S.}~\bibnamefont {Ciliberto}},
  \bibinfo {author} {\bibfnamefont {R.}~\bibnamefont {Tripiccione}}, \bibinfo
  {author} {\bibfnamefont {C.}~\bibnamefont {Baudet}}, \bibinfo {author}
  {\bibfnamefont {F.}~\bibnamefont {Massaioli}}, \ and\ \bibinfo {author}
  {\bibfnamefont {S.}~\bibnamefont {Succi}},\ }\href {\doibase
  10.1103/PhysRevE.48.R29} {\bibfield  {journal} {\bibinfo  {journal} {Phys.
  Rev. E}\ }\textbf {\bibinfo {volume} {48}},\ \bibinfo {pages} {R29} (\bibinfo
  {year} {1993})}\BibitemShut {NoStop}%
\bibitem [{\citenamefont {Chakraborty}\ \emph {et~al.}(2009)\citenamefont
  {Chakraborty}, \citenamefont {Frisch},\ and\ \citenamefont
  {Ray}}]{Chakraborty2009}%
  \BibitemOpen
  \bibfield  {author} {\bibinfo {author} {\bibfnamefont {S.}~\bibnamefont
  {Chakraborty}}, \bibinfo {author} {\bibfnamefont {U.}~\bibnamefont {Frisch}},
  \ and\ \bibinfo {author} {\bibfnamefont {S.~S.}\ \bibnamefont {Ray}},\
  }\href@noop {} {\  (\bibinfo {year} {2009})},\ \Eprint
  {http://arxiv.org/abs/0912.2406} {arXiv:0912.2406 [nlin.CD]} \BibitemShut
  {NoStop}%
\bibitem [{\citenamefont {Arad}\ \emph {et~al.}(1999)\citenamefont {Arad},
  \citenamefont {Biferale}, \citenamefont {Mazzitelli},\ and\ \citenamefont
  {Procaccia}}]{Arad1999}%
  \BibitemOpen
  \bibfield  {author} {\bibinfo {author} {\bibfnamefont {I.}~\bibnamefont
  {Arad}}, \bibinfo {author} {\bibfnamefont {L.}~\bibnamefont {Biferale}},
  \bibinfo {author} {\bibfnamefont {I.}~\bibnamefont {Mazzitelli}}, \ and\
  \bibinfo {author} {\bibfnamefont {I.}~\bibnamefont {Procaccia}},\ }\href
  {\doibase 10.1103/PhysRevLett.82.5040} {\bibfield  {journal} {\bibinfo
  {journal} {Phys. Rev. Lett.}\ }\textbf {\bibinfo {volume} {82}},\ \bibinfo
  {pages} {5040} (\bibinfo {year} {1999})}\BibitemShut {NoStop}%
\bibitem [{\citenamefont {Biferale}\ and\ \citenamefont
  {Procaccia}(2005)}]{Biferale05}%
  \BibitemOpen
  \bibfield  {author} {\bibinfo {author} {\bibfnamefont {L.}~\bibnamefont
  {Biferale}}\ and\ \bibinfo {author} {\bibfnamefont {I.}~\bibnamefont
  {Procaccia}},\ }\href {\doibase
  https://doi.org/10.1016/j.physrep.2005.04.001} {\bibfield  {journal}
  {\bibinfo  {journal} {Physics Reports}\ }\textbf {\bibinfo {volume} {414}},\
  \bibinfo {pages} {43 } (\bibinfo {year} {2005})}\BibitemShut {NoStop}%
\bibitem [{vid()}]{vid:stable_anisotropy_23}%
  \BibitemOpen
  \href@noop {} {\bibinfo  {journal} {See the movie in supplementary
  materials}\ }\BibitemShut {NoStop}%
\bibitem [{\citenamefont {Campbell}(2020)}]{shadedEb}%
  \BibitemOpen
\bibfield  {journal} {  }\bibfield  {author} {\bibinfo {author} {\bibfnamefont
  {R.}~\bibnamefont {Campbell}},\ }\href
  {https://github.com/raacampbell/shadedErrorBar), GitHub} {\  (\bibinfo {year}
  {2020})}\BibitemShut {NoStop}%
\end{thebibliography}%

\end{document}